\documentclass[aps,twocolumn,floatfix]{revtex4-2}
\usepackage[utf8]{inputenc}
\usepackage[T1]{fontenc}
\usepackage[english]{babel}
\usepackage{mathtools}
\usepackage[scaled=1.005]{newtxtext}
\usepackage{amsmath}
\usepackage{amsfonts}
\usepackage[scaled=1.005,varbb,varg]{newtxmath} 
\usepackage{bm,dsfont,eucal}

\AtBeginDocument{\renewcommand{\vec}[1]{\bm{#1}}}

\usepackage[dvipsnames,svgnames,x11names,hyperref]{xcolor}
\usepackage[hypertexnames=false]{hyperref}
\hypersetup{ 
    colorlinks=true, linkcolor=NavyBlue, urlcolor=NavyBlue, citecolor=NavyBlue
}

\usepackage{physics,braket,siunitx,slashed}

\begin{document}

\title{Singular Weak-Field Thermodynamics of 2D Superconductors}

\author{Guopeng Xu and Chunli Huang}
\affiliation{Department of Physics and Astronomy, University of Kentucky,  Lexington, Kentucky 40506-0055, USA}

\date{\today}

\begin{abstract}
In a bulk 3D type-II superconductor, the lower critical field at which an isolated vortex becomes thermodynamically favorable is a size-independent material property. We show that the situation is  different in 2D superconductors: the larger the superconductor, the weaker the field needed to create its first vortex. The  lower critical field in 2D is always size-dependent. For a disk of area $\mathcal A$, the lower critical field $B_v(\mathcal A)$ scales as $\mathcal A^{-1}\ln(\mathcal A/\mathcal A_0)$ in the weak-screening regime and as $\mathcal A^{-1/2}$ in the strong-screening regime. We derive these results from an analytically tractable microscopic model that admits many-body wavefunctions for both the uniform and singly quantized vortex states in a magnetic field, and incorporate screening by coupling their long-distance 2D supercurrents to 3D Maxwell equations. These results motivate organizing the weak-field ground-state of a 2D superconductor in the $(1/\mathcal A,B)$ plane. The origin represents the zero-field thermodynamic limit and it is singular. Approaching the origin along the $B$ axis leads to an increasingly dilute vortex lattice, whereas approaching along the $1/\mathcal A$ axis yields the uniform vortex-free state. Our theory shows that every trajectory carrying fixed finite flux ultimately approaches the vortex-free state in the thermodynamic limit and provides a firm microscopic foundation for the weak-field thermodynamics of 2D superconductors.
\end{abstract}


\maketitle 
\addtocontents{toc}{\protect\setcounter{tocdepth}{-10}}


\section{Introduction}

The Meissner effect, the expulsion of magnetic flux from the superconducting bulk, is a defining property of superconductivity. In a bulk 3D type-II superconductor, the vortex-free state remains thermodynamically favorable below the lower critical field $H_{c1}$, an intrinsic material scale determined by the London penetration depth and coherence length. This familiar picture relies on bulk screening, which confines the magnetic field and supercurrent of a vortex and makes its energy independent of the sample size in the bulk limit.

The electrodynamics of an atomically thin superconductor is fundamentally different. Unlike in a bulk material, every electron resides at the surface and is directly exposed to the applied electromagnetic environment; there is no deep interior shielded by a macroscopic superconducting bulk. The screening current is confined to the material plane, whereas the magnetic field it generates extends through the surrounding 3D space and decays algebraically rather than exponentially~\cite{Pearl1964}. Consequently, the sample boundary can remain relevant to the weak-field response even when the system is much larger than the screening length. This situation is realized in superconducting graphene moiré systems~\cite{Cao2018,Park2021,Matthew2019,Hoke2026,Stepanov2020,Arora2020,Oh2021,zhang2026imagingmeissnereffectlocal,Yiran2022,Kim2026,Zhang2026}, transition-metal dichalcogenides~\cite{Xia2025,Guo2025,Guo2026,delgado2026recordmagnetoresistanceenhancedsuperconductivity,xu2026signaturesunconventionalsuperconductivitynear}, and rhombohedral multilayer graphene~\cite{Choi2025,Han2025,Zhou2021,Kumar2026,hua2026multiknobswitchablechiralsuperconductivity,yang2026magneticfieldenhancedgraphenesuperconductivity,Seo_2026,dutta2026reconfigurablechiralsuperconductivity,kumar2026pervasivespintripletsuperconductivityrhombohedral,deng2026magneticfieldinducedsuperconductivityhexalayerrhombohedral,deng2026superconductivityferroelectricorbitalmagnetism,nguyen2026coexistingchargedensitywave,qin2026stripeordermetallicsuperconducting,Li2024}. In particular, Ref.~\cite{Han2025,hua2026multiknobswitchablechiralsuperconductivity,dutta2026reconfigurablechiralsuperconductivity,sheekey2026visualizing} reported superconductivity with spontaneous orbital magnetization in rhombohedral multilayer graphene.

This raises a basic thermodynamic question: does a 2D superconductor possess a unique zero-field thermodynamic limit? Or, do the limits $B\to0$ and $\mathcal A\to\infty$ commute? 
Consider the plane $(x,y)=(1/\mathcal A,B)$, where $\mathcal A$ is the
sample area and $B$ is the perpendicular magnetic field (Fig.~1). The origin corresponds to
the zero-field thermodynamic limit. Approaching the origin along the $y$ axis, by first taking $\mathcal A\to\infty$ at nonzero $B$, yields
an increasingly dilute vortex lattice. Approaching along the $x$ axis, by
setting $B=0$ before taking $\mathcal A\to\infty$, instead yields a uniform,
vortex-free condensate. Thus, distinct approaches to the same limiting
point lead to topologically distinct superconducting ground states.

\begin{figure}[t]
    \centering
    \includegraphics[width=0.85\linewidth]{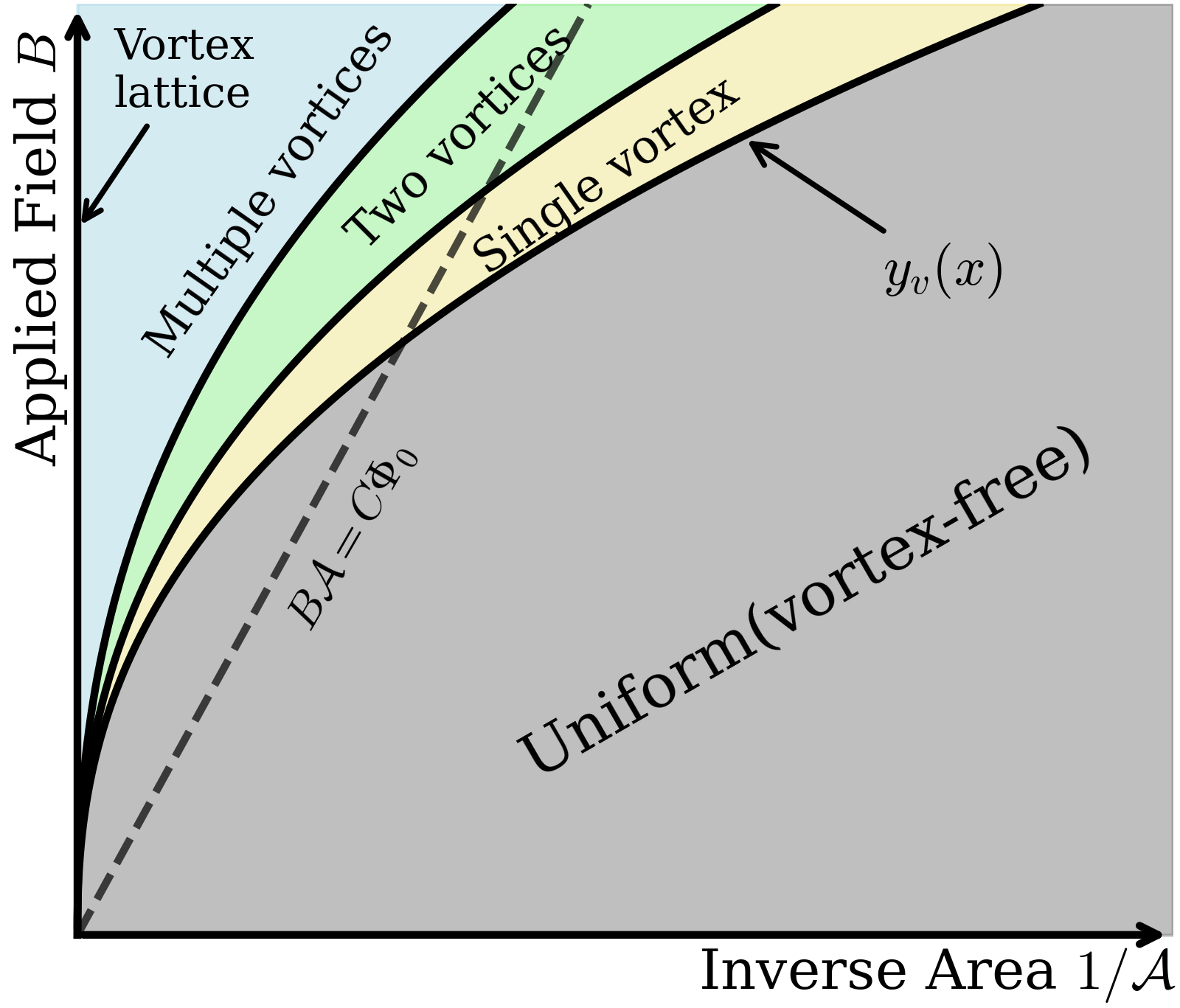}
    \caption{Ground state of a 2D superconductor in the weak-field thermodynamic limit. The solid boundary separating the vortex-free and single-vortex
states is described by Eq.~\eqref{eq:main_equation}, while the remaining boundaries
are inferred schematically by continuity. The black dashed line represents a fixed-flux trajectory where $\Phi_0=h/2e$.
    }
    \label{fig:main_schematic_phase_diagram}
\end{figure}

In this Letter, we provide a generic finite-size weak-field magnetic phase diagram of a 2D superconductor, Fig.~1. We show that the first-vortex boundary approaches the origin nonanalytically as
\begin{equation}
\label{eq:main_equation}
y_v(x)=
\begin{cases}
\displaystyle
c_w \, x
\ln\left(\frac{ x_0}{ x}\right)
& \; \text{weak screening},\\[1em]
\displaystyle
c_s \sqrt{x}
& \; \text{strong screening}.
\end{cases}
\end{equation}
Here, $c_w$, $c_s$, and $x_0$ are positive nonuniversal parameters independent of $\mathcal A$. The screening crossover is controlled by the ratio $\mathcal A/(\pi\Lambda^2)$, where $\Lambda=\Phi_0^2/( 2\pi^2\mu_0\rho_b)$ is the Pearl length. Strong screening means $\frac{\mathcal A}{\pi\Lambda^2}\gg1$, while weak screening means $\frac{\mathcal A}{\pi\Lambda^2}\ll1$. For the typical superfluid stiffness characteristic of many atomically thin superconductors \cite{zhang2026imaging,banerjee2025superfluid},
$\rho_b\sim0.1~\mathrm{meV}$, the Pearl length can reach $\Lambda\sim1~\mathrm{cm}$, much larger than the typical device dimensions of $1$--$10~\mu\mathrm{m}$. Most superconducting devices made of 2D materials therefore lie deep in the weak-screening regime, even though the strong-screening form determines the ultimate thermodynamic limit.
Because $y_v(x)/x$ diverges as $x\rightarrow0$, the vortex nucleation boundary eventually lies above every trajectory with finite slope. Consequently, an arbitrarily weak magnetic field need not create vortices: when the thermodynamic limit is taken at fixed total flux, the ground state remains vortex free. We now derive these results, beginning with a microscopic theory in the weak-screening regime and then incorporating electromagnetic screening.

\section{Landau-level toy model}

To study the singular weak-field limit microscopically, we introduce a
two-dimensional model in which the vortex-free and single-vortex
many-body states can be constructed at finite area $\mathcal A$ and
arbitrarily weak applied field $B$. The field cannot be treated by an
ordinary perturbative expansion. Although $B$ may be infinitesimal,
the corresponding vector potential grows linearly with distance, so
that its effect becomes nonperturbative in a sufficiently large
sample. We therefore retain the orbital effect of $B$ exactly by
Landau quantizing the kinetic energy
\cite{zhu2026microscopic,huang2026orbital,fukuyama1971theory}.

For this purpose, we use the adiabatic model of twisted
transition-metal dichalcogenides introduced in
Ref.~\cite{Nicol2024prl}. It consists of two time-reversed sets of
Landau levels (LL) with opposite chiralities, supplemented by a
phenomenological attractive contact interaction:
\begin{align}
H =\sum_{j=1}^N\frac{(\vec p_j+e(\sigma_j\vec a_j+\vec A_j))^2}{2m^*}
-J\pi\ell_b^2
\sum_{i\neq j}^N
\delta(\vec r_i-\vec r_j)\delta_{\sigma_i,\bar \sigma_j}
\label{eq:many_body_hamiltonian}
\end{align}
Here, $\sigma_j=\pm1$ labels the two spin sectors. The emergent
vector potential $\vec a$ generates a uniform $b$-field,
\begin{equation}
    \vec\nabla\times\vec a=b\hat z,
    \qquad
    \ell_b=\sqrt{\frac{\hbar}{eb}},
\end{equation}
where $\ell_b$ is the magnetic length associated with the emergent
field $b$. The opposite signs with which $\vec a$ couples to the two
spin sectors preserve time-reversal symmetry. By contrast, the
physical vector potential generates a uniform $B$-field,
\begin{equation}
    \vec\nabla\times\vec A=B\hat z,
\end{equation}
 which explicitly breaks time-reversal
symmetry. The parameter $J>0$ controls the attractive interaction
between opposite-spin electrons.

We consider a disk of area $\mathcal A$. It is useful to characterize
the physical magnetic field relative to the emergent field by the
dimensionless ratio
\begin{equation}
    \lambda\equiv\frac{B}{b}.
\end{equation}
Throughout this work, we focus on the weak-physical-field regime
$|B|\ll b$, or equivalently $|\lambda|\ll1$. The system size is controlled by
a cutoff in the guiding-center angular momentum. The
Landau-level basis allows the external field to be treated
nonperturbatively even as $B\rightarrow0$. We focus on pairing in the lowest Landau level; Landau-level mixing changes quantitative details but does not alter our main conclusions. Further
details of the model are given in
Refs.~\cite{xuexciton2026,xu2026two}.

We determine the ground state at fixed electron density by
solving the gap and number equations self-consistently. The 
nonlinear gap equation is
\begin{align}\label{eq:non_linear_gap}
        \!\Delta_m^{(\eta)} 
    & = -\!J\! \sum_{\!m'\!=\!0}\! V_{m + \eta ,m ,m' + \eta ,m '} \frac{\Delta^{(\eta)}_{m'} ( f(E_{m\uparrow})+f(E_{m\downarrow})-1 )}{\!2\!\sqrt{( \frac{\hbar \omega_b}{2}-\mu)^2 + |\Delta^{(\eta)}_{m'}|^2}}
\end{align}
Here, $f(E)$ is the Fermi--Dirac distribution, $E_{m\sigma}$ are the Bogoliubov quasiparticle excitation energies in the $m$th BdG block, and
$V_{m_1,m_2,m_3,m_4}$ is the matrix element of the LLL-projected contact
interaction. $\hbar\omega_{b}/2$ is the zero-point energy of the LL. The chemical
potential $\mu$ is determined simultaneously from the number equation 
\begin{equation} \label{eq:number_equation}
    \langle N\rangle=\nu N_\phi
\end{equation}
where $\nu\in(0,2)$ is the filling fraction and $N_\phi=b\mathcal A/\phi_0$ is the
orbital degeneracy of the LL at zero external
field. 
Here $\eta$ is the integer winding number and we focus only on $\eta=0,1$ in this work. The converged superconducting ground state is characterized by the self-energy,
\begin{align}\label{eq:gap_def}
\Delta(\mathbf r)
&=\sum_m \Delta_m^{(\eta)} \phi_{(m+\eta)\uparrow}(\vec r)\phi_{m\downarrow}(\vec r)
\end{align}
where $\phi_{m\uparrow}(\vec r)\propto ((x+iy)/\ell_\uparrow)^m e^{-r^2/4\ell_\uparrow^2}$ and $\phi_{m\downarrow}(\vec r)\propto ((x-iy)/\ell_\downarrow)^me^{-r^2/4\ell_\downarrow^2}$ are the zeroth LL orbitals that have opposite chirality. $\ell_\sigma$ is the spin-dependent magnetic length:
\begin{align}\label{eq:spin_dependent_magnetic_length}
     \ell_\sigma = \frac{\ell_b}{\sqrt{1+\sigma \lambda}}
 \end{align}

\subsection{Uniform zero-field superconductor}

When $B=0$, time-reversal and
magnetic-translation symmetries favor a uniform, zero-winding solution,
$\Delta_m^{(0)}=\Delta$.
Substituting the uniform solution  into
Eq.~\eqref{eq:non_linear_gap} and solving it together with
Eq.~\eqref{eq:number_equation} self-consistently for $\Delta$ and $\mu$ gives the mean-field ground state 
\begin{equation} \label{eq:BCS_wavefunction_uniform}
    |\Phi_{\rm u}\rangle
    =
    \prod_{m=0}^{N_\phi-1}
    \left(
        u+v\,c^\dagger_{m\uparrow}c^\dagger_{m\downarrow}
    \right)|0\rangle,
\end{equation}
where the coherence factors $u$ and $v$ are independent of the
guiding-center index $m$.
This independence has an important physical consequence: changing the
electron density does not select a subset of guiding-center orbitals to
participate in pairing. Instead, every orbital in the macroscopically
degenerate LL contributes equally to the condensate for all
$0<\nu<2$; the filling changes only their common coherence factors and
pairing amplitude \cite{sup-mat}. This equal-weight participation is enforced by
magnetic-translation symmetry and is required for the order parameter
to remain spatially uniform. Thus, unlike conventional BCS
superconductivity, in which pairing is concentrated near a Fermi
surface, the uniform LL superconductor pairs the entire LL at any
partial filling.

The gap and number equations can then be solved analytically, giving the closed form of the zero-temperature gap $\Delta$, critical temperature $T_c$, and chemical potential $\mu$, 
\begin{equation}
\begin{split} 
\label{eq:zero_field_results}
    T_c
    =
    \frac{1-\nu}
    {2\ln(2/\nu-1)}
    \frac{J}{k_B}\;,\;
    \Delta=
    \frac{J}{2}\sqrt{\nu(2-\nu)},\\
    \mu(T<T_c)
    =
    \frac{\hbar\omega_b}{2}
    -\frac{J}{2}(1-\nu).
\end{split}
\end{equation}
where $\hbar \omega_b/2$ is the zero-point energy. 
Both the gap amplitude and critical temperature are maximal at half filling,
$\nu=1$, and are symmetric under the particle-hole transformation
$\nu\leftrightarrow2-\nu$ due to LL projection. 
An interesting property is that for $T<T_c$, $\mu$ is temperature-independent, which is different from a dispersive metal where the chemical potential shifts with temperature to maintain a fixed density as particles redistribute across the band. The
projected lowest LL, however, has no dispersion over which
particles can redistribute. At fixed filling, the number equation is instead maintained through the
temperature dependence of $\Delta(T)$, which modifies the coherence
factors and allows $\mu$ to remain constant below $T_c$.

We further find that the lowest-energy Cooper pair lies in the
$\mathbf q=0$ channel. Because the two electrons experience opposite $b$, their center-of-mass motion can be labeled by
a conserved momentum $\mathbf q$. As in the magnetoexciton problem of Ref.~\cite{Kallin1984}, this momentum determines
the relative displacement of their guiding centers $\vec d=\hat{\vec z}\times\vec q\,\ell_{b}^2.$ The contact attraction is maximized when the
opposite-spin electrons coincide, i.e. $d=0$. The interaction therefore condensed in the $\vec q=0$ pairing channel.
Despite the complete absence of single-particle dispersion in the LL, the paired state $|\Phi_{\rm u}\rangle$ has a finite superfluid stiffness $\rho_b(T=0)$. As shown in Sup-Mat.~\cite{sup-mat}, $\rho_b(T=0) =\frac{3|\Delta|^2}{2\pi J}$. This finite phase rigidity shows that the uniform
paired state is a genuine superconductor rather than merely a collection of localized bound pairs.

\subsection{Landau quantization of the Cooper-pair condensate}
\begin{figure*}[t]
    \centering
    \includegraphics[width=1\linewidth]{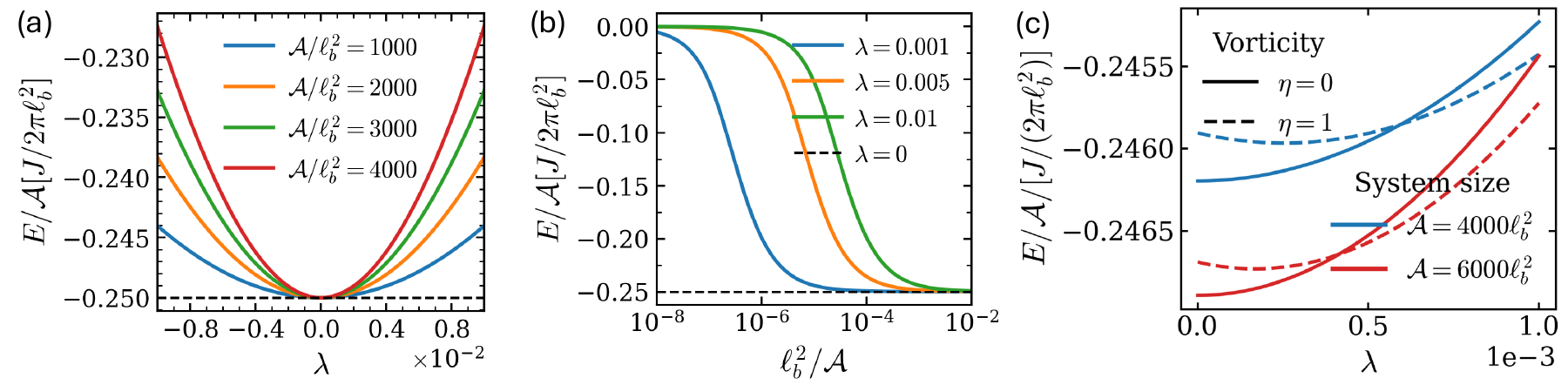}
    \caption{
\textbf{(a)} Energy density dependence on $\lambda$ for several system areas. The narrowing valley with increasing $\mathcal{A}$ signals the emergence of a nonanalyticity at $\lambda=0$ in the thermodynamic limit.
\textbf{(b)} Energy density dependence on inverse area for several fixed values of $\lambda$. For any finite $\lambda$, the energy density vanishes as $\mathcal{A}\to\infty$, reflecting saturation of the total energy, whereas for $\mathcal{A}\ll\pi\ell_C^2$ it approaches the uniform zero-field bulk value indicated by the dashed line.
\textbf{(c)} Comparison of energy density between the uniform and single-vortex states as a function of $\lambda$ for several system sizes.}
    \label{fig:zero_field_result}
\end{figure*}

We now study how an external magnetic field modifies the vortex-free
Cooper-pair condensate.  For clarity, we focus on $T=0$ and half
filling, $\nu=1$, where superconductivity is strongest. Deviations
from half filling do not alter the conclusions below
\cite{sup-mat}. Figure~\ref{fig:zero_field_result} summarizes the main results. The
field-induced mismatch between the two Landau levels confines the
vortex-free condensate (Eq.~\eqref{eq:finite_field_gap_profile}), produces a singular weak-field
thermodynamic limit
[Figs.~\ref{fig:zero_field_result}(a) and (b)], and controls the
competition between the vortex-free and single-vortex states
[Fig.~\ref{fig:zero_field_result}(c)].

For $\sigma=\pm1$, the total magnetic field experienced by
the electrons with spin $\sigma$ is $ B_\sigma
= \sigma b+B$. Consequently, the external field modifies the magnetic lengths shown in Eq.~\eqref{eq:spin_dependent_magnetic_length}, LL
degeneracies, and zero-point energies,
\begin{align}
         N_{\phi,\sigma}=
    (1+\lambda \sigma)N_\phi,\quad
    \varepsilon_{b,\sigma}=
    \frac{\hbar\omega_b(1+\lambda \sigma)}{2}.
\end{align}
The degeneracy imbalance creates excess orbitals near the boundary of a
finite disk. The difference in zero-point energies introduces a Zeeman-like 
single-particle energy mismatch. For the bulk order-parameter profile,
however, the dominant effect is the change in the interaction matrix
elements caused by $\ell_\uparrow\neq\ell_\downarrow$. In this situation, any isotropic interaction
can be expressed in terms of generalized Haldane pseudopotentials in the
center-of-charge basis introduced in Ref.~\cite{xu2026two}.

Accounting for the mismatch in the magnetic lengths, the coherence factors $u$ and $v$, which are guiding-center independent in the zero-field uniform BCS wavefunction in Eq.~\eqref{eq:BCS_wavefunction_uniform}, become guiding-center dependent. Furthermore, the order parameter $\Delta_{\rm u}(\bf r)$ acquires a Gaussian envelope with a field-dependent decay length:

\begin{equation}\label{eq:finite_field_gap_profile}
\begin{split}
\Delta_{\rm u}(\mathbf r)
&=\Delta(\lambda)e^{-r^2/\ell_C^2},
\qquad
\Delta(\lambda)
=\frac{J}{2}
\frac{(1-\lambda^2)^{3/2}}
     {2-\sqrt{1-\lambda^2}},
\\[-2pt]
\ell_C &= \sqrt{2} \left( 1-\frac{1-\lambda^2}{2-\sqrt{1-\lambda^2}} \right)^{-\frac{1}{2}} \ell_{b}
\end{split}
\end{equation} 

This suppression of the gap has a simple microscopic origin. The orbital
$\phi_{m\sigma}$ is concentrated near $    r_{m\sigma}\simeq\sqrt{2m}\,\ell_\sigma
$.  Pairing orbitals with the same guiding-center index therefore produces a radial mismatch
$\delta r_m \simeq \sqrt{2m}\,\left|\ell_\uparrow-\ell_\downarrow\right|\simeq \sqrt{2m}|\lambda|\ell_b$.
The mismatch grows
with distance from the disk center, reducing the orbital
overlap and weakening the corresponding pairing matrix elements.

In the lowest Landau level, the condensation energy is equal to
the total energy up to a state-independent constant \cite{sup-mat}.
For a disk of area $\mathcal A$, the total energy is therefore
\begin{align}\label{eq:zero_winding_energy}
    E_{\rm u}(\mathcal A,\lambda)
    &=
    -\frac{1}{2\pi J\ell_b^2}
    \int_{\mathcal A}d^2r\,
    \left|\Delta_{\rm u}(\mathbf r)\right|^2
    \nonumber\\
    &=
    -\frac{|\Delta(\lambda)|^2\ell_C^2}
    {4J\ell_b^2}
\left(1-e^{-2\mathcal A/\pi \ell_C^{2}}\right).
\end{align}

The behavior of Eq.~\eqref{eq:zero_winding_energy} is controlled by the
ratio $\mathcal A/\ell_C^2$ and therefore depends singularly on how the
weak-field thermodynamic limit $(1/\mathcal A,B)\to(0,0)$ is
approached. To make this structure explicit, we write the zero-winding
energy density as
\begin{equation}
\begin{split}
\frac{E_{\rm u}(\mathcal A,\lambda)}{\mathcal A}
=
-\epsilon(\lambda)F(z),
\qquad
F(z)=\frac{1-e^{-z}}{z},\\
z=\frac{2\mathcal A}{\pi\ell_C^2},
\qquad
\epsilon(\lambda)=
\frac{|\Delta(\lambda)|^2}{2\pi J\ell_b^2}.
\end{split}
\end{equation}
In the weak-field limit,
$\ell_C^2\simeq4\ell_b^2/3\lambda^2$ and
$\epsilon(\lambda)\simeq\epsilon_0=J/(8\pi\ell_b^2)$, so that
\begin{equation}\label{eq:zero_winding_scaling}
\frac{E_{\rm u}(\mathcal A,\lambda)}{\mathcal A}
\simeq
-\epsilon_0
F\left(
\frac{\mathcal A \lambda^2}{\pi \ell_b^2}
\right).
\end{equation}

Fig.~\ref{fig:zero_field_result}(a) and (b) show complementary cuts of
Eq.~\eqref{eq:zero_winding_scaling}. At fixed area, the energy density as a
function of $B$ forms a valley centered at $B=0$ [Fig.~\ref{fig:zero_field_result}(a)]. As
$\mathcal A$ increases, the depth of the valley remains fixed at
$-\epsilon_0$, while its width shrinks as
$B_\ast\sim\mathcal A^{-1/2}$. Consequently, its curvature at $B=0$,
which determines the diamagnetic magnetic susceptibility, diverges in the thermodynamic
limit. The complementary fixed-$B$ cuts in Fig.~\ref{fig:zero_field_result}(b) show that, for any
nonzero $B$, the zero-winding energy density vanishes as
$\mathcal A\rightarrow\infty$, whereas at $B=0$ it remains equal to the
bulk condensation-energy density $-\epsilon_0$. Thus, the energy landscape
develops a singular, infinitely narrow valley at the origin (diverging diamagnetic susceptibility), demonstrating
that the weak-field and thermodynamic limits do not commute:
\begin{equation}\label{eq:noncommuting_limits}
\begin{split}
\lim_{B\to0}\lim_{\mathcal A\to\infty}
\frac{E_{\rm u}}{\mathcal A}=0,\;\;
\lim_{\mathcal A\to\infty}\lim_{B\to0}
\frac{E_{\rm u}}{\mathcal A}=-\epsilon_0.
\end{split}
\end{equation}

This singular behavior has a simple physical origin. At fixed nonzero
field, only an area of order $\ell_C^2$ contributes appreciably to the
zero-winding condensation energy. The vortex-free solution therefore
resembles a finite superconducting droplet rather than an extensive
superconducting phase, and its energy density vanishes in the
thermodynamic limit. This does not describe the equilibrium
thermodynamics at fixed $B$: because the total flux grows extensively
with $\mathcal A$, the ground state is instead expected to contain an
extensive number of vortices and approach a vortex lattice.

This non-commutativity motivates us to consider a different joint limit
in which the total flux is held fixed, $B\mathcal A=C\Phi_{0}$ where $C$ is finite as $\mathcal A\to\infty$ and $B\to0$.
Using $B=\lambda b$ and $\ell_b^2=\hbar/(eb)$, this condition becomes $\mathcal A=\frac{\pi C\ell_b^2}{\lambda}.$
Since $\ell_C\propto\lambda^{-1}$ at small $\lambda$,  we have $ \frac{\mathcal A}{\ell_C^2}
\propto C\lambda \longrightarrow0$. Thus, the localization length becomes parametrically larger than the radius of the sample, and the vortex-free order parameter becomes
asymptotically uniform. Expanding Eq.~\eqref{eq:zero_winding_energy} in
this limit gives
\begin{equation}
    E_{\rm u}(\lambda,\mathcal A)
    \simeq
    -\frac{|\Delta(\lambda)|^2}{2\pi J\ell_b^2}\,
    \mathcal A.
\end{equation}
Hence, every trajectory with finite total flux asymptotically recovers
the uniform bulk condensation-energy density and the zero-winding state remains a
well-defined thermodynamic superconducting state. Whether it is the
ground state must be determined by comparison with the finite-winding
sectors.

\subsection{BCS wavefunction of the single-vortex state}

In the vortex-free state, the Landau level orbital $m\downarrow$ pairs with
$m\uparrow$ so that the net angular momentum of the Cooper pair vanishes. In our toy model, the Cooper pair made up of
$m\downarrow$ and $(m+1)\uparrow$ electrons instead has angular momentum $\pm1$. The 
BCS wavefunction for the single-vortex state can then be written as
\begin{equation} \label{eq:BCS_wavefunction_vortex}
|\Phi_{v}\rangle =\prod_m (u_m(\lambda) +v_m(\lambda) c_{m+1\uparrow}^\dagger c_{m\downarrow}^\dagger)|0\rangle 
\end{equation} 
Solving for $u_m$ and $v_m$ self-consistently at fixed density, we arrived at the spatially non-uniform order parameter $\Delta_{\rm v}(\vec r)$,
\begin{align}
  \Delta_{\rm v}(|\mathbf r| \!\ll\! \xi)\! \sim \!\frac{r e^{i\theta}}{\xi} ,\! \quad   \!\Delta_{\rm v}(|\mathbf r|\!\gg\! \xi) \!\sim\! (1\!+\!\frac{3\lambda}{2})\Delta_{\rm u}(\vec{r})e^{i\theta}(1\!-\!\frac{3\xi^2}{8r^2})
\end{align}
where $\xi=\sqrt{2}\ell_b$ is the coherence length or vortex core size of the present model and $\Delta_{\rm u}(\vec r)$ is the vortex-free order parameter in Eq.~\eqref{eq:finite_field_gap_profile}. Thus, the order parameter vanishes linearly near the vortex core while it has the same profile as the vortex-free order parameter at long distances. 
This common long-distance
behavior reflects the field-induced mismatch between paired Landau-level
orbitals, whereas the winding modifies only the structure near the vortex
core.

\begin{table*}[t]
\centering
\renewcommand{\arraystretch}{1.4}
\begin{tabular}{c|c|c|c|c}
\hline
Quantity
& Weak screening, $R\ll\Lambda$
& Character
& Strong screening, $R\gg\Lambda$
& Character \\
\hline
$G_0$
& $\ln(R/\xi)$
& Subextensive
& $\ln(\Lambda/\xi)+\mathcal{O}(1)$
& Intensive \\
\hline
$M_v$
& $R^2\propto\mathcal A$
& Extensive
& $\Lambda R\propto\Lambda\sqrt{\mathcal A}$
& Subextensive \\
\hline
$\chi$
& $R^4\propto\mathcal A^2$
& Superextensive
& $R^3\propto\mathcal A^{3/2}$
& Superextensive \\
\hline
$B_v$
& $R^{-2}\ln(R/\xi)$
& $\mathcal A^{-1}\ln\mathcal A$
& $(\Lambda R)^{-1}$
& $\mathcal A^{-1/2}$ \\
\hline
\end{tabular}\label{tab:size_scaling}
\caption{Magnetic properties of a single vortex state. System-size dependence of the zero-field vortex energy ($E_v^{(0)}$), vortex magnetic moment ($M_v$), magnitude of the diamagnetic susceptibility ($\chi$), and lower critical vortex field ($B_v=E_v^{(0)}/M_v$). System-size-independent prefactors and nonuniversal core contributions are omitted.}
\end{table*}

\subsection{First-vortex field in the weak-screening limit}
The first vortex enters when the magnetic energy gained from its
orbital moment compensates its excitation energy.
To make
this competition explicit, we first keep the system area $\mathcal A$
finite and expand the energies of the vortex-free and single-vortex
sectors about $B=0$:
\begin{align}
E_{u}(B,\mathcal A)
&=
E_{\rm bulk}(\mathcal A)
+\frac{1}{2}\chi_0(\mathcal A)B^2+\cdots,
\nonumber\\
E_{v}(B,\mathcal A)
&=
E_{\rm bulk}(\mathcal A)
+E_v^{(0)}(\mathcal A)
-M_v(\mathcal A)B
+\frac{1}{2}\chi_1(\mathcal A)B^2+\cdots .
\label{eq:vortex_energy_expansion}
\end{align}

This expansion is valid before taking the thermodynamic limit $\mathcal{A}\rightarrow\infty$; otherwise, the energy density is non-analytic.
Here, $E_v^{(0)}(\mathcal A)$ is the energy required to create a
vortex at zero field and $M_v(\mathcal A)$ is its orbital magnetic
moment. The extensive bulk condensation energy $E_{\rm bulk}$ is
common to the two sectors and therefore cancels in their difference. 
Their expressions are
\begin{align}
    \chi_0&= \frac{\pi^3\rho_b R^4}{2\Phi_0^2}
    ,\quad \chi_1=  \chi_0+O(\xi/R)\\
    M_v&=\frac{\pi^2\rho_b R^2}{\Phi_0},\quad
    E^{(0)}_v= \frac{\pi\rho_b}{2} \ln{\frac{\mathcal{A}}{\pi \xi^2}}\label{eq:orbital_moment}
\end{align}

Since the (Meissner) diamagnetic screening is the same in the two sectors when the disk radius is much greater than the coherence length $R\gg \xi$, the energy difference between the two sectors is given by
$\Delta E_v(B,\mathcal A)
=E_v^{(0)}(\mathcal A)-M_v(\mathcal A)B$.
The first-vortex entry field is consequently determined by
\begin{equation}
    B_v(\mathcal A)=\frac{E_v^{(0)}(\mathcal A)}{M_v(\mathcal A)}.
\end{equation}
 Since the vortex
current retains its unscreened $1/r$ form throughout the sample, its
phase-stiffness energy therefore grows logarithmically with the system
size. By contrast, the \textit{total}
vortex magnetic moment is extensive, $M_v(\mathcal A)\propto\mathcal
A$.
Therefore,
\begin{equation}
B_v(\mathcal A)
=
\frac{\Phi_{0}}{\mathcal A}
\left[
\frac{1}{2}
\ln\left(
\frac{\mathcal A}{\pi\xi^2}
\right)
+
\frac{2\varepsilon_{\rm core}}{3} 
\right].
\label{eq:single_vortex_boundary}
\end{equation}
where $\varepsilon_{\rm core}>0$ is a dimensionless, nonuniversal contribution from the vortex core.
Thus, the thermodynamic magnetic field $B_v(\mathcal A)\propto \mathcal{A}^{-1}\ln\mathcal A$, i.e.~the field at which the energy level crossing occurs,  decreases with increasing
system size.  In the next section, we show that electromagnetic screening changes this scaling to $B_v(\mathcal A)\propto \mathcal A^{-1/2}$, but does not restore a size-independent lower critical field. The crossover of the vortex energy, magnetic moment, diamagnetic susceptibility, and first-vortex field is summarized in Table I.

\section{Electromagnetic screening}
The microscopic wavefunctions of the uniform state, Eq.~\eqref{eq:BCS_wavefunction_uniform}, and the singly quantized vortex state, Eq.~\eqref{eq:BCS_wavefunction_vortex}, teach us three lessons. First, at finite field $B$, Landau quantization gives both states a common field-induced amplitude envelope, $\Delta_{\eta}(r)\simeq
\Delta_0 e^{-r^2/\ell_C^2}$, apart from the vortex core.  Second, far outside the vortex core, their amplitudes become identical and the only distinction between them is the additional $2\pi$ phase winding of the vortex state. Third, for every finite sample area $\mathcal A$, the energies of both states are analytic functions of $B$.

These results suggest that, as long as the Cooper pair magnetic localization length $\ell_C$ is much larger than the system size and superconducting coherence length, the energy difference between the two states is controlled by the supercurrent associated with the vortex phase winding and its self-consistent electromagnetic field. In principle, one could determine this field by solving the microscopic BCS gap equation and Maxwell equations simultaneously, iterating the vector potential and order parameter until full self-consistency is reached. Such a calculation would provide microscopic information such as modifications of the vortex-core structure and the Bogoliubov excitation spectrum. For the present purpose, however, a fully microscopic treatment is unnecessary. The system-size dependence of the finite-size first-vortex field $B_v$ is determined by the screening of the long-distance supercurrent and can therefore be obtained from the leading-gradient Ginzburg-Landau theory coupled to 3D Maxwell equations. We thus study the electrodynamics of a superconducting disk of area $\mathcal A$. One can verify that, at the critical field where the single vortex becomes the ground state, the field-induced variation of the order-parameter amplitude is negligible across the disk, justifying this long-wavelength treatment.

\begin{figure*}
    \centering
    \includegraphics[width=1.01\linewidth]{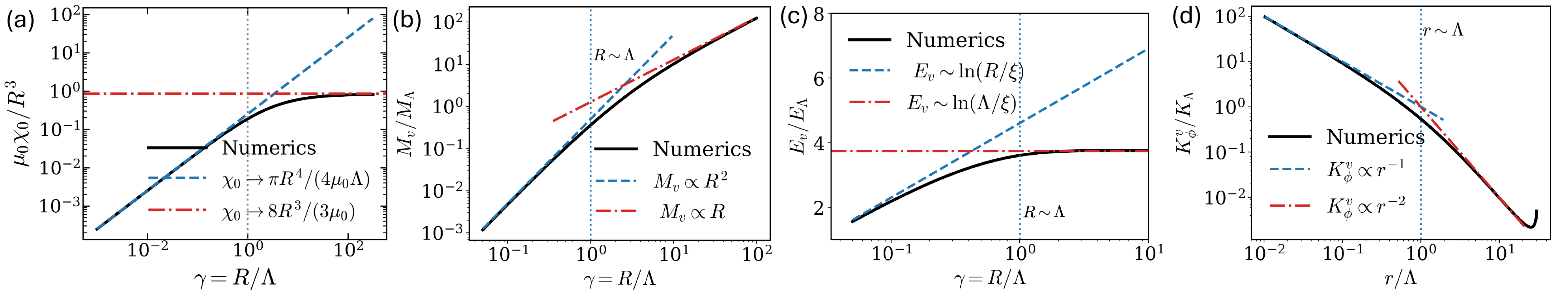}
    \caption{Screening crossover. 
\textbf{(a)} Diamagnetic response as a function
of $\gamma=R/\Lambda$. The numerical result approaches the weak screening
 behavior $\chi_{0}=R^4/(4\mu_0\Lambda)$ for
$R\ll\Lambda$, while saturating to the strong-screening limit
$\chi_{0}=8R^3/(3\mu_0)$ for $R\gg\Lambda$.
\textbf{(b)} Orbital magnetic moment of a centered vortex, showing the crossover from
$M_v\propto R^2$ in the weak-screening regime to $M_v\propto R$ in the
strong-screening regime. $M_\Lambda\equiv \Phi_0\Lambda/\mu_0$.
\textbf{(c)} Vortex stiffness energy. For $R\ll\Lambda$ it grows as
$E_v\sim E_\Lambda\ln(R/\xi)$ where $E_\Lambda\equiv \Phi_0^2/(2\pi\mu_0\Lambda)$, whereas for $R\gg\Lambda$ Pearl
screening cuts off the logarithmic growth at the scale $\Lambda$, giving
$E_v\sim E_\Lambda\ln(\Lambda/\xi)$.
\textbf{(d)} Vortex sheet-current profile for $R\gg\Lambda$, exhibiting the crossover
from $K_\phi^v\propto r^{-1}$ at $r\ll\Lambda$ to
$K_\phi^v\propto r^{-2}$ at $r\gg\Lambda$. $K_\Lambda\equiv \Phi_0/(\pi\mu_0 \Lambda^2)$. 
Dashed and dash-dotted lines denote the corresponding weak- and
strong-screening asymptotic results, respectively, and vertical dotted lines
mark the Pearl crossover scales $R\sim\Lambda$ or $r\sim\Lambda$.
}
    \label{fig:fig3}
\end{figure*}

We write the applied vector potential in the symmetric gauge as
$\mathbf A_{\rm ext}=(Br/2)\hat{\boldsymbol\phi}$ and denote the vector potential generated by the screening current by $\vec a$. The Gibbs free energy, measured relative to the vacuum energy of the applied field, is
\begin{align}
G_\eta[\vec a]
={}&
\frac{1}{\mu_0\Lambda}
\int_{\xi<r<R}d^2r
\left[
a_\phi(r,0)+\frac{Br}{2}
-\frac{\eta\Phi_0}{2\pi r}
\right]^2
\nonumber\\
&+
\frac{1}{2\mu_0}
\int d^3r 
\left|\nabla\times\mathbf a\right|^2
+
E_{\rm core}^{(\eta)} .
\label{eq}
\end{align}
where the Pearl length $\Lambda$ is defined as:
\begin{align}
   \Lambda\equiv \frac{\Phi_0^2}
    {2\pi^2\mu_0\rho_b}
\end{align}
The first term is the material term representing the kinetic energy of the supercurrent. The second term is the magnetic-field energy in the surrounding 3D space. This energy functional has been previously studied in Ref.~\cite{Magnetization_scaling,Fetter1980} in different contexts.

We find that strong-screening physics, i.e.~physics at distances $r\gg\Lambda$, can be quite easily understood from flux
conservation. By construction, a vortex carries one flux quantum
through the superconducting plane, $\int d^2r\mathcal B_z^v(r,0)=\Phi_0$, where $\boldsymbol{\mathcal B}^{v}$ is the magnetic field produced by the screened vortex.
Because $\vec{\nabla}\cdot\boldsymbol{\mathcal B}^{v}=0$, the same amount of flux must
flow through a large hemisphere in the upper half-space. The flux
through the corresponding lower hemisphere is $-\Phi_0$, where the
sign is defined relative to its outward normal. 
 Since the area of a
hemisphere of radius $r$ is $2\pi r^2$, flux conservation gives
\begin{equation}
\left|\boldsymbol{\mathcal{B}^{\,v}}(r)\right|
\simeq
\frac{\Phi_0}{2\pi r^2},
\qquad
r\gg\Lambda.
\label{eq:pearl_far_field}
\end{equation}
The field is approximately radial, pointing outward in the upper
half-space and inward in the lower half-space. Thus, the field in the
upper half-space resembles that of a positive magnetic monopole,
whereas that in the lower half-space resembles a negative magnetic
monopole. This analogy applies to the two half-spaces separately. The
complete magnetic field of course remains divergence-free.

Immediately above and below the superconducting sheet, the radial
fields are $\mathcal B_r^v(r,0^+)\simeq\Phi_0/(2\pi r^2)$ and
$\mathcal B_r^v(r,0^-)\simeq-\Phi_0/(2\pi r^2)$ for $r\gg\Lambda$. The discontinuity of the radial magnetic field is generated by the azimuthal sheet current according to
Amp\`ere's boundary condition:
\begin{equation}
    K_\phi^v(r)=\frac{\mathcal B_r^v(r,0^+)- \mathcal B_r^v(r,0^-)}{\mu_0}
\simeq\frac{\Phi_0}{\pi\mu_0r^2}.\label{eq:vortex_current}
\end{equation}
This is precisely the $1/r^2$ current tail obtained
by Pearl~\cite{Pearl1964}. Because it decays more rapidly than the
unscreened vortex current, $K_\phi^v\propto 1/r$, its contribution to
the vortex self-energy is infrared convergent:
$\int_\Lambda^R r\,dr\,[K_\phi^v(r)]^2\propto
\int_\Lambda^R dr/r^3$. The logarithmic contribution instead arises
from the intermediate region $\xi\ll r\ll\Lambda$, where
$K_\phi^v(r)\propto 1/r$. Consequently,
\begin{equation}
E_v(R)
=
\frac{\Phi_0^2}{2\pi\mu_0\Lambda}
\left[
\ln\left(\frac{\Lambda}{\xi}\right)+c_E
\right]
+
\mathcal O\left(
\frac{\Phi_0^2}{\mu_0R}
\right),
\label{eq:vortex_energy_strong}
\end{equation}
for $R\gg\Lambda\gg\xi$. Here, $c_E$ is an order-one constant that contains the vortex-core
energy and the crossover contribution from $r\sim\Lambda$.
Electromagnetic screening therefore makes the vortex energy
independent of the sample size in the thermodynamic limit.

The total vortex magnetic moment behaves differently. The sheet current has dimensions of current/length. Thus, an annulus of radius $r$ and width $dr$ carries a current
$dI=K_\phi^v(r)\,dr$. It gives rise to an orbital magnetic moment
\begin{equation}
    dM_v=\pi r^2dI=\pi r^2K_\phi^v(r)\,dr\sim \frac{\Phi_0}{\mu_0}dr.
\end{equation}
Integrating this to
the sample boundary gives
\begin{equation}
    M_v(R)=\pi\int_\xi^Rdr\,r^2K_\phi^v(r)
\simeq\frac{\Phi_0R}{\mu_0}[1+o(1)]\label{eq:orbital_moment_strong_screening}
\end{equation}
for $R\gg\Lambda$. 
Using $-M_v(R)B=E_v(R)$ with the asymptotic vortex
energy and magnetic moment, we obtain
\begin{equation}
B_v(R)
\simeq
\frac{\Phi_0}{2\pi\Lambda R}
\left[
\ln\left(\frac{\Lambda}{\xi}\right)+c_E
\right],
\qquad
R\gg\Lambda\gg\xi.
\label{eq:first_vortex_field_strong}
\end{equation}
Thus, although screening cuts off the logarithmic growth of the vortex
energy at the Pearl length, the field required to create the first
vortex continues to vanish as $B_v(R)\propto 1/R$. Therefore, in both the weak- and strong-screening regimes, the field at which the energy of the single-vortex state crosses that of the uniform state decreases with increasing system size. The full crossover from weak to strong screening can only be obtained numerically, and the results are shown in Fig.~\ref{fig:fig3}. Figure~\ref{fig:fig3}(a) shows the crossover of the diamagnetic susceptibility from $\chi_{0}\sim R^4/\Lambda$ for $R\ll\Lambda$ to $\chi_{0}\sim R^3$ for $R\gg\Lambda$ as a function of $\gamma=R/\Lambda$. Since $\chi_{0}/\mathcal A\sim R$ still diverges in the strong-screening regime, the uniform vortex-free state cannot remain thermodynamically stable in the thermodynamic limit in either the unscreened or strong-screening case.  Fig.~\ref{fig:fig3}(d) describes the full screening crossover of the vortex circulating current $K_\phi^v$ as a function of $r/\Lambda$ for $R = 30\Lambda$. For $r\ll\Lambda$, the vortex current retains the unscreened behavior $K_\phi^v\sim r^{-1}$, whereas for $r\gg\Lambda$ it crosses over to the Pearl tail $K_\phi^v\sim r^{-2}$, as shown in Eq.~\eqref{eq:vortex_current}. This crossover directly modifies both the vortex magnetic moment and stiffness energy. As shown in Fig.~\ref{fig:fig3}(b), the orbital magnetic moment scales as $M_v\propto R^2\propto\mathcal A$ in the weak-screening regime, consistent with Eq.~\eqref{eq:orbital_moment}, while the faster decay of the current for $R\gg\Lambda$ changes the scaling to $M_v\propto R\propto\sqrt{\mathcal A}$, as given in Eq.~\eqref{eq:orbital_moment_strong_screening}. Similarly, Fig.~\ref{fig:fig3}(c) shows that the vortex stiffness energy grows logarithmically as $\ln(R/\xi)$ for $R\ll\Lambda$, whereas for $R\gg\Lambda$ the logarithmic growth is cut off at the Pearl length, giving $E_v\sim E_\Lambda\ln(\Lambda/\xi)$ as in Eq.~\eqref{eq:vortex_energy_strong}.

\section{Summary and Outlook}

Our work provides a microscopic foundation for the weak-field thermodynamics of 2D superconductors and shows that their magnetic properties depend sensitively on how the thermodynamic limit is taken. Fundamentally, because the static transverse response is finite in a superconductor $\rho_s=\lim_{q\rightarrow0}Q_T(q,\omega=0)$, a uniform perpendicular magnetic field, whose vector potential grows linearly with distance, probes the system on the scale of its full extent. 
Consequently, thermodynamic quantities such as the total magnetic moment, $M=-\partial E/\partial B$, and the magnetic susceptibility, $\chi=\partial^2E/\partial B^2$, exhibit the unusual system-size dependencies summarized in Table I. Importantly, the energy density of the uniform state is nonanalytic at
$(1/\mathcal A,B)=(0,0)$. The weak-field response therefore cannot be
obtained by first sending $\mathcal A\rightarrow\infty$ and then
expanding the thermodynamic energy density in powers of $B$. Instead, the energy must be expanded at finite $\mathcal A$ before the
thermodynamic limit is taken. As we have shown in this work, the time-reversal-breaking vortex states will eventually cross the uniform
state in energy as $\mathcal A\rightarrow\infty$ for a fixed external $B$. The field at which this energy crossing occurs, i.e., the
lower critical field, decreases as $\mathcal A^{-1}\ln\mathcal A$
in the weak-screening regime and as $1/\sqrt{\mathcal A}$ in the
strong-screening regime. 
The size dependence of flux penetration in thin
superconducting disks has been studied previously
~\cite{Magnetization_scaling,Fetter1980}.  Our contribution is to place this behavior
within the broader thermodynamic structure of a strictly
two-dimensional superconductor. We show that
$(1/\mathcal A,B)=(0,0)$ is a singular point at which the weak-field
and thermodynamic limits do not commute, and demonstrate how to compute weak-field magnetic properties both in
microscopic BCS theory and Ginzburg-Landau theory. Several extensions are natural. The boundaries between higher-vorticity sectors in Fig.~1 are schematic. Determining them requires treating vortex-vortex interactions and spatially nonuniform multivortex configurations, which would connect the first-vortex transition studied here to the dilute vortex-lattice regime~\cite{Fetter1980}. The present analysis is restricted to the zero-temperature $(1/\mathcal A,B)$ plane. Extending it to the three-dimensional parameter space $(1/\mathcal A,B,T)$ would reveal how vortex entropy and thermally induced phase fluctuations reshape the singular weak-field thermodynamic limit.


\textit{Acknowledgment:} We are supported by the DOE, Office of Basic Energy Sciences under Award Number DE-SC-0024346 and the National Science Foundation CAREER grant No.~DMR-2541471.

\newpage
\pagebreak
\bibliographystyle{ieeetr}
\bibliography{reference}

\clearpage
\onecolumngrid

\setcounter{section}{0}
\setcounter{subsection}{0}
\setcounter{subsubsection}{0}
\setcounter{equation}{0}
\setcounter{figure}{0}
\setcounter{table}{0}
\setcounter{footnote}{0}

\makeatletter
\let\sm@orig@addcontentsline\addcontentsline
\renewcommand{\addcontentsline}[3]{%
  \def\sm@tocname{toc}%
  \def\sm@arg{#1}%
  \ifx\sm@arg\sm@tocname
    \sm@orig@addcontentsline{smtoc}{#2}{#3}%
  \else
    \sm@orig@addcontentsline{#1}{#2}{#3}%
  \fi
}
\newcommand{\smtableofcontents}{%
  \begin{center}
    {\normalsize\bfseries CONTENTS\par}
  \end{center}
  \vspace{0.25em}%
  \begingroup
    \small
    \setcounter{tocdepth}{2}
    \renewcommand{\@pnumwidth}{2.4em}%
    \renewcommand{\@tocrmarg}{3.2em}%
    \renewcommand{\@dotsep}{4.5}
    \renewcommand*{\l@section}[2]{%
      \addvspace{0.30em}%
      \@dottedtocline{1}{0em}{2.8em}{\bfseries ##1}{\bfseries ##2}}%
    \renewcommand*{\l@subsection}{\@dottedtocline{2}{2.8em}{2.6em}}%
    \@starttoc{smtoc}%
  \endgroup
}
\makeatother

\begin{center}
    {\normalsize\bfseries Supplemental Material for: Singular Weak-Field Thermodynamics of 2D Superconductors\par}
    \vspace{0.35em}
    {\small\today\par}
\end{center}
\vspace{0.9em}

\hypersetup{linkcolor=black}
\smtableofcontents
\hypersetup{linkcolor=NavyBlue}

\setcounter{section}{0}
\setcounter{subsection}{0}
\setcounter{subsubsection}{0}

\section{Model and Lowest Landau level projected BdG Hamiltonian}

In this section, we first introduce the model in Sec.~\ref{sec:model}, including the spin-dependent single-particle eigenstates and the interaction matrix elements required for the subsequent analysis. We then formulate the superconducting state within mean-field theory in Sec.~\ref{sec:mean_field_SC} and derive the corresponding self-consistent nonlinear gap equation.

\subsection{Model Introduction}\label{sec:model}
Here, we introduce the model studied in the main text. It consists of a pair of time-reversed Landau levels generated by a time-reversal-symmetric effective magnetic field, 
\begin{align}\label{eq:many_body_Hamiltonian}
    H =\sum_\sigma \int d^2 \mathbf r \psi^\dagger_\sigma(\mathbf r) h_\sigma \psi_\sigma(\mathbf r) - J 2\pi \ell_b^2 \int d^2 r :\psi^\dagger_\uparrow(\mathbf{r}) \psi_\uparrow(\mathbf{r}) \psi^\dagger_{\downarrow}(\mathbf{r}) \psi_{\downarrow}(\mathbf{r}) :
\end{align}
where the single-particle Hamiltonian $h_\sigma$ is:
\begin{align}\label{eq:single_particle_SM}
    h_\sigma = \frac{(\vec p+ e \sigma \vec a+e \vec{A})^2}{2m^*}, \quad \nabla \times \vec a =  b \hat{z}, \quad \nabla \times \vec A = B\hat{z}
\end{align}
Here, $\sigma=\pm1$ labels the spin sector, and $J>0$ is the strength of the attractive contact interaction. 
The effective gauge field $\mathbf a$ produces a magnetic field that reverses sign between the two spin sectors and therefore preserves time-reversal symmetry. The vector potential $\mathbf A$ represents the externally applied magnetic field and breaks time-reversal symmetry. The magnetic length associated with the effective field is
\begin{align}\label{eq:magnetic_length_sky}
    \ell_b = \sqrt{\frac{\hbar}{e b}}
\end{align}
This model can describe twisted transition-metal-dichalcogenides (TMDs) in the adiabatic limit\cite{Nicol2024prl}.
We focus on the regime $B\ll b$ and define the ratio as $\lambda$:
\begin{align}
    \lambda = \frac{B}{b} \ll 1 
\end{align}

A key feature of this model is that, even for a weak external magnetic field, the single-particle states are already strongly Landau quantized by the time-reversal-symmetric effective field. The external field breaks time-reversal symmetry and makes the total fields experienced by the two spin sectors different in both magnitude and direction. In the remainder of the Supplemental Material, we work in symmetric gauge and project onto the lowest Landau level (LLL), $n=0$. 

The external magnetic field modifies three properties of the time-reversed LLs: the single-particle energy, orbital degeneracy, and magnetic length. First, the LLL zero-point energy is
\begin{align}
    \varepsilon_{\sigma} =s_\sigma\hbar \omega_b/2 , \quad s_\sigma \equiv 1+\sigma\lambda
\end{align}
where $\omega_b=eb/m^*$ is the cyclotron frequency set by the effective field and $n$ is the Landau level index. Thus, the single-particle energy acquires a Zeeman-like term. 

Next, the LL degeneracies, i.e., the numbers of guiding-center orbitals, are
\begin{align}\label{eq:orbital_degenercy}
    N_{\phi\sigma} = s_\sigma N_\phi, \quad N_\phi = \frac{\mathcal A}{2\pi \ell_b^2}
\end{align}
where $\mathcal{A}$ is the system area. For a finite disk, the number of guiding-center orbitals in each LL must be an integer and changes by approximately $\pm\operatorname{int}(\lambda N_\phi)$. The total number of states summed over the two LLs remains unchanged, while the spin-majority LL gains states at the expense of the spin-minority LL. This spectral flow is proportional to the net external flux through the system:
\begin{align}\label{eq:SM_net_flux}
  B\mathcal A
= B 2\pi\ell_b^2N_\phi
&= 2\lambda N_\phi\Phi_0\\
&= (N_{\phi\uparrow} - N_{\phi\downarrow}) \Phi_0
\end{align}
where $\Phi_0=h/(2e)$ is the superconducting flux quantum. Thus, the spectral flow is proportional to the net external flux through the system.

Last but not least, the magnetic length in spin $\sigma$ sector is given by:
\begin{align}
    \ell_\sigma = \frac{\ell_b}{\sqrt{s_\sigma}}
\end{align}
The spin-dependent magnetic lengths modify the orbital wavefunctions in the two spin sectors. The LLL orbital wavefunction is
\begin{align}\label{eq:orbital_wavefunction}
    \phi_{m\sigma} = \frac{1}{\sqrt{2\pi m!} \ell_\sigma} \left(\frac{re^{i\sigma\theta}}{\sqrt{2}\ell_\sigma}\right)^m e^{-\frac{r^2}{4\ell_\sigma^2}}
\end{align}
The factor $\sigma$ in the phase reflects the opposite chirality of the two spin sectors. Hence, the orbitals in the spin-majority LL become more localized, whereas those in the spin-minority LL become more extended. 

Finally, we present the contact-interaction matrix elements in the spin-dependent orbital basis, which will be used in the nonlinear gap equation:
\begin{align}\label{eq:SM_matrix_elements}
    V_{m_1,m_2,m_3,m_4} &\equiv \langle m_1\uparrow, m_2 \downarrow | 2\pi \ell_b^2\delta(\vec r-\vec r')|m_3\uparrow, m_4\downarrow\rangle \\  & = 2\pi \ell_b^2 \int d^2 \mathbf{r}  \phi_{m_1\uparrow}^*(\vec r) \phi_{m_2\downarrow}^*(\vec r) \phi_{m_4\downarrow}(\vec r)\phi_{m_3\uparrow}(\vec r) \\
    & = \frac{1}{2}
\cdot
\frac{
\Gamma\!\left(\frac{S + 2}{2}\right)
}{
2^{\frac{S}{2}}
\sqrt{m_1!\, m_2!\, m_3!\, m_4!}
}
\cdot
(1+\lambda)^{\frac{m_1 + m_3 + 2}{2}}
(1-\lambda)^{\frac{m_2 + m_4 + 2}{2}}\delta_{m_1+m_4,m_2+m_3} 
\end{align}
where $\Gamma(a)$ is the gamma function and $S = m_1+m_2+m_3+m_4$. The Kronecker delta gives the angular-momentum selection rule. Its form differs from the usual same-chirality Landau-level problem because the two spin sectors have opposite chirality. 

Using the spin-dependent LLL orbitals as a basis, the many-body Hamiltonian can be written as
\begin{align}\label{eq:H_LLL}
    H = \sum_{m_1}^{N_{\phi\uparrow}-1} \varepsilon_{\uparrow}c_{m_1\uparrow}^\dagger c_{m_1\uparrow} + \sum_{m_2}^{N_{\phi\downarrow}-1} \varepsilon_{\downarrow}c_{m_2\downarrow}^\dagger c_{m_2\downarrow}  - J \sum_{m_1,m_2,m_3,m_4} V_{m_1,m_2,m_3,m_4} c^\dagger_{m_1\uparrow} c^\dagger_{m_2\downarrow}c_{m_4\downarrow}c_{m_3\uparrow}
\end{align}
where in the last summation, the $m_1$, $m_3$ range from $(0,N_{\phi\uparrow})$ and $m_2$, $m_4$ range from $(0,N_{\phi\downarrow})$.

\subsection{BdG Hamiltonian}\label{sec:mean_field_SC}

In this section, we derive the Bogoliubov-de-Gennes Hamiltonian within mean-field theory and the nonlinear gap equation we present in the main text. 

We define the superconducting order parameter as
\begin{align}\label{eq:SM_gap_def}
    \Delta(\mathbf{r}) = -J 2\pi \ell_b^2 \langle \psi_\downarrow(\mathbf{r}) \psi_\uparrow(\mathbf{r})\rangle = -J 2\pi \ell_b^2 \sum_{m_1,m_2} \phi_{m_1\uparrow}\phi_{m_2\downarrow} \langle c_{m_2\downarrow} c_{m_1\uparrow}\rangle 
\end{align}
The corresponding gap matrix in the guiding center space is given by:
\begin{align}
     \Delta_{m',m} \equiv \int d^2r \Delta(\mathbf{r}) \phi_{m'\uparrow}^*(r) \phi^*_{m\downarrow}(r) &= - J 2\pi \ell_b^2 \sum_{m_1,m_2} \int d^2 \mathbf{r}\phi_{m'\uparrow}^*(r) \phi^*_{m\downarrow}(r) \phi_{m_1\uparrow}\phi_{m_2\downarrow} \langle c_{m_2\downarrow} c_{m_1\uparrow}\rangle \\
     & = -J \sum_{m_1,m_2} V_{m',m,m_1,m_2} \langle c_{m_2\downarrow} c_{m_1\uparrow}\rangle\label{eq:gap_matrix_elements}
\end{align}
where $V_{m',m,m_1,m_2}$ is given by Eq.~\eqref{eq:SM_matrix_elements}. Within the mean-field approximation, the four fermion term in the many-body Hamiltonian in Eq.~\eqref{eq:H_LLL} is given by: 
\begin{align}
    c^\dagger_{m_1\uparrow} c^\dagger_{m_2\downarrow}c_{m_4\downarrow}c_{m_3\uparrow} \approx  \langle c^\dagger_{m_1\uparrow} c^\dagger_{m_2\downarrow} \rangle c_{m_4\downarrow}c_{m_3\uparrow}+  c^\dagger_{m_1\uparrow} c^\dagger_{m_2\downarrow}\langle c_{m_4\downarrow}c_{m_3\uparrow}\rangle -  \langle c^\dagger_{m_1\uparrow} c^\dagger_{m_2\downarrow}\rangle \langle c_{m_4\downarrow}c_{m_3\uparrow} \rangle
\end{align}
We work in the grand canonical ensemble and define the grand-canonical
Hamiltonian
\begin{align}
    K = H-\mu N,
\end{align}
where $\mu$ is the chemical potential and $N$ is the total particle-number
operator. For notational simplicity, we will henceforth denote $K$ by $H$
and refer to it loosely as the Hamiltonian.
The many-body Hamiltonian can then be written as:
\begin{align}\label{eq:mean_field_H}
    H &\approx \sum_{m_1}^{N_{\phi\uparrow}-1} \xi_\uparrow c_{m_1\uparrow}^\dagger c_{m_1\uparrow} + \sum_{m_2}^{N_{\phi\downarrow}-1} \xi_\downarrow c_{m_2\downarrow}^\dagger c_{m_2\downarrow} +  \sum_{m_1,m_2} \Delta_{m_1,m_2}^\dagger c_{m_2\downarrow}c_{m_1\uparrow} +J \sum_{m_1,m_2} \Delta_{m_1,m_2}c^\dagger_{m_1\uparrow} c^\dagger_{m_2\downarrow}- J \sum_{m_1,m_2} \Delta_{m_1,m_2}  \langle c^\dagger_{m_1\uparrow} c^\dagger_{m_2\downarrow}\rangle  
\end{align}
where $\xi_\sigma = \varepsilon_\sigma-\mu$ and we have relabeled the summation index for clarity. 
Since we are primarily interested in the vortex-free and single-vortex sectors, we impose the rotationally invariant ansatz
\begin{align}
    \arg{\Delta(\mathbf{r})} = \eta \theta
\end{align}
where $\eta=0$ corresponds to the zero-winding state and $\eta=1$ to a single vortex. The $\eta=-1$ sector describes a vortex of the opposite chirality to the applied field and is therefore higher in energy, so we do not consider it further.   Angular-momentum conservation then constrains the gap matrix elements in guiding-center space to be:
\begin{align}\label{eq:SM_matrix_elements_gap}
    \Delta_{m',m}\equiv \Delta_{m}^{(\eta)}\delta_{m',m+\eta}
\end{align}
For simplicity, we will write $\Delta_m^{(\eta)}=\Delta_m$ in the BdG Hamiltonian but restore the $\eta$ dependence in the nonlinear gap equation. The Kronecker delta implies that for sector $\eta$, the pairing orbitals are between $(m+\eta,\uparrow)$ and $(m,\downarrow)$. A Cooper pair formed from $(m+\eta,\uparrow)$ and $(m,\downarrow)$ carries net angular momentum $\eta$: the two orbitals contribute $m+\eta$ and $-m$, respectively, because the spin sectors have opposite chirality. Angular-momentum conservation therefore reduces Eq.~\eqref{eq:mean_field_H} to

\begin{align}\label{eq:mean_f_H}
    H
    & = \sum_{m=0}^{N_{\phi\uparrow}-1} \xi_\uparrow c^\dagger_{m\uparrow} c_{m\uparrow} + \sum_{m'=0}^{N_{\phi\downarrow}-1} \xi_\downarrow c^\dagger_{m'\downarrow}c_{m'\downarrow} + \sum_{m=0}^{M} \Delta_{m}c^\dagger_{(m+\eta)\uparrow}c^\dagger_{m\downarrow} + \sum_{m=0}^{M} \Delta_{m}^\dagger c_{m\downarrow}c_{(m+\eta)\uparrow}- \sum_{m=0}^{M} \Delta_{m}\langle c^\dagger_{(m+\eta)\uparrow} c^\dagger_{m\downarrow}\rangle 
\end{align}
where
\begin{align}
M=\min(N_{\phi\downarrow}-1, N_{\phi\uparrow}-1-\eta),
\qquad
\xi_\sigma=\varepsilon_{0\sigma}-\mu .
\end{align}
The last term in the Hamiltonian contributes an overall energy shift. We omit it when diagonalizing the BdG Hamiltonian and restore it when evaluating the total energy. 

The Hamiltonian can be further written as  $2\times2$ BdG form:

\begin{align}
    H_{\rm BdG} &=\sum_{m=0}^{M}  \begin{pmatrix}
        c^\dagger_{(m+\eta)\uparrow} & c_{m\downarrow}
    \end{pmatrix} \begin{pmatrix}
        \xi_\uparrow & \Delta_{m} \\
        \Delta^*_{m} & -\xi_\downarrow 
    \end{pmatrix} \begin{pmatrix}
        c_{(m+\eta)\uparrow}\\ c^\dagger_{m\downarrow} 
    \end{pmatrix} + H_{\rm unpaired}+ \text{constant}
\end{align}
where the $H_{\rm unpaired}$ is the Hamiltonian for unpaired orbitals:
\begin{align}\label{eq:unpaired_H}
    H_{\rm unpaired} =
\sum_{m\in
\{0,\dots,\eta-1\}
\cup
\{\eta+M+1,\dots,N_{\phi\uparrow}-1\}}
\xi_\uparrow c^\dagger_{m\uparrow}c_{m\uparrow}+\sum_{m\in\{M+1,\dots,N_{\phi\downarrow}-1\}}
\xi_\downarrow c^\dagger_{m\downarrow}c_{m\downarrow}.
\end{align}
These unpaired states are localized in their corresponding guiding-center orbitals. Since the unpaired Hamiltonian is diagonal in guiding-center space, we only need to diagonalize the BdG Hamiltonian of the paired orbitals.

The BdG Hamiltonian is solved by the following eigenvalue matrix:
\begin{align}
    \begin{pmatrix}
        \xi_\uparrow & \Delta_{m} \\
        \Delta^*_{m} & -\xi_\downarrow 
    \end{pmatrix} \begin{pmatrix}
        u_m \\ v_m
    \end{pmatrix} = E_m \begin{pmatrix}
        u_m \\ v_m
    \end{pmatrix}
\end{align}
The eigenvalue and corresponding eigenvector are given by:
\begin{align}\label{eq:two_energy_branch}
    E_{m\pm} &= \frac{\lambda \hbar \omega_b}{2} \pm \sqrt{|\Delta_m|^2+\xi^2}, \quad \psi_{+} = \begin{pmatrix}
        u_m \\
        v_m \\
    \end{pmatrix}, \quad \psi_{-} = \begin{pmatrix}
        v_m^* \\
        -u_m^* \\
    \end{pmatrix}
\end{align}
Here, $ \xi = \frac{\hbar \omega_b}{2} -\mu$. The coherence factor $u_m$ and $v_m$ are given by
\begin{align}\label{eq:SM_coherence_factor}
    |u_m|^2 = \frac{1}{2} \left( 1+\frac{\xi}{\sqrt{\xi^2 + |\Delta_m|^2}} \right), |v_m|^2 = \frac{1}{2} \left( 1- \frac{\xi}{\sqrt{\xi^2 + |\Delta_m|^2}}  \right), \quad u_m v_m^* = \frac{\Delta_m}{2\sqrt{\xi^2+|\Delta_m|^2}}
\end{align}
Thus, apart from the dependence of $\Delta_m$ itself, the coherence factors do not explicitly depend on the external magnetic field.
We define the Bogoliubov quasiparticle operators by
\begin{align}\label{eq:quasi_particle_operator_def}
    \gamma_{m\uparrow} = u_m^*c_{(m+\eta)\uparrow} + v_m^* c^\dagger_{m\downarrow}, \quad \gamma_{m\downarrow}^\dagger = -v_m c_{(m+\eta)\uparrow} + u_m c^\dagger_{m\downarrow}
\end{align}
In terms of these operators, the paired BdG Hamiltonian takes the diagonal form up to some constant energy shift, 
\begin{align}
    H_{\rm BdG} = \sum_{m=0}^{M}\sum_{\sigma} E_{m\sigma}\gamma^\dagger_{m\sigma}\gamma_{m\sigma}
\end{align}
where $\sigma=\uparrow,\downarrow$ and the $E_{m\sigma}$ are given by
\begin{align}
    E_{m\uparrow} = E_{m+} = \sqrt{|\Delta_m|^2+\xi^2}+\frac{\lambda \hbar \omega_b}{2}, \quad E_{m\downarrow} = - E_{m-} = \sqrt{|\Delta_m|^2+\xi^2}-\frac{\lambda \hbar \omega_b}{2}
\end{align}
The Bogoliubov quasiparticle occupations satisfy

\begin{align}\label{eq:quasi_particle_occ}
    \langle \Phi_\eta |\gamma_{m\sigma}^\dagger \gamma_{m'\sigma'}|\Phi_\eta \rangle = \delta_{mm'}\delta_{\sigma\sigma'}f(E_{m\sigma}), \quad f(E) = \frac{1}{e^{\frac{E}{k_BT}}+1}
\end{align}
where $f(E)$ is the Fermi-distribution function. Here, $T$ is the temperature. In the limit of $T=0$, for weak field $\lambda$ such that $\sqrt{|\Delta_m|^2+\xi^2}>\lambda \hbar \omega_b/2$, 
\begin{align}
\langle\Phi_\eta|\gamma^\dagger_{m\sigma}\gamma_{m\sigma}|\Phi_\eta\rangle=0
\end{align}
where the ground state is
\begin{align}\label{eq:BCS_groundstate_eta}
    |\Phi_{\eta}\rangle = \prod_m(u_m^*-v_m^*c^{\dagger}_{(m+\eta)\uparrow}c^{\dagger}_{m\downarrow})|0\rangle 
\end{align}
The difference from the familiar expression comes from the notation convention in the eigenvectors in Eq.~\eqref{eq:two_energy_branch}.

We next derive the self-consistent gap equation. Let's first express the electron creation and annihilation operator in terms of the quasi-particle operator using Eq.~\eqref{eq:quasi_particle_operator_def}, 
\begin{align}\label{eq:inver_trans}
    c_{(m+\eta)\uparrow} = u_m\gamma_{m\uparrow}-v_m^* \gamma_{m\downarrow}^\dagger, \quad c_{m\downarrow}^\dagger = v_m\gamma_{m\uparrow}+u_m^* \gamma_{m\downarrow}^\dagger
\end{align}
Therefore, the anomalous average takes the following form:
\begin{align}
   \left\langle c_{m\downarrow}c_{(m+\eta)\uparrow}\right\rangle
={}&
-u_m v_m^*
+
u_m v_m^*
\left(
\langle \gamma_{m\uparrow}^\dagger\gamma_{m\uparrow}
+
\gamma_{m\downarrow}^\dagger\gamma_{m\downarrow}\rangle 
\right)-(v_m^*)^2
\langle \gamma_{m\uparrow}^\dagger\gamma_{m\downarrow}^\dagger\rangle 
+
u_m^2
\langle\gamma_{m\downarrow}\gamma_{m\uparrow}\rangle \\
&=
u_{m}v_{m}^*
\left[
f(E_{m\uparrow})+f(E_{m\downarrow})-1
\right] \\
&=  \frac{\Delta_m}{2\sqrt{\xi^2+|\Delta_m|^2}}\left[
f(E_{m\uparrow})+f(E_{m\downarrow})-1
\right]
\end{align}
where in the first line, the last two terms vanish in the ground state and we used Eq.~\eqref{eq:quasi_particle_occ} and Eq.~\eqref{eq:SM_coherence_factor} in the second and third line separately. 
Substituting the above expression into Eq.~\eqref{eq:gap_matrix_elements} and restoring the $\eta$ dependence, we obtain the self-consistent nonlinear gap equation:
\begin{align}
    \Delta^{(\eta)}_{m}
    & =-J  \sum_{m'=0}^M V_{m+\eta,m,m'+\eta,m'} \frac{\Delta^{(\eta)}_{m'}}{2\sqrt{\xi^2+|\Delta^{(\eta)}_{m'}|^2}} \left[
f(E_{m\uparrow})+f(E_{m\downarrow})-1
\right]\label{eq:SM_self_consistent_equation}
\end{align}
The nonlinear gap equation depends explicitly on the chemical potential $\mu$ through $\xi$, which is determined simultaneously from the number equation:

\begin{align}\label{eq:SM_number_equation}
    \sum_{m\sigma}\langle  c_{m\sigma}^\dagger c_{m\sigma}\rangle=N \equiv \nu N_\phi
\end{align}
where $\nu\in(0,2)$ is the total filling fraction of the LLL.
We can group the $\langle c^\dagger_{m\sigma}c_{m\sigma}\rangle $ into paired orbitals and unpaired orbitals:
\begin{align}
    \sum_{m} \langle c^\dagger_{m\sigma}c_{m\sigma} \rangle= \sum_{\rm paired}\langle c^\dagger_{m\sigma}c_{m\sigma} \rangle+\sum_{\rm unpaired} \langle c^\dagger_{m\sigma}c_{m\sigma}\rangle
\end{align}
For the paired orbitals, the occupation is determined by the quasi-particle operator. Using Eq.~\eqref{eq:inver_trans}, the occupation of the paired orbitals is given by:
\begin{align}
     \langle c_{(m+\eta)\uparrow}^{\dagger}
c_{(m+\eta)\uparrow}\rangle 
={}&
|v_m|^2
+|u_m|^2
\langle \gamma_{m\uparrow}^\dagger\gamma_{m\uparrow}\rangle 
-|v_m|^2
\langle \gamma_{m\downarrow}^\dagger\gamma_{m\downarrow}\rangle-
u_m^*v_m^*
\langle\gamma_{m\uparrow}^\dagger
\gamma_{m\downarrow}^\dagger\rangle
-
u_mv_m
\langle\gamma_{m\downarrow}
\gamma_{m\uparrow}\rangle, \\
\langle c_{m\downarrow}^{\dagger}
c_{m\downarrow}\rangle
={}&
|v_m|^2
-|v_m|^2
\langle\gamma_{m\uparrow}^\dagger\gamma_{m\uparrow}\rangle
+|u_m|^2
\langle\gamma_{m\downarrow}^\dagger\gamma_{m\downarrow}\rangle-
u_m^*v_m^*
\langle\gamma_{m\uparrow}^\dagger
\gamma_{m\downarrow}^\dagger\rangle
-
u_mv_m
\langle\gamma_{m\downarrow}
\gamma_{m\uparrow}\rangle.
\end{align}
So the contribution for the number expectation from the paired orbitals is given by:
\begin{align}\label{eq:occupation}
    \sum_{m=0}^M \left(\langle c_{(m+\eta)\uparrow}^\dagger c_{(m+\eta)\uparrow} \rangle + \langle c_{m\downarrow}^\dagger c_{m\downarrow}\rangle \right) &= \sum_{m=0}^M (2| v_m|^2+(|u_m|^2-|v_m|^2)(f(E_{m\uparrow})+f(E_{m\downarrow})))\\
    & = \sum_{m=0}^{M} \left(1+\frac{\xi}{\sqrt{\xi^2+|\Delta_m|^2}}(f(E_{m\uparrow})+f(E_{m\downarrow})-1)\right)
\end{align}
As for the unpaired orbitals, the occupation is
determined separately from its single-particle energy. Their
contribution to the number equation is
\begin{align}
    \sum_{m\in
\{0,\dots,\eta-1\}
\cup
\{\eta+M+1,\dots,N_{\phi\uparrow}-1\}}
\langle c_{m\uparrow}^\dagger c_{m\uparrow}\rangle + \sum_{m\in\{M+1,\dots,N_{\phi\downarrow}-1\}}
\langle c^\dagger_{m\downarrow}c_{m\downarrow} \rangle = \max{(2\lambda N_\phi,\eta)}f(\xi_\uparrow)+\max{(0,\eta-2\lambda N_\phi)}f(\xi_\downarrow)
\end{align}
Together, the nonlinear gap equation and the number equation determine $\Delta_m$ and $\mu$ self-consistently. 

Last but not least, let's consider the total energy. Here, we include both the paired and unpaired orbitals and demonstrate that the contribution from unpaired orbitals is negligible.  The canonical energy is obtained from the expectation value of the
grand-canonical Hamiltonian in Eq.~\eqref{eq:mean_f_H} by adding back
the chemical-potential contribution:
\begin{align}
    E &= \langle H\rangle+\mu \langle N \rangle \\
    &\approx N_\uparrow \xi_\uparrow + N_\downarrow \xi_\downarrow+ \mu \nu N_\phi  - \sum_m \Delta_m \langle c^\dagger_{(m+\eta)\uparrow} c^\dagger_{m\downarrow}\rangle \\
    & = \frac{\hbar \omega_b}{2} N + \frac{\lambda \hbar\omega_b}{2}(N_\uparrow-N_\downarrow)  - \frac{1}{J 2\pi \ell_b^2}  \int d^2\mathbf{r} |\Delta(\mathbf r)|^2 \label{eq:total_energy}
\end{align}
where in the second line we have rewritten the guiding-center sum in real space.  As for the second term $N_\uparrow-N_\downarrow$, the $N_\uparrow$ and $N_\downarrow$ in the $T=0$ is given by: 
\begin{align}
    N_\uparrow &= \sum_{m}\langle c^\dagger_{m\uparrow}c_{m\uparrow}\rangle = \sum_m | v_m|^2 +  \max{(2\lambda N_\phi,\eta)}f(\xi_\uparrow)  \\
    N_\downarrow & = \sum_m \langle c_{m\downarrow}^\dagger c_{m\downarrow} \rangle = \sum_m ( 1-|u_m|^2) + \max{(0,\eta-2\lambda N_\phi)}f(\xi_\downarrow) = \sum_m |v_m|^2  + \max{(0,\eta-2\lambda N_\phi)}f(\xi_\downarrow)
\end{align}
The first term comes from paired orbitals while the second term comes from unpaired orbitals.  The second term in the energy vanishes in the weak-field thermodynamic limit $\lambda(N_\uparrow-N_\downarrow) \sim O(\lambda^2 N_\phi)\ll1$ and the first term is fixed by the number expectation equation
. Thus, the Landau-level kinetic contribution is identical between the paired bulk sectors of the uniform and vortex states. Therefore, the state-dependent part of the on-shell superconducting energy can be written as 
\begin{align}\label{eq:cond_energy}
    E_{\rm cond} = - \frac{1}{J 2\pi \ell_b^2}  \int d^2\mathbf{r} |\Delta(\mathbf r)|^2 
\end{align}
Therefore, when comparing the energies of different self-consistent superconducting states at fixed particle number, it is sufficient to consider only this state-dependent superconducting-energy contribution.

\section{Zero Field}
In this section, we investigate superconductivity in a flat band at zero magnetic field. We first consider the spatially uniform superconducting state in Sec.~\ref{sec:uniform_zero_field}, where analytical expressions are derived for the ground state, the zero-temperature gap amplitude, and the critical temperature. We then study the single-vortex solution in Sec.~\ref{sec:vortex_zero_field}, followed by a calculation of the phase stiffness in Sec.~\ref{sec:phase_stiffness}. The purpose of this section is to establish analytical results in the absence of a magnetic field, which will serve as a reference for the finite-field analysis presented in Sec.~\ref{sec:finite_field_study}.
\subsection{Uniform solution: $\eta=0$}\label{sec:uniform_zero_field}

When $\lambda=0$ and $\eta=0$, the self-consistent gap equation Eq.~\eqref{eq:SM_self_consistent_equation} and the number expectation equation Eq.~\eqref{eq:SM_number_equation} reduces to the following form:
\begin{align}
    &\Delta_{m,\eta=0}(\lambda=0) 
     = -J\sum_{m'=0}^{N_\phi-1} V_{m,m,m',m'}(\lambda=0) \frac{\Delta_{m',\eta=0}}{2\sqrt{\xi^2 + |\Delta_{m',\eta=0}|^2}} (f(E_{m\uparrow})+f(E_{m\downarrow})-1) \\
    & \sum_{m=0}^{N_\phi-1}\left(1+\frac{\xi}{\sqrt{\xi^2+|\Delta_{m}|^2}}(f(E_{m\uparrow})+f(E_{m\downarrow})-1)\right) = \nu N_\phi
\end{align}
At zero external field, the paired electrons experience zero net magnetic field, so the natural ansatz is a uniform superconducting state with constant gap matrix $\Delta_{m,\eta=0}(\lambda=0)=\Delta$. As a result, the energy spectrum of the BdG Hamiltonian is also independent of the guiding-center index, $E_{m\sigma}=E_{\sigma}$. Thus the gap and number equations become 
\begin{align}
    &\Delta = -J\frac{\Delta}{2\sqrt{\xi^2+|\Delta|^2}} (f(E_\uparrow)+f(E_\downarrow)-1) \\
    &1+\frac{\xi}{\sqrt{\xi^2+|\Delta|^2}}(f(E_{\uparrow})+f(E_{\downarrow})-1) = \nu
\end{align}
where we have used $\sum_{m'} V_{m,m,m',m'}=1$ that is independent of $m$ for large $N_\phi$. The same thermal factor appears in both equations and can be eliminated, giving
\begin{align}\label{eq:chemical_potential}
    \xi = \frac{J}{2} (1-\nu), \quad \mu = \frac{\hbar \omega_b}{2} - \frac{J}{2}(1-\nu)
\end{align}

\begin{figure}[t]
    \centering
    \includegraphics[width=1\linewidth]{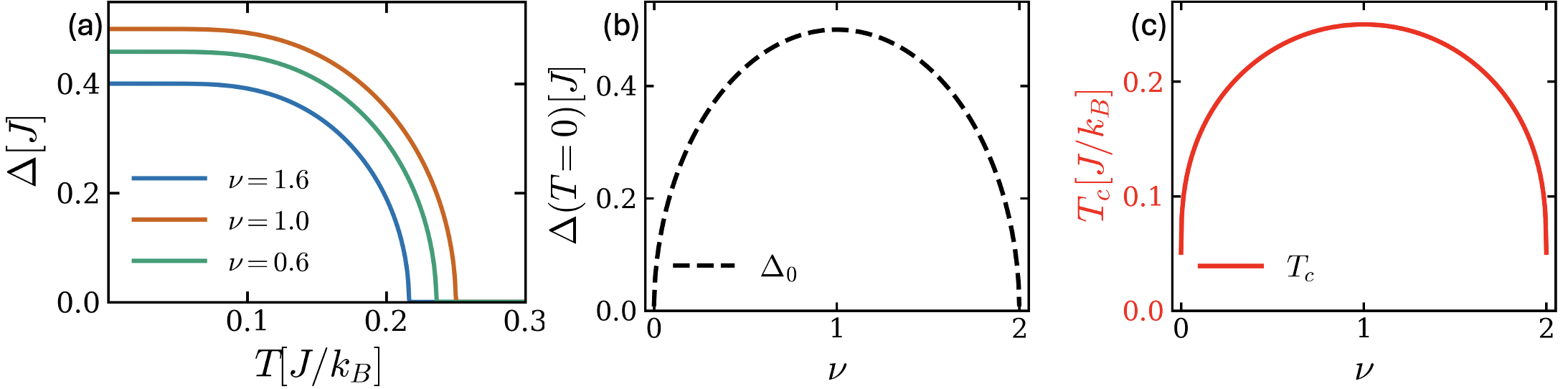}
    \caption{(a) The self-consistent gap solution as function of temperature. (b) The zero T superconducting gap as a function of total filling fraction. (c) The critical temperature $T_c$ as a function of total filling fraction.}
    \label{fig:SM_gap_Tc}
\end{figure}
Thus the chemical potential is independent of $T$ below the critical temperature. 
In an ordinary metal, increasing temperature broadens the Fermi--Dirac distribution and the chemical potential shifts to keep the particle number fixed. In the projected LLL, however, there is no dispersive spectrum over which particles can redistribute. Instead, the temperature dependence of $\Delta(T)$ changes the coherence factors and maintains the number constraint, allowing $\xi$ and hence $\mu$ to remain temperature independent below $T_c$. 
For $T>T_c$, the chemical potential is determined by
\begin{equation}
    \nu=2f(\xi)
\end{equation}
Inverting this relation gives
\begin{equation}
    \mu_N(T)=\frac{\hbar\omega_b}{2}+k_BT\,\ln\frac{\nu}{2-\nu}
\end{equation}
The subscript $N$ denotes the normal state. One can verify that $\mu_N(T\rightarrow T_c)=\mu$, with $\mu$ given by Eq.~\eqref{eq:chemical_potential}. Thus the chemical potential is continuous across the superconducting phase transition even though it does not change for $T<T_c$.

Fig.~\ref{fig:SM_gap_Tc}(a) shows the self-consistent solution to the gap equation. The gap decreases with increasing temperature, and vanishes at the critical temperature. 
Since the chemical potential is temperature independent below $T_c$, the zero-temperature gap follows directly from the gap equation:
\begin{align}\label{eq:zero_T_gap}
    \Delta(T\rightarrow 0) = \frac{J}{2} \sqrt{\nu(2-\nu)}
\end{align}
The real-space order parameter is
\begin{align}
    \Delta_{\rm u}(\vec r) &= 2\pi \ell^2_{b}\sum_{m=0}^{N_\phi-1} \Delta \phi_{m\uparrow}(\vec r) \phi_{m\downarrow}(\vec r)
=\Delta
\end{align}
Here, the subscript "u" stands for uniform.

Fig.~\ref{fig:SM_gap_Tc}(b) shows the zero temperature gap as a function of filling fraction. 
The gap is maximal at $\nu=1$, where charge fluctuations are largest, and is symmetric under $\nu\leftrightarrow2-\nu$, reflecting particle-hole symmetry in the projected LL. Moreover, the gap is linear in the interaction strength $J$, in contrast to the exponentially small weak-coupling gap of a dispersive BCS metal.
With the zero temperature $T=0$ gap expression, the BdG spectrum is given by
\begin{align}
    E_{\sigma} &= \sqrt{|\Delta|^2 + \xi^2} = \frac{J}{2}
\end{align}
so the energy cost to create a quasi-particle is independent of density even though the superconducting gap $\Delta$ depend on density.In the $T=0$, the coherence factor is given by:
\begin{align}
    |u|^2 = 1-\frac{\nu}{2},\quad 
    |v|^2 = \frac{\nu}{2}
\end{align}
The BCS ground state at $T=0$ is obtained in Eq.~\eqref{eq:BCS_groundstate_eta}, with guiding center independent coherence factor given above, 
\begin{align}
    |\Phi_{\rm u}\rangle = \prod_m ( u^*-v^* c^\dagger_{m\uparrow}c^\dagger_{m\downarrow} )|0\rangle
\end{align}

The guiding center independent coherent factors reflect the fact that for the uniform superconductor, the pairing amplitude are the same across all the guiding center. This equal-weight participation is enforced by
magnetic-translation symmetry and is required for the order parameter
to remain spatially uniform.
When $\nu=0$, the state just represent vacuum while when $\nu=2$, the state represent fully occupied LLL which is a trivial Slater determinant. The reason is that at $\nu=2$, the Slater determinant state is incompressible and has charge gap, in the projected LL, the charge gap can be considered as infinite, so there is no low energy charge fluctuations,  the superconducting state is not favored and the state becomes trivial quantum hall insulator. 

Next, let's consider the energy of the superconducting state. The total energy is given in Eq.~\eqref{eq:total_energy} and the energy per orbital (i.e. $E/N_\phi$) is therefore given by:
\begin{align}
   \frac{E_{SC}}{N_\phi} =  \frac{\langle H_{\text{BdG}} \rangle}{N_\phi} + \mu \frac{N}{N_\phi} = \frac{\hbar \omega_b }{2} \nu  -\frac{\nu(2-\nu)}{4} J
\end{align}
The first term is the LLL kinetic energy which equals to the normal state energy. The second term is the superconducting condensation energy. This energy indicates that as long as $0<\nu<2$, the superconducting state is always lower in energy than the normal state.

Finally, 
the critical temperature can also be found by setting $\Delta\rightarrow0$ in the gap equation, the result is given by:
\begin{align}\label{eq:SM_critical_temperature}
T_c = \frac{1-\nu}{2\ln(\frac{2}{\nu}-1)}\frac{J }{k_B}
\end{align}
Fig.~\ref{fig:SM_gap_Tc}(c) shows critical temperature as a function of the total filling fraction. 
The critical temperature is likewise linear in the interaction strength and maximal at $\nu=1$. Physically, this linear dependence arises because all guiding-center orbitals participate equally in superconducting pairing, strongly enhancing superconductivity in the flat band.

\subsection{Single-vortex solution: $\eta=1$}\label{sec:vortex_zero_field}

Here, we derive the single-vortex solution in the zero-external-field limit. We focus on the zero-temperature limit $T=0$. The single-vortex state is obtained by setting $\eta=1$ in the gap equation, giving the corresponding self-consistent gap and number equations:
\begin{align}
   &\Delta_{m,\rm \eta=1}(\lambda=0) = J  \sum_{m'=0}^M V_{m,m+1,m',m'+1}(\lambda=0) \frac{\Delta_{m',\eta=1}}{2\sqrt{\xi^2+|\Delta_{m',\eta=1}|^2}}\label{eq:SM_single_vortex_gap_equation}\\
   & \sum_{m=0}^{M}\left(1-\frac{\xi}{\sqrt{\xi^2+|\Delta_{m}|^2}}\right)+2\theta(-\xi) = \nu N_\phi
\end{align}
\begin{figure}[h]
    \centering
    \includegraphics[width=1\linewidth]{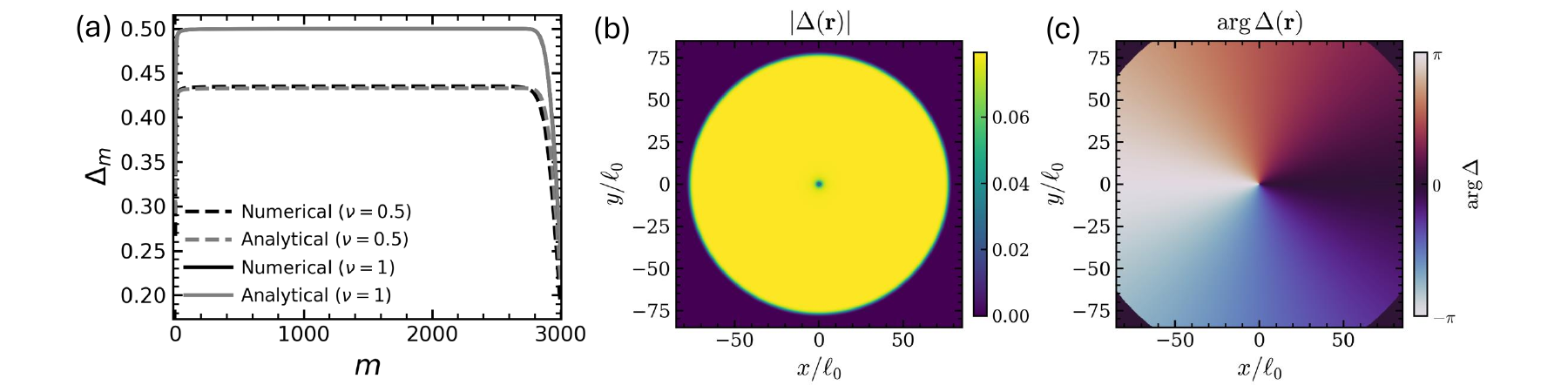}
    \caption{(a) The superconducting gap matrix of single-vortex state as a function of the guiding center at different filling fraction.  (b) The real space distribution of the gap amplitude for the single-vortex solution.(c) The phase winding of the single-vortex solution. Here we take $N_\phi=3000$. }
    \label{fig:SM_single_vortex}
\end{figure}
where $M= N_\phi-2$ and $\theta(x)$ is the step function. The first term in the number equation comes from the paired orbitals while the second term is from the unpaired orbitals.  For large orbital index $m$ , $\Delta_m$ should approach the uniform gap of Eq.~\eqref{eq:zero_T_gap} because the larger the $m$ the more far the orbitals away from the origin, and both the gap and number equations are dominated by contributions from large orbital index $m$. Replacing $\Delta_{m'}$ and $\xi$ inside the sums by their uniform-state values (Eqs.~\eqref{eq:zero_T_gap} and \eqref{eq:chemical_potential}) then gives a closed-form approximation for the converged gap matrix:
\begin{align}\label{eq:SM_single_vortex_half}
    \Delta_{m,\eta=1} &\approx \Delta \sum_{m'=0}^{N_\phi-2} V_{m,m+1,m',m'+1} \\
    & \equiv \Delta S_{m,\eta=1}(\lambda=0) 
\end{align}
where $\Delta$ is given by Eq.~\eqref{eq:zero_T_gap} and $S_{m,\eta}$ is defined as:
\begin{align}
    S_{m,\eta}(\lambda) = \sum_{m'} V_{m,m+\eta,m',m'+\eta}(\lambda)
\end{align}
When $\eta=0$, the $S_{m,0}(\lambda=0)=1$ correspond to the uniform gap matrix.
Fig.~\ref{fig:SM_single_vortex}(a) compares this approximation (gray) against the full numerical solution of the self-consistent equations (black) as a function of guiding center. At half filling the two agree very well, since $\xi=0$ solves the number equation exactly and $\Delta_{m'}$ cancels out between numerator and denominator. This cancellation is valid because the interaction matrix elements are all real as shown in Eq.~\eqref{eq:SM_matrix_elements}. Thus, the self-consistent equations reduce identically to Eq.~\eqref{eq:SM_single_vortex_half}. Away from half filling (e.g. $\nu=0.5$), the analytic profile still tracks the numerical solution closely, for the reason given above.

The real-space gap function follows from the inverse transformation of Eq.~\eqref{eq:SM_matrix_elements_gap}:
\begin{align}
    \Delta_{\rm v}(\vec r) &= 2\pi \ell^2_{b}\sum_{m=0}^{N_\phi-2} \Delta_{m,\eta=1} \phi_{(m+1)\uparrow}(\vec r) \phi_{m\downarrow}(\vec r) \\
    & = \Delta\frac{r e^{i \theta }}{\sqrt{2}\ell_b} \sum_{m=0}^{N_\phi-2}\frac{S_{m,1}(0)}{\sqrt{(m+1)!m!}} \left( \frac{r^2}{2\ell_b^2}  \right)^m e^{-\frac{r^2}{2\ell_b^2}} \label{eq:real_space_single_vortex}
\end{align}
Here the subscript "v" stands for vortex.
Fig.~\ref{fig:SM_single_vortex}(b) and (c) show the amplitude and phase of the above gap function in the real space. The dark point in (b) at the center indicates vortex core where the gap amplitude goes to zero and the phase winding around it is $2\pi$. For small radius $r\ll \sqrt{2}\ell_b$,  the majority contribution of the orbital summation  comes from the $m=0$, so the gap function is proportional to $re^{i\theta}$ at small $r$. For $r\gg \sqrt{2}\ell_b$, note that the summation is a Poisson average of $S_{m,1}/\sqrt{m+1}$:
\begin{align}\label{eq:S_{m,1}_eval}
    \sum_m \frac{S_{m,1}(0)}{\sqrt{m!(m+1)!}} \left(\frac{r^2}{2\ell_b^2}\right)^m e^{-\frac{r^2}{2\ell_b^2}} &= \sum_m \frac{S_{m,1}(0)}{\sqrt{m+1}} \frac{1}{m!}(\frac{r^2}{2\ell_b^2})^m e^{-\frac{r^2}{2\ell_b^2}}\\
    & \equiv \langle \frac{S_{m,1}(0)}{\sqrt{1+m}}\rangle_{\text{Poisson}}
\end{align}
The main contribution for large $r$ is from $m \sim \frac{r^2}{2\ell_b^2}$ and note that $S_{m,1}(0)\approx 1-1/4m$ for large $m$, the average is approximately given by:
\begin{align}
    \langle \frac{S_{m,1}(0)}{\sqrt{m+1}} \rangle_{\text{Poisson}} \approx \frac{\sqrt{2}\ell_b}{r} - \frac{3\sqrt{2}\ell_b^3}{4r^3}
\end{align}
So the leading order for the gap function is the uniform superconducting gap that we solved in the previous section. The asymptotic expression for the single vortex gap function is given by the following:
\begin{align}\label{eq:vortex_R_r_form}
  \Delta_{\rm v}(\vec r) \approx \begin{cases}
      &\Delta_{\rm u} \frac{re^{i\theta}}{\sqrt{2}\ell_b} ,\quad r\ll \sqrt{2} \ell_b \\
      & \Delta_{\rm u} e^{i\theta}(1-\frac{3\ell_b^2}{4r^2}+O(r^{-4})), \quad r\gg \sqrt{2} \ell_b
  \end{cases} 
\end{align}
Therefore, the gap vanishes linearly near the vortex core and remain uniform away from the vortex core. The size of the vortex core is therefore dictated by the  magnetic length of the internal field,
\begin{align}
    \xi \equiv \sqrt{2} \ell_b. 
\end{align}
In contrast to a conventional BCS vortex, whose core radius $\xi_{sc}=\hbar v_F/\pi\Delta$ emerges from a dynamical competition between kinetic and condensation energy, here the gap amplitude enters Eq.~\eqref{eq:SM_single_vortex_half} only as an overall multiplicative prefactor, while the spatial profile of the vortex is entirely determined by the pairing form factor $S_{m,1}$, which is built from guiding-center orbitals of fixed size $\ell_b$. The vortex core size is therefore set kinematically by $\ell_b$ and is independent of $\Delta$.

\subsection{Phase Stiffness}\label{sec:phase_stiffness}
In this section, we compute the phase stiffness of the uniform superconducting state and the result is shown in Eq.~\eqref{eq:phase_stiffness_result}.  

The phase stiffness is defined as the logarithmic energy cost of twisting the superconducting phase; on a disk, imposing a single global phase winding is equivalent to placing a single vortex at the origin. We therefore define the phase stiffness $\rho_b$ through
\begin{equation}\label{eq:stiffness_def}
    E_{\rm v}-E_{\rm u} \equiv \frac{\pi \rho_b}{2} \ln (\frac{\mathcal{A}}{2 \pi \ell^2_b})+E_{\rm core}
\end{equation}
where $E_{\eta}$ is the total energy, $\mathcal{A}$ is the sample area and the $E_{\rm core}$ is the core energy.  Their difference is  determined
by the on-shell superconducting energy in Eq.~\eqref{eq:cond_energy}, 
\begin{align}
    E_{\rm v}- E_{\rm u} = -\frac{1}{J2\pi \ell_b^2} \int _{\mathcal{A}} d^2 \left( |\Delta_{\rm v}|^2-|\Delta_{\rm u}|^2  \right)
\end{align}
To extract the logarithmic term which is long distance behavior, we use the asymptotic expression in Eq.~\eqref{eq:vortex_R_r_form}, so the difference of the square of the order parameter in the long distance is given by:
\begin{align}
    |\Delta_{\rm v}|^2 - |\Delta_{\rm u}|^2 \approx -|\Delta|^2 \frac{3\ell_b^2}{2r^2}
\end{align}
where we used $|\Delta_{\rm u}|=\Delta$ where $\Delta$ is given by Eq.~\eqref{eq:zero_T_gap}. Therefore, the logarithmic energy is given by:
\begin{align}
E_{\rm v}-E_{\rm u}
=
\frac{3|\Delta|^2}{4J}
\ln\left(
\frac{\mathcal A}{2\pi\ell_b^2}
\right)
+E_{\rm core}
\end{align}

So the phase stiffness $\rho_b$ is given by:
\begin{align}\label{eq:phase_stiffness_result}
    \rho_b = \frac{3|\Delta|^2}{2\pi J}
\end{align}
Therefore, even though the fermi velocity vanish in the flat band limit, but the stiffness is nonzero. The core energy value is not universal and depends on the convention of the logarithmical scaling denominator. A full estimation of the  energy difference  gives $E_{\rm core} = 0.746\pi \rho_b$.


\section{Finite Field}\label{sec:finite_field_study}
In this section, we study the response of the vortex-free (Sec.~\ref{sec:uniform_finite_field}) and single-vortex (Sec.~\ref{sec:vortex_finite_field}) states to a weak external field, $0<\lambda\ll1$, and analyze different paths to the weak-field thermodynamic limit. Unless otherwise stated, we focus on $T=0$ and $\nu=1$, where the superconducting gap is maximal [Fig.~\ref{fig:SM_gap_Tc}(b)].
\subsection{Vortex-free Solution: $\eta=0$}\label{sec:uniform_finite_field}
In this section, we show that a weak external magnetic field gives the vortex-free state a Gaussian envelope whose decay length scales as $1/\lambda$, rather than the $1/\sqrt{\lambda}$ scale obtained from the linearized GL theory in Sec.~\ref{sec:GL_theory}. Depending on the ratio between the sample area and the Gaussian localization area, the $\eta=0$ solution can therefore remain nearly uniform across the disk or become a localized magnetic pairing droplet, reflecting the singular weak-field thermodynamic limit. We use ``vortex-free and ``uniform interchangeably for the $\eta=0$ sector.

\begin{figure}
    \centering
    \includegraphics[width=0.5\linewidth]{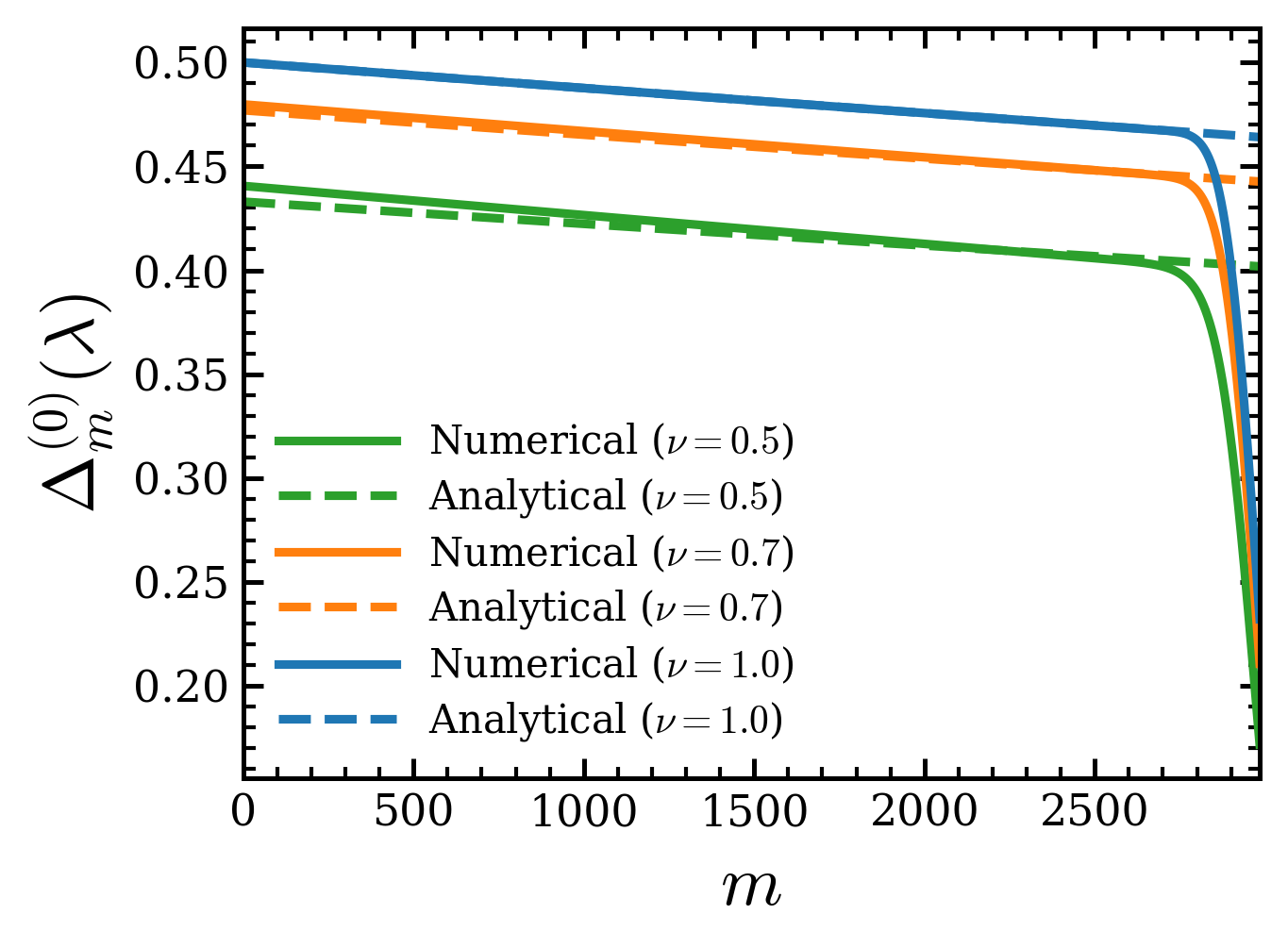}
    \caption{This graph compares the numerically converged gap matrix with the closed form in Eq.~\eqref{eq:SM_closed_gap_matrix} under finite magnetic field $\lambda=0.005$.}
    \label{fig:SM_filling_dependence}
\end{figure}

When $\lambda\neq0$, the self-consistent gap and number equations with $\eta=0$ become the following:
\begin{align}
    &\Delta_{m,\eta=0}(\lambda) 
     = J\sum_{m'=0}^{M} V_{m,m,m',m'}(\lambda) \frac{\Delta_{m',\eta=0}}{2\sqrt{\xi^2 + |\Delta_{m',\eta=0}|^2}} \\
    & \sum_{m=0}^{M}\left(1-\frac{\xi}{\sqrt{\xi^2+|\Delta_{m,\eta=0}|^2}}\right)+2\lambda N_\phi f(\xi_\uparrow) = N_\phi
\end{align}
where $M = (1-\lambda)N_\phi-1$ and $\xi_\uparrow = (1+\lambda)\hbar \omega_b/2-\mu$.
In the number equations, since $\lambda\ll1$, the contribution from the unpaired orbitals is much smaller than the paired orbitals, so the $\xi\approx 0$ is still an approximated converged solution to the number equations, and the gap equations leads to the following gap matrix:
\begin{align}
    \Delta_{m,\eta=0}(\lambda) = \Delta \sum_{m'=0}^{(1-\lambda)N_\phi-1} V_{m,m,m',m'}(\lambda) & = \Delta S_{m,0}(\lambda)\\
    & \approx \Delta \frac{(\sqrt{1-\lambda^2})^{m+1}}{(2-\sqrt{1-\lambda^2})^{m+2}}.
    \label{eq:SM_closed_gap_matrix}
\end{align}
In the second equality, we take the large $N_\phi$ limit in the summation and neglect the boundary effect. Although we focus on $\nu=1$, the solution at other filling fractions also satisfies the expression above, with $\Delta$ given by Eq.~\eqref{eq:zero_T_gap}. This is because, for small $\lambda$, the gap matrix varies slowly with the guiding center. The guiding-center dependence of $\Delta_m'/\sqrt{\xi^2+|\Delta_m'|^2}$ largely cancels. We may therefore approximate both $\Delta_m'$ and $\xi$ on the right-hand side of the summation by their zero-field values, thereby obtaining the expression above.
A comparison between the analytical expression and the numerical results is shown in Fig.~\ref{fig:SM_filling_dependence}. The dashed curves represent the closed-form expression, whereas the solid curves represent the numerical results. At half-filling, $\nu=1$, the numerical result agrees exactly with the closed-form expression because $\xi\approx 0$. The guiding center dependence in the $\Delta_m'$ cancels out in the ratio.  Away from half-filling, the analytical results deviate only slightly from the numerical results. Therefore, varying the filling fraction does not alter our conclusions, and we focus on $\nu=1$ in the following.

The magnetic field affects the gap matrix in two ways: it changes the overall pairing amplitude and introduces a guiding-center dependence that modifies the real-space profile. As a result, the vortex-free state is no longer strictly uniform. Its real-space order parameter is
\begin{align}
    \Delta_{\rm u}(\vec r) = 2\pi \ell^2_{b} \sum_{m} \Delta_{m,\eta=0}(\lambda) \phi_{m\uparrow}(\vec r)\phi_{m\downarrow}(\vec r) & = \frac{J}{2} \sqrt{1-\lambda^2} \sum_{m} \frac{S_{m,0}(\lambda)}{m!} \left( \frac{\sqrt{1-\lambda^2}r^2}{2\ell_b^2} \right)^m e^{-\frac{r^2}{2\ell_b^2}} \label{eq:uniform_order_parameter}  \\
    &= \frac{J}{2}\frac{(1-\lambda^2)^{\frac{3}{2}}}{2-\sqrt{1-\lambda^2}} e^{-\left( 1- \frac{1-\lambda^2}{2-\sqrt{1-\lambda^2}} \right) \frac{r^2}{2\ell_b^2}}
\end{align}
where we have used $\Delta=J/2$ for $\nu=1$. 
Thus, the real-space gap acquires a Gaussian envelope because of the guiding-center dependence of the gap matrix and the mismatch between the spin-$\uparrow$ and spin-$\downarrow$ magnetic lengths. We define the decay length $\ell_C$ by
\begin{align}
    \ell_C \equiv \sqrt{2} \left( 1-\frac{1-\lambda^2}{2-\sqrt{1-\lambda^2}} \right)^{-\frac{1}{2}} \ell_b \approx \frac{2}{\sqrt{3}} \frac{\ell_b}{\lambda}
\end{align}
As $\lambda\to0$, $\ell_C$ diverges and the order parameter becomes
uniform. At finite $\lambda$, the mismatch between the magnetic lengths
of the paired Landau-level orbitals produces a Gaussian envelope, which
may be viewed as a manifestation of Landau quantization of the
Cooper-pair center-of-mass motion. Nevertheless, $\ell_C$ is not the
magnetic length of a pointlike charge-$2e$ particle: it is determined
self-consistently by the overlap of two field-dependent Landau-level
orbitals. Accordingly, $\ell_C\sim |\lambda|^{-1}$, whereas
$\ell_{2e}\sim |\lambda|^{-1/2}$. This difference in scaling originates
from the nonlinear self-consistency of the gap equation. As shown later
in the Ginzburg--Landau section, the linearized gap equation instead
admits a Gaussian solution whose width is set by the conventional
charge-$2e$ magnetic length.
\begin{figure}
    \centering
    \includegraphics[width=0.5\linewidth]{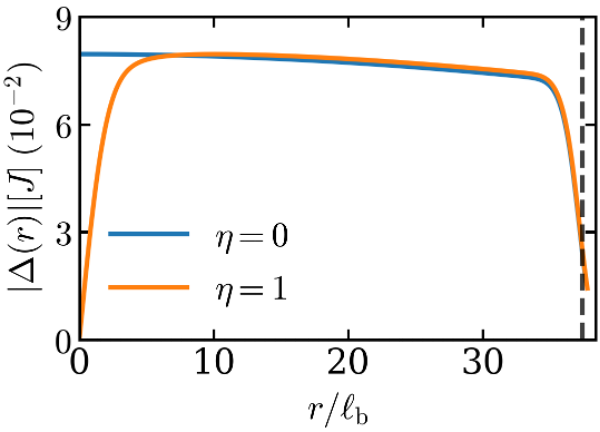}
    \caption{Radial profiles of the vortex-free (blue) and single-vortex (orange) states. The black dashed line marks the system edge; the sharp drop near the boundary is a finite-size boundary effect. }
    \label{fig:SM_radial_distribution}
\end{figure}
At fixed nonzero magnetic field, taking the thermodynamic limit causes the order parameter to decay away from the center. Thus, the vortex-free solution becomes a magnetic pairing droplet rather than a bulk superconducting state; the equilibrium state at fixed field is instead expected to be vortex-lattice-like. 
This can be seen from the  condensation energy of this state, which is given by:
\begin{align}
    E_{\rm u}(\lambda) = -\frac{1}{J 2\pi \ell^2_{b}} \int_{0}^{r_b} dr 2\pi r|\Delta_{\rm u}(r)|^2 &=
-\frac{J}{16}
\frac{\ell_C^2}{\ell_b^2}
\frac{(1-\lambda^2)^3}
{\left(2-\sqrt{1-\lambda^2}\right)^2}
\left[
1-\exp\left(-\frac{2\mathcal{A}}{\pi\ell_C^2}\right)
\right]\label{eq:finite_field_energy_uniform_exact} 
\end{align}
As a result, at fixed nonzero $\lambda$, the localization
length $\ell_C$ remains finite. The total condensation energy
therefore saturates, and its density vanishes as
$\mathcal A\to\infty$:
\begin{align}
   \lim_{\lambda\rightarrow0}   \lim_{\mathcal{A}\rightarrow\infty} \frac{E_{\rm u}(\lambda)}{\mathcal{A}} \rightarrow 0
\end{align}
The situation is different when the total external flux is held fixed while taking the thermodynamic limit. In this case, the sample area and magnetic field satisfy
\begin{align}\label{eq:fixed_flux_area}
    B \mathcal{A} = C \Phi_0 \rightarrow \mathcal{A} =  \frac{C \pi}{\lambda} \ell_b^2
\end{align}
Therefore, the effective area of the Gaussian decay is always much larger than the sample area as function of magnetic field:
\begin{align}
    \mathcal{A} \ll \pi \ell^2_C, \quad \rm for \,  \lambda \ll1
\end{align}
Thus, the vortex-free order parameter remains asymptotically uniform across the sample.
\begin{align}
    E_{\rm u}(\lambda) & =
-\frac{J}{16}
\frac{\ell_C^2}{\ell_b^2}
\frac{(1-\lambda^2)^3}
{\left(2-\sqrt{1-\lambda^2}\right)^2}
\left[
1-\exp\left(-\frac{2\mathcal{A}}{\pi\ell_C^2}\right)
\right] \\
& \approx
-\frac{J}{8}
\frac{(1-\lambda^2)^3}
{\left(2-\sqrt{1-\lambda^2}\right)^2}
\left[
\frac{\mathcal A}{\pi\ell_b^2}
-
\frac{\mathcal A^2}{\pi^2\ell_b^2\ell_C^2}
\right]
+O\!\left[
\left(\frac{\mathcal A}{\ell_C^2}\right)^3
\right]. \\
& \approx  -\frac{J}{4}\frac{\mathcal{A}}{2\pi\ell_b^2} + \frac{3J}{32} C^2 +O(\lambda C)\label{eq:finite_field_energy_uniform} 
\end{align}
where in the second line, we used $\mathcal{A}\ll \pi \ell_C^2$ as in the fixed flux limit and substituted $\mathcal{A}$ in Eq.~\eqref{eq:fixed_flux_area} to obtain the third equation.
The first term is exactly the uniform condensation energy at $\nu=1$ in the zero field limit.  The energy density in the thermodynamic limit is given by:
\begin{align}
    \lim_{\substack{\mathcal A\to\infty,\;\lambda\to0\\
                     B\mathcal A=C\Phi_0}}
    \frac{E_{\rm u}}{\mathcal A}
    =
    -\frac{J}{4}\frac{1}{2\pi\ell_b^2}.
\end{align}
Thus, along a fixed-flux trajectory, the total condensation
energy remains extensive and its density approaches the
zero-field value. This path dependence reflects the
nonanalyticity at the origin of the $(1/\mathcal A,B)$ plane.

Finally, the leading finite-flux correction in the microscopic uniform-state energy can be identified with the conventional diamagnetic phase-stiffness energy of the vortex-free state. In the long-wavelength description,
\begin{align}
E_{\rm dia}
=
\frac{\rho_b}{2}
\int_{\mathcal A} d^2r\,
\left(
\frac{2\pi}{\Phi_0}\mathbf A
\right)^2 .
\end{align}
For a uniform perpendicular field in symmetric gauge,
$\mathbf A=(Br/2)\hat{\boldsymbol\phi}$, on a disk of area
$\mathcal A=\pi R^2$, this gives
\begin{align}
E_{\rm dia}
=
\frac{\pi\rho_b}{4}
\left(
\frac{B\mathcal A}{\Phi_0}
\right)^2.
\end{align}
Along a fixed-flux trajectory, $B\mathcal A=C\Phi_0$. Using
$\rho_b=3J/(8\pi)$ at $\nu=1$, we obtain
\begin{align}
E_{\rm dia}
=
\frac{3J}{32}C^2,
\end{align}
which exactly reproduces the leading finite-flux correction obtained from the microscopic uniform-state energy. Thus, this term is the conventional bulk diamagnetic contribution.

\subsection{Single-vortex solution:$\eta=1$}\label{sec:vortex_finite_field}

In this section, we study the single-vortex solution in a weak external magnetic field and compare its energy with the vortex-free state. We show that the vortex orbital magnetization lowers the single-vortex energy and can favor this sector once the total flux is sufficiently large.

The single-vortex solution is obtained by setting $\eta=1$ and solving the nonlinear gap equation in Eq.~\eqref{eq:SM_self_consistent_equation} with $\lambda\neq0$. At half filling, $\nu=1$, and $T=0$, the converged gap matrix is
\begin{align}
    \Delta_{m,\eta=1}(\lambda) &= \frac{J}{2} \sum_{m'} V_{m+1,m,m'+1,m'}(\lambda) \\
    & =\frac{J}{2} \frac{1-\lambda^2}{2} \frac{1+\lambda}{2} \sum_{m'=0} \frac{\Gamma(m+m'+2)}{\sqrt{(m+1)!m!m'!(m'+1)!}} (\frac{\sqrt{1-\lambda^2}}{2})^{m+m'} \\
    & \equiv \frac{J}{2} S_{m,1}(\lambda)
\end{align}

The corresponding superconducting order parameter is
\begin{align}
    \Delta_{\rm v}(\vec r) &=2\pi \ell_b^2 \sum_{m=0} \Delta_{m,\eta=1}(\lambda) \phi_{m+1,\uparrow}(\vec r) \phi_{m\downarrow} \\
    & = \frac{J}{2} (\sqrt{1+\lambda}) \sqrt{1-\lambda^2} \left(\frac{r}{\sqrt{2}\ell_b}\right) e^{i \theta } \sum_{m=0}\frac{S_{m,\eta=1}(\lambda)}{\sqrt{(m+1)!m!}} \left( \frac{ \sqrt{1-\lambda^2} r^2}{2\ell_b^2}  \right)^m e^{-\frac{r^2}{2\ell_b^2}} \label{eq:vortex_gap_real_space} 
\end{align}
where $\phi_{m\sigma}$ is given by Eq.~\eqref{eq:orbital_wavefunction}.
The radial profile of this order parameter is shown by the orange curve in Fig.~\ref{fig:SM_radial_distribution}. It vanishes linearly at the vortex core and approaches the uniform bulk profile far from the core, but with a slightly enhanced amplitude. 

This enhancement is a genuine feature of the single-vortex solution rather than a numerical artifact. Extending the large-distance expansion used in Eqs.~\eqref{eq:S_{m,1}_eval}--\eqref{eq:vortex_R_r_form} to finite $\lambda$ gives
\begin{align}
\Delta_{\rm v}(|\mathbf r|\gg \ell_b)
\simeq{}&
\left(\frac{1+\lambda}{1-\lambda}\right)^{3/4}
\Delta_{\rm u}(\mathbf r)e^{i\theta}
\left[
1-
\frac{3+\lambda^2}{4(1-\lambda^2)^{3/2}}
\frac{\ell_b^2}{r^2}
+O\!\left(\frac{\ell_b^4}{r^4}\right)
\right].
\label{eq:vortex_gap_large_r}
\end{align}
Thus the single-vortex and vortex-free states share the same field-induced Gaussian envelope, while the vortex state has a weak-field bulk enhancement $[(1+\lambda)/(1-\lambda)]^{3/4}=1+3\lambda/2+O(\lambda^2)$ together with the familiar $1/r^2$ vortex tail. The latter reduces to $1-3\ell_b^2/(4r^2)$ as $\lambda\to0$. 
Next, we compute the on-shell superconducting energy along the fixed-flux thermodynamic trajectory. The system area is $\mathcal{A} = C \pi \ell^2_b/\lambda$ and  the on-shell superconducting energy is given by:
\begin{align}
     E_{\rm v}(\lambda) &= -\frac{1}{J 2\pi \ell^2_{b}} \int_\mathcal{A} d^2r|\Delta_{\rm v}(r)|^2 \\
     & \approx -\frac{J}{4}\frac{\mathcal{A}}{2\pi\ell_b^2}
    -\frac{3JC}{8}
    +\frac{3JC^2}{32}
    +\frac{3J}{16}
    \ln\left(
        \frac{\mathcal A}{2\pi\ell_b^2}
    \right)
    +E_{\rm core}
    +O\!\left[J\lambda\ln\!\left(1/\lambda\right)\right].
    \label{eq:SM_vortex_energy_expansion}
\end{align}
The first term is the uniform condensation energy, while the third term is the bulk diamagnetic contribution identified in the previous section. The second term is linear in the total flux, $C=B\mathcal A/\Phi_0$, and corresponds to the orbital-magnetization energy of the single-vortex state. In the long-wavelength description,
\begin{align}
E_{\rm mag}
=-\rho_b\frac{2\pi}{\Phi_0}\int_{\mathcal A}d^2r\,\nabla\theta\cdot\mathbf A
=-\rho_b\frac{2\pi}{\Phi_0}\int_{\mathcal A}d^2r\,\frac{1}{r}\frac{Br}{2}
=-\pi\rho_b\frac{B\mathcal A}{\Phi_0}.
\end{align}
Using $B\mathcal A=C\Phi_0$ and $\rho_b=3J/(8\pi)$ at $\nu=1$, this becomes
\begin{align}
E_{\rm mag}
=
-\pi\rho_b C
=
-\frac{3J}{8}C,
\end{align}
in exact agreement with the microscopic result.
\begin{figure}[h]
    \centering
    \includegraphics[width=0.75\linewidth]{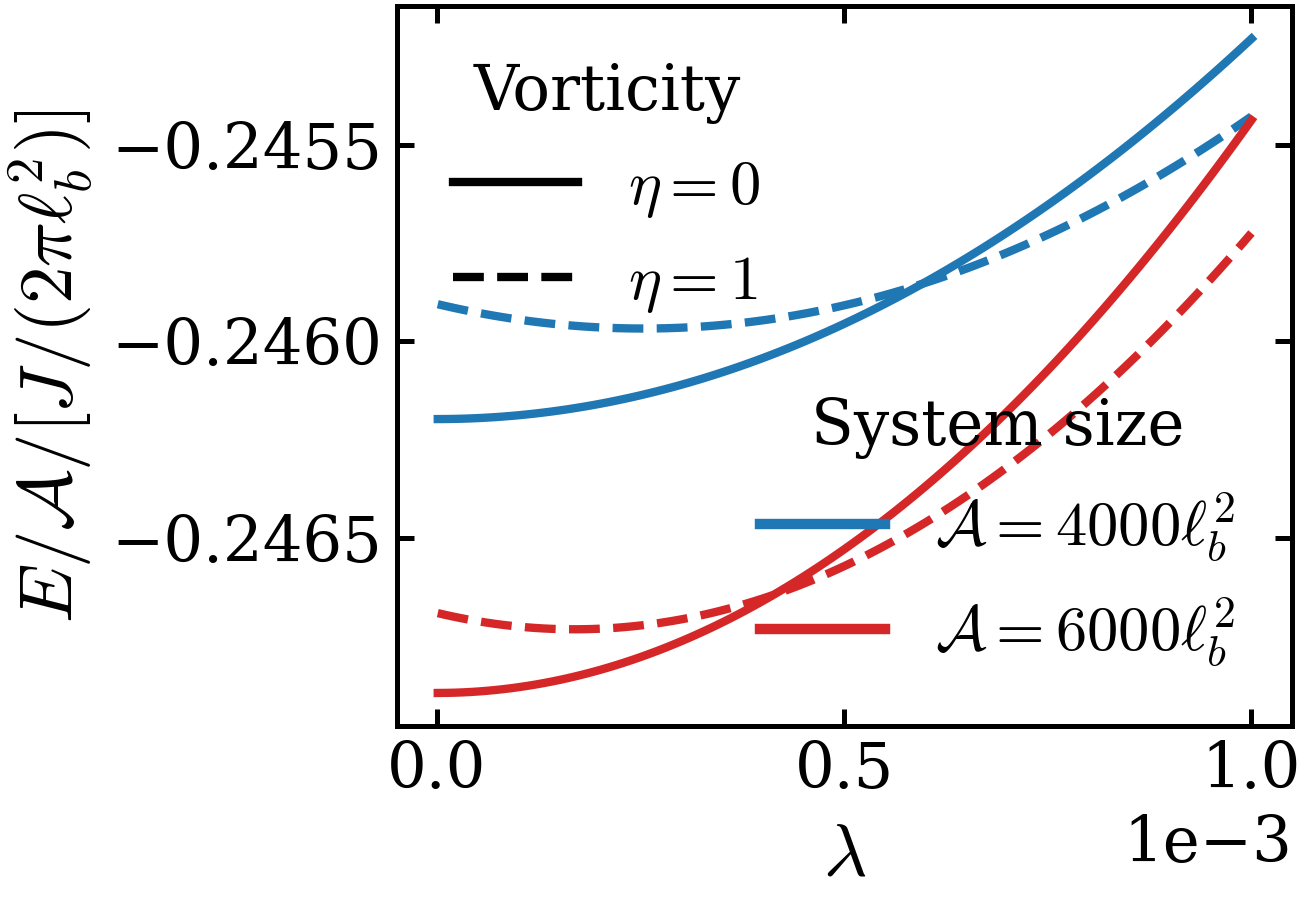}
    \caption{Condensation energy of the uniform ($\eta=0$) and single-vortex ($\eta=1$) states as a function of magnetic field for several sample areas, obtained numerically.}
    \label{fig:two_energy_compare}
\end{figure}
Figure~\ref{fig:two_energy_compare} shows the energy densities of the vortex-free state in Eq.~\eqref{eq:finite_field_energy_uniform} and the single-vortex state discussed above as functions of the magnetic field $\lambda$ for several sample areas. At $\lambda=0$, the energy-density difference between the uniform and single-vortex states scales as $\ln \mathcal{A}/\mathcal{A}$, which vanishes in the thermodynamic limit $\mathcal{A}\rightarrow\infty$.

The energy density of the vortex-free state increases quadratically with $\lambda$ because of the diamagnetic contribution. As the system size increases, the energy density rises more rapidly because the coefficient of the quadratic term is proportional to the sample area, as shown in Eq.~\eqref{eq:finite_field_energy_uniform}. As discussed in the main text, this increasingly sharp dependence reflects the divergence of the magnetic susceptibility and signals the instability of the uniform state in the fixed-field thermodynamic limit.

For the single-vortex states with $\eta=1$, the field-dependent energy contains not only a diamagnetic contribution but also an orbital-magnetization term, since these states already break time-reversal symmetry at $\lambda=0$. Because the orbital-magnetization contribution is linear in $\lambda$, it dominates the field dependence for sufficiently weak magnetic fields. Consequently, as shown in the figure, the energy density of the single-vortex state initially decreases with increasing $\lambda$, before turning upward once the quadratic diamagnetic contribution becomes dominant. Because the diamagnetic contribution is the same to leading order in the two sectors, the orbital-magnetization term drives a level crossing between the vortex-free and single-vortex states, corresponding to first-vortex entry. As the system size grows, the critical field decreases because the zero-field energy-density difference scales as $\ln(\mathcal A)/\mathcal A$. This does not imply that an arbitrarily weak field necessarily nucleates a vortex in the thermodynamic limit; the outcome depends on the total flux $B\mathcal A$, or equivalently on the path by which the origin of the $(1/\mathcal A,B)$ plane is approached. 

The energy difference between the single-vortex and vortex-free solutions is
\begin{align}
E_{\rm v}-E_{\rm u} &= -\frac{3JC}{8}
    +\frac{3J}{16}
    \ln\left(
        \frac{\mathcal A}{2\pi\ell_b^2}
    \right)
   +E_{\rm core} \\
&=
-\frac{3J B\mathcal{A}}{8\Phi_0}
+\frac{3J}{16}
\ln\left({\frac{\mathcal{A}}{2\pi \ell_b^2}}\right) + E_{\rm core} \label{eq:energy_differ_finite_field}
\end{align}

\begin{figure}[t]
    \centering
    \includegraphics[width=0.75\linewidth]{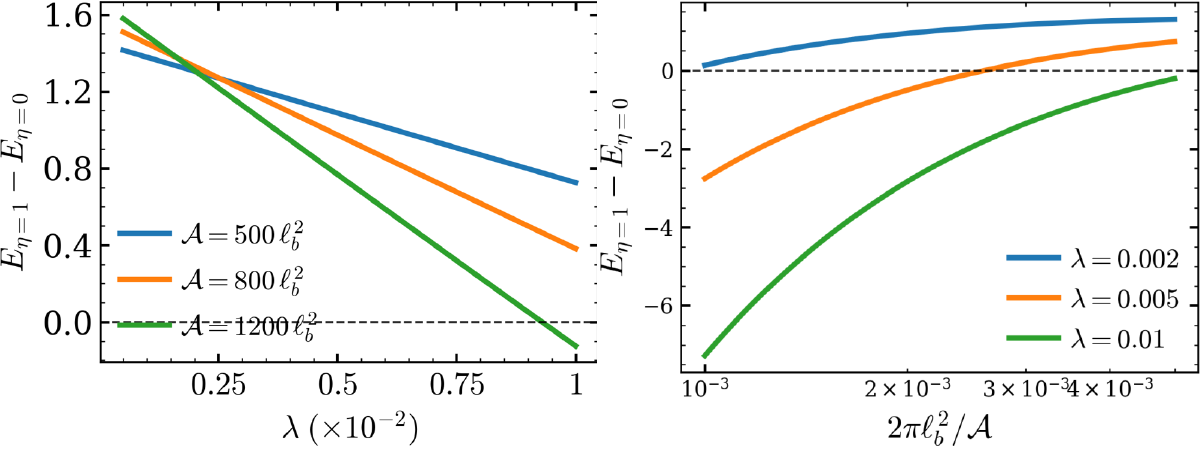}
    \caption{The energy difference between the single-vortex ($\eta=1$) and vortex-free ($\eta=0$) states as a function of magnetic field (left panel) and inverse area (right panel)}
    \label{fig:SM_energy_differ_finite_field}
\end{figure}

As the total flux increases with the
sample area, the bulk condensation-energy gain grows linearly with
\(C\), whereas the phase-stiffness cost grows only logarithmically with
\(\mathcal A\). Consequently, beyond a critical flux, the orbital magnetization energy
gain overcomes the logarithmic stiffness energy cost and the single-vortex state
becomes energetically favored. Figure~\ref{fig:SM_energy_differ_finite_field} shows the energy difference as a function of magnetic field (left) and inverse area (right). The energy difference decreases linearly with field, with a steeper slope for larger samples, consistent with Eq.~\eqref{eq:energy_differ_finite_field}, whose slope is $-3J\mathcal A/(8\Phi_0)$. At fixed field, increasing the area likewise favors the single-vortex state because the linear orbital-magnetization gain grows faster than the logarithmic stiffness cost.

The equal-energy boundary is
\begin{align}
    B = \frac{\Phi_0}{\mathcal{A}} \left( \frac{1}{2}\ln{\frac{\mathcal{A}}{2\pi \ell_b^2}}+ \frac{2\varepsilon_{\rm core}}{3}    \right)
\end{align}
where $\varepsilon_{\rm core} = 4 E_{\rm core}/J$ is a non-universal constant. 
This boundary expression takes the following form in the $(B, 1/\mathcal{A})$ plane where the origin represents the weak-field thermodynamic limit:
\begin{align}
    y_v(x)=c_1 x\left[\ln(x_0/x)+c_2\right],
\end{align}
where $c_1$,$c_2$ are non-universal positive constants.
As a result, this line is nonanalytic near the origin. As shown in Fig.~1 of the main text, since $-\log x\rightarrow \infty$ as $x\rightarrow0$, the vortex boundary eventually lies above every fixed-flux straight line. Therefore, any finite total flux is insufficient to nucleate a vortex in the thermodynamic limit and the system ultimately approaches a vortex-free, uniform superconducting state.

\section{Ginzburg--Landau theory}\label{sec:GL_theory}

In this section, we derive the long-wavelength Ginzburg--Landau (GL) description associated with the microscopic Landau-level model. We first obtain the quadratic Cooper-pair kernel and use the corresponding linearized gap equation to identify the magnetic length that controls the superconducting instability. We then expand the same kernel at long wavelength and recover the standard charge-$2e$ gauge-covariant GL functional. This analysis connects the microscopic Landau-level problem to the phase-only electrodynamics used for Pearl screening in Sec.~\ref{sec:screening_finite_disk}.

We start from the Euclidean action
\begin{align}
S={}&\int d\tau d^2r\sum_{\sigma}
\bar\psi_\sigma(\mathbf r,\tau)
\left(\partial_\tau+h_\sigma-\mu\right)
\psi_\sigma(\mathbf r,\tau)
-g\int d\tau d^2r\,
\bar\psi_\uparrow\bar\psi_\downarrow\psi_\downarrow\psi_\uparrow,
\qquad g=2\pi J\ell_b^2,
\label{eq:GL_action}
\end{align}
where $h_\sigma$ is defined in Eq.~\eqref{eq:single_particle_SM}. Introducing a complex Hubbard--Stratonovich pairing field $\Delta$ and integrating out the fermions gives
\begin{align}
S_{\rm eff}[\Delta]
=
\int d1\,\frac{|\Delta(1)|^2}{g}
-\Tr\ln[-\mathcal G^{-1}].
\label{eq:GL_effective_action}
\end{align}
Expanding Eq.~\eqref{eq:GL_effective_action} to quadratic order in a static pairing field gives
\begin{align}
F^{(2)}
=
\int d^2r\,\frac{|\Delta(\mathbf r)|^2}{g}
-\int d^2r\,d^2r'\,
\Delta^*(\mathbf r)\Pi(\mathbf r,\mathbf r')\Delta(\mathbf r'),
\label{eq:GL_quadratic_kernel}
\end{align}
with the normal-state Cooper-pair susceptibility
\begin{align}
\Pi(\mathbf r,\mathbf r')
=
T\sum_{\omega_n}
G_{\uparrow,0}(\mathbf r,\mathbf r';i\omega_n)
G_{\downarrow,0}(\mathbf r,\mathbf r';-i\omega_n).
\label{eq:GL_pair_susceptibility}
\end{align}

Projecting the normal-state Green functions onto the LLL yields
\begin{align}
G_\sigma(\mathbf r,\mathbf r';i\omega_n)
&=\frac{P_\sigma(\mathbf r,\mathbf r')}{i\omega_n-\xi_\sigma},
\qquad
P_\sigma(\mathbf r,\mathbf r')
=
\frac{e^{-|\mathbf r-\mathbf r'|^2/(4\ell_\sigma^2)}}{2\pi\ell_\sigma^2}
\exp\left[-\frac{i\sigma}{2\ell_\sigma^2}(\mathbf r\times\mathbf r')_z\right].
\label{eq:GL_LLL_green}
\end{align}
Substituting Eq.~\eqref{eq:GL_LLL_green} into Eq.~\eqref{eq:GL_pair_susceptibility} gives
\begin{align}
\Pi(\mathbf r,\mathbf r')
=
\Pi_0
\exp\left[-\frac{|\mathbf r-\mathbf r'|^2}{2\ell_b^2}\right]
\exp\left[-\frac{i\lambda}{\ell_b^2}(\mathbf r\times\mathbf r')_z\right],
\label{eq:GL_pair_kernel}
\end{align}
where
\begin{align}
\Pi_0
=
\frac{1-\lambda^2}{4\pi^2\ell_b^4}
\frac{1-f(\xi_\uparrow)-f(\xi_\downarrow)}
{\xi_\uparrow+\xi_\downarrow}.
\label{eq:Pi_0}
\end{align}
The Gaussian factor in Eq.~\eqref{eq:GL_pair_kernel} reflects the finite extent of the LLL orbitals, while the phase factor describes the center-of-mass coupling of a Cooper pair to the external magnetic field.

\subsection{Linearized gap equation and charge-$2e$ magnetic length}

We first use the quadratic kernel to determine the leading superconducting instability. Minimizing Eq.~\eqref{eq:GL_quadratic_kernel} gives
\begin{align}
\Delta(\mathbf r)
=
g\int d^2r'\,\Pi(\mathbf r,\mathbf r')\Delta(\mathbf r').
\label{eq:linearized_gap}
\end{align}
For a uniform perpendicular field, a Gaussian ansatz $\Delta(\mathbf r)=\Delta_0e^{-r^2/r_w^2}$ solves Eq.~\eqref{eq:linearized_gap}. Performing the Gaussian integral gives
\begin{align}
(1-\lambda^2)a^2-2a+1=0,
\qquad
1=g\Pi_0\,2\pi\ell_b^2a,
\qquad
a\equiv\frac{r_w^2}{r_w^2+2\ell_b^2}.
\label{eq:GL_Gaussian_matching}
\end{align}
The physical root $0<a<1$ is $a=(1+\lambda)^{-1}$, and therefore
\begin{align}
r_w^2=\frac{2\ell_b^2}{\lambda},
\qquad
r_w=\frac{\sqrt{2}\ell_b}{\sqrt{\lambda}}
=\sqrt{\frac{2\hbar}{eB}}
=2\ell_{2e},
\qquad
\ell_{2e}\equiv\sqrt{\frac{\hbar}{2eB}}.
\label{eq:rw_result}
\end{align}
Thus the linearized order parameter has the conventional charge-$2e$ Landau-level envelope, $\Delta(\mathbf r)\propto e^{-r^2/(4\ell_{2e}^2)}$, and its width scales as $r_w\propto B^{-1/2}\propto\lambda^{-1/2}$. This is parametrically different from the nonlinear solution in Sec.~\ref{sec:uniform_finite_field}, where $\ell_C\propto\lambda^{-1}$. The former describes the instability of the normal state, whereas the latter follows from the fully self-consistent superconducting state.

The second condition in Eq.~\eqref{eq:GL_Gaussian_matching} determines the transition line. In the zero-field limit, $a\to1$ and $\xi_\uparrow=\xi_\downarrow=\xi$, giving
\begin{align}
1=J\frac{\tanh(\xi/2k_BT)}{2\xi}.
\label{eq:linear_gap_zero_field}
\end{align}
At $T=T_c$, using $\xi=J(1-\nu)/2$, Eq.~\eqref{eq:linear_gap_zero_field} reproduces Eq.~\eqref{eq:SM_critical_temperature}.

\subsection{Long-wavelength limit}

We next extract the local GL functional from the same nonlocal kernel. The purpose is to identify the universal long-distance coupling between the superconducting phase and the vector potential, rather than to replace the fully nonlinear microscopic solution. Introducing center-of-mass and relative coordinates $\mathbf X=(\mathbf r+\mathbf r')/2$ and $\mathbf d=\mathbf r-\mathbf r'$, we expand the order parameter and the field-dependent phase of Eq.~\eqref{eq:GL_pair_kernel} to second order in gradients and $\mathbf A$. Performing the Gaussian moments of $\mathbf d$ gives
\begin{align}
\mathcal F^{(2)}(\mathbf X)
={}&
\left(\frac{1}{g}-2\pi\ell_b^2\Pi_0\right)|\Delta|^2
+\pi\ell_b^4\Pi_0|\nabla\Delta|^2
-2\pi\lambda\ell_b^2\Pi_0|\Delta|^2
(\hat{\mathbf z}\times\mathbf X)\cdot\nabla\theta
+\pi\Pi_0\lambda^2X^2|\Delta|^2
+\cdots,
\label{eq:GL_local_expansion}
\end{align}
where $\Delta=|\Delta|e^{i\theta}$. In symmetric gauge, $\mathbf A=(B/2)\hat{\mathbf z}\times\mathbf X$, the terms in Eq.~\eqref{eq:GL_local_expansion} combine into
\begin{align}
F^{(2)}
=\int d^2X\left[
\alpha_{\rm GL}|\Delta|^2
+K_{\rm GL}|\mathbf D\Delta|^2
\right]
+O(D^4),
\qquad
\mathbf D\equiv\nabla-i\frac{2\pi}{\Phi_0}\mathbf A,
\label{eq:conventional_covariant_GL_limit}
\end{align}
with
\begin{align}
\alpha_{\rm GL}=\frac{1}{g}-2\pi\ell_b^2\Pi_0,
\qquad
K_{\rm GL}=\pi\ell_b^4\Pi_0,
\qquad
\Phi_0=\frac{h}{2e}.
\label{eq:GL_coefficients}
\end{align}
Thus the charge-$2e$ covariant derivative emerges directly from the microscopic LLL Cooper-pair kernel. Along a fixed-flux trajectory, $R^2=C\ell_b^2/\lambda$, so $\lambda R/\ell_b=\sqrt{C\lambda}\to0$ and the long-wavelength expansion becomes asymptotically controlled throughout the sample. Deep in the superconducting phase, the amplitude is already established and the magnetic response is more naturally expressed through the microscopic phase stiffness $\rho_b$ obtained in Sec.~\ref{sec:phase_stiffness}. We use this phase-only description below to include the self-consistent electromagnetic field.

\section{Electromagnetic screening in a finite disk}\label{sec:screening_finite_disk}

The microscopic analysis above determines the condensate structure and its long-wavelength phase rigidity, but it does not yet include the magnetic field generated by the supercurrent itself. In this section, we couple the phase stiffness to three-dimensional Maxwell electrodynamics and study Pearl screening in a finite disk. Related finite-size effects in the magnetic response and flux penetration of thin superconducting films have been studied previously~\cite{Magnetization_scaling,Fetter1980}.   

We organize the analysis into four steps. We first derive the self-consistent finite-disk Pearl equation for the vortex-free state and use it to obtain the weak- and strong-screening limits together with the resulting magnetic-field profiles. We then use these solutions to determine how the screening energy is distributed and how the diamagnetic Gibbs energy scales with field and system size. Next, we extend the same electromagnetic kernel to a centered vortex and derive the screened vortex profile and zero-field self-energy. Finally, we combine the vortex and Meissner sectors to obtain the field-dependent vortex energy, its weak- and strong-screening limits, and the crossover observed numerically. This organization separates the microscopic input---$\rho_b$, $\xi$, and the vortex core energy---from the long-distance electrodynamics.

\subsection{Meissner screening in a finite disk}

We consider a strictly two-dimensional superconducting disk of radius $R$ in the plane $z=0$, with a perpendicular applied field $\mathbf B_{\rm ext}=B_{\rm ext}\hat{\mathbf z}$. In this subsection we derive the self-consistent screening equation, identify the control parameter $R/\Lambda$, and extract both the limiting vector potentials and the corresponding magnetic-field profiles. Deep in the superconducting phase we neglect amplitude fluctuations and retain only the phase stiffness. For the vortex-free state, $\nabla\theta=0$, and in symmetric gauge $\mathbf A_{\rm ext}=(B_{\rm ext}r/2)\hat{\boldsymbol\phi}$. We write the total vector potential as
\begin{align}
A_\phi(r,z)=\frac{B_{\rm ext}r}{2}+a_\phi(r,z),
\label{eq:Aphi_total_SI}
\end{align}
where $a_\phi$ is generated by the screening current. Cylindrical symmetry and charge conservation imply an azimuthal sheet current $\mathbf K=K_\phi(r)\hat{\boldsymbol\phi}$ and allow the Coulomb-gauge choice $\mathbf a=a_\phi(r,z)\hat{\boldsymbol\phi}$.

With the energy of the external source subtracted, the Gibbs functional is
\begin{align}
G[a_\phi]
={}&
\frac{\rho_b}{2}\int_{r<R}d^2r\left(\frac{2\pi}{\Phi_0}A_\phi(r,0)\right)^2
+\frac{1}{2\mu_0}\int d^3r\,|\nabla\times\mathbf a|^2
\nonumber\\
={}&
\frac{1}{\mu_0\Lambda}\int_{r<R}d^2r\,A_\phi(r,0)^2
+\frac{1}{2\mu_0}\int d^3r\,|\nabla\times\mathbf a|^2,
\label{eq:Gibbs_Pearl_SI}
\end{align}
where
\begin{align}
\Lambda\equiv\frac{\Phi_0^2}{2\pi^2\mu_0\rho_b}
\label{eq:Pearl_length_SI}
\end{align}
is the Pearl length in our stiffness convention. Varying the superconducting term gives the two-dimensional London relation
\begin{align}
K_\phi(r)=-\frac{2}{\mu_0\Lambda}A_\phi(r,0),
\qquad r<R.
\label{eq:London_Pearl_SI}
\end{align}
The current responds to the \emph{total} vector potential on the sheet, not to the applied vector potential alone. The remaining task is therefore to determine $a_\phi$ self-consistently from Maxwell's equation.

Varying the magnetic contribution and integrating by parts gives the Maxwell
equation
\begin{equation}
    \nabla\times(\nabla\times\mathbf a)
    =
    \mu_0K_\phi(r)\delta(z)\Theta(R-r)\hat{\boldsymbol\phi}.
    \label{eq:Maxwell_sheet_SI}
\end{equation}
For $\mathbf a=a_\phi\hat{\boldsymbol\phi}$, one has
$\nabla\cdot\mathbf a=0$, and therefore
\begin{equation}
    \left(
    \partial_r^2
    +\frac{1}{r}\partial_r
    +\partial_z^2
    -\frac{1}{r^2}
    \right)a_\phi(r,z)
    =
    -\mu_0K_\phi(r)\delta(z)\Theta(R-r).
    \label{eq:a_PDE_SI}
\end{equation}
Using Eq.~\eqref{eq:London_Pearl_SI}, we obtain the closed self-consistent
equation
\begin{equation}
    \left(
    \partial_r^2
    +\frac{1}{r}\partial_r
    +\partial_z^2
    -\frac{1}{r^2}
    \right)a_\phi(r,z)
    =
    \frac{2}{\Lambda}
    \left[
    \frac{B_{\rm ext}r}{2}+a_\phi(r,0)
    \right]
    \delta(z)\Theta(R-r),
    \label{eq:selfconsistent_PDE_SI}
\end{equation}
with $a_\phi\rightarrow0$ at spatial infinity. Equation~\eqref{eq:selfconsistent_PDE_SI}
makes the dimensional structure of the Pearl problem explicit: the source is
confined to a two-dimensional sheet, while the electromagnetic response lives
in three-dimensional space.

The differential equation above is useful conceptually, while the Green-function
form is more convenient for the finite disk and for numerical calculations. We
therefore reduce the three-dimensional Maxwell problem to an integral equation
for the vector potential on the superconducting sheet. The resulting Pearl
equation is a two-dimensional boundary equation for a fully three-dimensional
electromagnetic field.

The magnetostatic Green function gives
\begin{equation}
    a_\phi(r,z)
    =
    \frac{\mu_0}{4\pi}
    \int_0^Rdr'\,r'K_\phi(r')Q(r,r';z),
    \label{eq:a_BiotSavart_3D_SI}
\end{equation}
where
\begin{equation}
    Q(r,r';z)
    =
    \int_0^{2\pi}d\theta\,
    \frac{\cos\theta}
    {\sqrt{r^2+r'^2-2rr'\cos\theta+z^2}}.
    \label{eq:Q_3D_SI}
\end{equation}
Using Eq.~\eqref{eq:London_Pearl_SI},
\begin{equation}
    a_\phi(r,z)
    =
    -\frac{1}{2\pi\Lambda}
    \int_0^Rdr'\,r'Q(r,r';z)A_\phi(r',0).
    \label{eq:a_3D_from_Aplane_SI}
\end{equation}
For numerical evaluation it is useful to write the kernel in terms of complete
elliptic integrals. Defining
\begin{equation}
    D_+=\sqrt{(r+r')^2+z^2},
    \qquad
    m=\frac{4rr'}{(r+r')^2+z^2},
\end{equation}
we obtain
\begin{equation}
    Q(r,r';z)
    =
    \frac{2}{rr'}
    \left[
    \frac{r^2+r'^2+z^2}{D_+}K(m)
    -D_+E(m)
    \right].
    \label{eq:Q_elliptic_3D_SI}
\end{equation}
For $z\neq0$ the kernel is nonsingular. The only singularity occurs on the
sheet as $r\rightarrow r'$, where it is logarithmic and integrable.

Restricting Eq.~\eqref{eq:a_3D_from_Aplane_SI} to $z=0$ and using
Eq.~\eqref{eq:Aphi_total_SI} gives the finite-disk Pearl equation,
\begin{equation}
    A_\phi(r,0)
    +
    \frac{1}{2\pi\Lambda}
    \int_0^Rdr'\,r'Q(r,r';0)A_\phi(r',0)
    =
    \frac{B_{\rm ext}r}{2},
    \qquad 0<r<R.
    \label{eq:Pearl_2D_SI}
\end{equation}
Once $A_\phi(r,0)$ is determined from Eq.~\eqref{eq:Pearl_2D_SI},
Eq.~\eqref{eq:a_3D_from_Aplane_SI} reconstructs the induced vector potential and
magnetic field everywhere in space.

Introducing $s=r/R$ and
$\widetilde A_\phi(s)=A_\phi(Rs,0)/(B_{\rm ext}R)$, together with the dimensionless kernel
$\mathcal Q(s,s';\zeta)\equiv RQ(Rs,Rs';R\zeta)$,
Eq.~\eqref{eq:Pearl_2D_SI} becomes
\begin{align}
\widetilde A_\phi(s)
+\frac{\gamma}{2\pi}
\int_0^1 ds'\,s'\,
\mathcal Q(s,s';0)\widetilde A_\phi(s')
=
\frac{s}{2}.
\label{eq:Atilde_Pearl_gamma}
\end{align}
The finite-disk Pearl problem is controlled by the single dimensionless
ratio
\begin{align}
    \gamma\equiv\frac{R}{\Lambda}.
    \label{eq:gamma_definition}
\end{align}
Thus the normalized screening profile is independent of $B_{\rm ext}$
within linear London theory and is determined entirely by $R/\Lambda$.
The limits $\gamma\ll1$ and $\gamma\gg1$ correspond respectively to
weak and strong Pearl screening.

We next extract the two limiting solutions directly from
Eq.~\eqref{eq:Atilde_Pearl_gamma}. The weak-screening limit leaves the total
vector potential close to the imposed one, whereas the strong-screening limit
makes the total vector potential on the sheet parametrically small through
cancellation between the imposed and induced contributions. These two limits
will also determine the field and energy profiles below.

\textit{Weak screening: $\gamma\ll1$:}
For $\gamma\ll1$, the integral term in Eq.~\eqref{eq:Atilde_Pearl_gamma} is a
small correction. To leading order,
\begin{equation}
    \widetilde A_\phi(s)
    =
    \frac{s}{2}+O(\gamma).
    \label{eq:Atilde_weak_leading}
\end{equation}
Keeping the first correction gives
\begin{equation}
    \widetilde A_\phi(s)
    =
    \frac{s}{2}
    -
    \frac{\gamma}{4\pi}
    \int_0^1ds'\,s'^2\mathcal Q(s,s';0)
    +O(\gamma^2).
    \label{eq:Atilde_weak_first}
\end{equation}
Thus the induced vector potential is only $O(\gamma)$, while the total vector
potential remains $O(1)$ in units of $B_{\rm ext}R$. In physical units,
\begin{equation}
    A_\phi(r,0)
    \simeq
    \frac{B_{\rm ext}r}{2},
\end{equation}
so the superconducting sheet responds essentially to the bare external vector
potential.

The corresponding current is
\begin{equation}
    K_\phi(r)
    =
    -\frac{2B_{\rm ext}}{\mu_0}\,
    \gamma\widetilde A_\phi(s),
    \label{eq:K_dimensionless_SI}
\end{equation}
and therefore
\begin{equation}
    K_\phi(r)
    \simeq
    -\frac{B_{\rm ext}}{\mu_0}\,\gamma s,
    \qquad \gamma\ll1.
    \label{eq:K_weak_SI}
\end{equation}
The current and its magnetic field are both perturbatively small in
$R/\Lambda$.

\textit{Strong screening: $\gamma\gg1$:}
For $\gamma\gg1$, Eq.~\eqref{eq:Atilde_Pearl_gamma} implies
$\widetilde A_\phi=O(\gamma^{-1})$ at fixed $s<1$. Defining
$F(s)=\lim_{\gamma\to\infty}\gamma\widetilde A_\phi(s)$, the leading equation is
\begin{align}
\frac{1}{2\pi}\int_0^1ds'\,s'\mathcal Q(s,s';0)F(s')=\frac{s}{2}.
\label{eq:strong_disk_integral}
\end{align}
This is the standard ideal thin-disk screening equation, whose solution is
\begin{align}
F(s)=\frac{2}{\pi}\frac{s}{\sqrt{1-s^2}},
\qquad\text{or equivalently}\qquad
\gamma\widetilde A_\phi(s)\longrightarrow
\frac{2}{\pi}\frac{s}{\sqrt{1-s^2}}.
\label{eq:Atilde_strong}
\end{align}
The inverse-square-root behavior is the familiar edge singularity of the ideal
zero-thickness screening current. Away from the edge,
\begin{equation}
    A_\phi(r,0)
    \simeq
    \frac{2B_{\rm ext}\Lambda}{\pi}
    \frac{r/R}{\sqrt{1-r^2/R^2}}.
    \label{eq:Aphysical_strong}
\end{equation}
The important change from the weak-screening regime is the replacement of the
scale $B_{\rm ext}R$ by $B_{\rm ext}\Lambda$. The total vector potential on the
sheet becomes parametrically small as $\Lambda/R\rightarrow0$.

This does not mean that the induced vector potential vanishes. Using Eq.~\eqref{eq:Aphi_total_SI},
\begin{equation}
    a_\phi(r,0)
    \simeq
    -\frac{B_{\rm ext}r}{2}
    +
    \frac{2B_{\rm ext}\Lambda}{\pi}
    \frac{r/R}{\sqrt{1-r^2/R^2}}.
    \label{eq:aphysical_strong_sheet}
\end{equation}
The induced vector potential remains finite and approaches
$-A_\phi^{\rm ext}$ over the interior. It is precisely this finite induced
potential that cancels the imposed vector potential and makes the total
$A_\phi$ small.

The limiting off-plane induced vector potential follows by substituting
Eq.~\eqref{eq:Atilde_strong} into Eq.~\eqref{eq:a_3D_from_Aplane_SI}.  The screening current also remains finite,
because the factor $1/\Lambda$ in the London relation compensates the small
$A_\phi\propto\Lambda$:
\begin{equation}
    K_\phi(r)
    \longrightarrow
    -\frac{4B_{\rm ext}}{\pi\mu_0}
    \frac{s}{\sqrt{1-s^2}}.
    \label{eq:K_strong_SI}
\end{equation}
Thus strong screening is not a limit of vanishing current. It is a limit in
which a finite current produces an induced vector potential that nearly cancels
the applied one on the superconducting sheet.

The limiting vector potentials also determine the magnetic-field response.
There are two complementary signatures of screening in a finite disk: $B_z$
measures flux expulsion and focusing, while $B_r$ measures the bending of field
lines around the sample. The weak- and strong-screening profiles follow directly
from whether the induced vector potential is respectively weak or of the same
order as the applied one. For the field profiles below we use $\zeta=z/R$ and
$\widetilde a_\phi(s,\zeta)=a_\phi(Rs,R\zeta)/(B_{\rm ext}R)$.

For an axisymmetric azimuthal vector potential,
\begin{equation}
    B_r(r,z)=-\partial_z a_\phi(r,z),
    \label{eq:Br_SI}
\end{equation}
and
\begin{equation}
    B_z(r,z)
    =
    B_{\rm ext}
    +
    \frac{1}{r}\partial_r\left[r a_\phi(r,z)\right].
    \label{eq:Bz_SI}
\end{equation}
In dimensionless form,
\begin{equation}
    \frac{B_r}{B_{\rm ext}}
    =
    -\partial_\zeta\widetilde a_\phi,
    \qquad
    \frac{B_z}{B_{\rm ext}}
    =
    1+\frac{1}{s}\partial_s
    \left[s\widetilde a_\phi\right].
    \label{eq:B_dimensionless_tilde}
\end{equation}
On the superconducting sheet,
\begin{equation}
    \frac{B_z(r,0)}{B_{\rm ext}}
    =
    \frac{1}{s}\frac{d}{ds}
    \left[s\widetilde A_\phi(s)\right].
    \label{eq:Bz_sheet_tilde}
\end{equation}
Reflection symmetry gives
\begin{equation}
    B_z(r,-z)=B_z(r,z),
    \qquad
    B_r(r,-z)=-B_r(r,z).
\end{equation}

We first consider the vertical component, which directly displays the redistribution of magnetic flux. For
$\gamma\ll1$, Eq.~\eqref{eq:Atilde_weak_first} gives
\begin{equation}
    \frac{B_z^{\rm ind}}{B_{\rm ext}}=O(\gamma),
\end{equation}
so the total field remains close to the applied field. The induced field is
negative in the central region and positive near and outside the rim, reflecting
the small amount of flux displaced from the disk interior.

At any fixed $z$, the induced vertical flux vanishes globally,
\begin{align}
    2\pi\int_0^\infty r\,dr\,B_z^{\rm ind}(r,z)
    &=
    2\pi\int_0^\infty dr\,
    \frac{\partial}{\partial r}
    \left[r a_\phi(r,z)\right]
    \nonumber\\
    &=
    2\pi\left[r a_\phi(r,z)\right]_0^\infty
    =0,
\end{align}
provided $a_\phi$ decays sufficiently rapidly at large radius. Negative induced
flux in the interior must therefore be accompanied by positive induced flux
elsewhere.

For $\gamma\gg1$, Eq.~\eqref{eq:aphysical_strong_sheet} shows that
$a_\phi\simeq-A_\phi^{\rm ext}$ over the interior. Consequently,
\begin{equation}
    B_z^{\rm ind}\simeq-B_{\rm ext},
\end{equation}
and the total field is strongly suppressed. More explicitly, substituting
Eq.~\eqref{eq:Atilde_strong} into Eq.~\eqref{eq:Bz_sheet_tilde} gives, away from
the edge,
\begin{equation}
    \frac{B_z(r,0)}{B_{\rm ext}}
    \simeq
    \frac{2}{\pi\gamma}
    \frac{2-s^2}{(1-s^2)^{3/2}},
    \qquad \gamma\gg1.
    \label{eq:Bz_strong_interior}
\end{equation}
Thus the field tends to zero at any fixed interior point as
$\gamma\rightarrow\infty$. The limit is not uniform near $s=1$: the ideal
zero-thickness disk develops the familiar edge singularity, and the expelled
flux is concentrated near the rim.

\begin{figure}[t]
    \centering
    \includegraphics[width=0.8\linewidth]
    {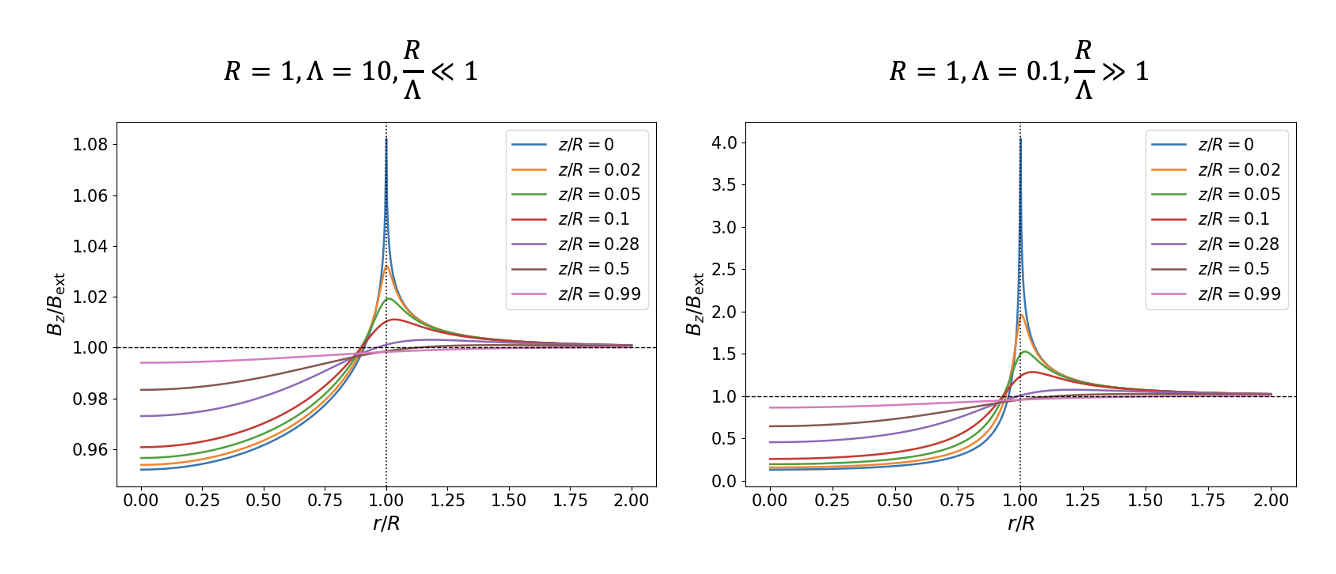}
    \caption{
    Normalized vertical magnetic field $B_z/B_{\rm ext}$ at several heights
    above the disk for weak and strong Pearl screening. In the weak-screening
    regime the field remains close to the imposed field, whereas in the
    strong-screening regime the interior field is suppressed and the expelled
    flux is focused near the disk edge.
    }
    \label{fig:Bz_profile_SI}
\end{figure}

Far from the disk,
\begin{equation}
    \mathbf B_{\rm ind}\rightarrow0,
    \qquad
    \mathbf B\rightarrow\mathbf B_{\rm ext}.
\end{equation}
Because the exterior is vacuum, the induced field decays algebraically rather
than exponentially. At sufficiently large distance the finite current
distribution behaves as a magnetic dipole.

The radial field gives a complementary view of the same screening process. The
imposed field has no radial component, so
\begin{equation}
    B_r=B_r^{\rm ind}=-\partial_z a_\phi.
\end{equation}
Immediately above the sheet, the magnetostatic boundary condition is
\begin{equation}
    B_r(r,0^+)-B_r(r,0^-)
    =
    \mu_0K_\phi(r).
\end{equation}
Using reflection symmetry,
\begin{equation}
    B_r(r,0^+)=\frac{\mu_0}{2}K_\phi(r),
\end{equation}
and Eq.~\eqref{eq:K_dimensionless_SI} gives
\begin{equation}
    \frac{B_r(r,0^+)}{B_{\rm ext}}
    =
    -\gamma\widetilde A_\phi(s).
    \label{eq:Br_boundary_tilde}
\end{equation}
For weak screening,
\begin{equation}
    \frac{B_r(r,0^+)}{B_{\rm ext}}
    \simeq
    -\frac{\gamma}{2}s,
    \qquad \gamma\ll1,
\end{equation}
whereas the strong-screening limit is
\begin{equation}
    \frac{B_r(r,0^+)}{B_{\rm ext}}
    \longrightarrow
    -\frac{2}{\pi}
    \frac{s}{\sqrt{1-s^2}}.
    \label{eq:Br_strong}
\end{equation}
The same asymptotic solution that suppresses $B_z$ in the interior therefore
produces an $O(B_{\rm ext})$ radial field and a strongly enhanced edge current.
At any finite $z>0$, the field is a nonlocal superposition of currents over the
whole disk and the sharp edge structure is smoothed.

\begin{figure}[t]
    \centering
    \includegraphics[width=0.8\linewidth]
    {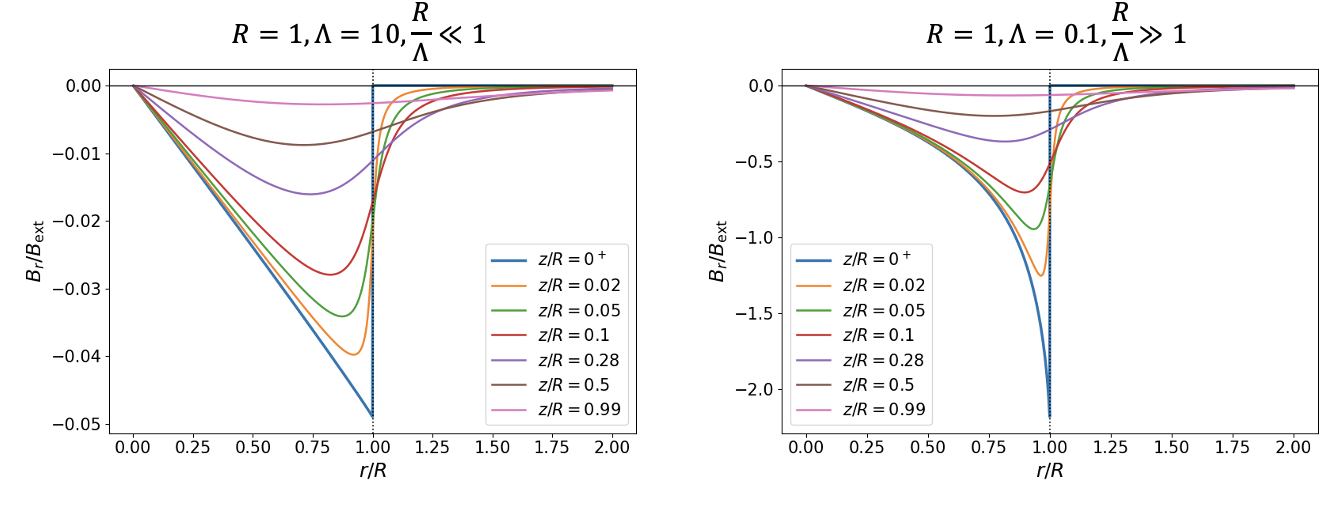}
    \caption{
    Normalized radial magnetic field $B_r/B_{\rm ext}$ at several heights above
    the disk for weak and strong Pearl screening. The surface radial field is
    directly proportional to the sheet current and becomes strongly enhanced
    near the edge in the ideal-screening limit.
    }
    \label{fig:Br_profile_SI}
\end{figure}

\subsection{Screening energy and diamagnetic response}

Having solved the self-consistent screening problem, we now use the resulting
fields to determine where the screening energy is stored and how the total
field-dependent Gibbs energy scales. We first compare the two-dimensional
phase-stiffness contribution with the three-dimensional magnetic stray-field
energy, and then evaluate the diamagnetic Gibbs energy in the weak- and
strong-screening limits. For weak screening the total vector potential on the
sheet remains close to $A_\phi^{\rm ext}$ while the induced field is small, so
the phase-stiffness energy dominates. For strong screening the total vector
potential on the sheet is reduced to $O(B_{\rm ext}\Lambda)$ while the induced
magnetic field becomes $O(B_{\rm ext})$, transferring the dominant energy cost
to the surrounding three-dimensional vacuum.

The superconducting contribution per unit area is
\begin{equation}
    \epsilon_{\rm sc}^{2D}(r)
    =
    \frac{1}{\mu_0\Lambda} A_\phi(r,0)^2\Theta(R-r),
    \label{eq:eps_sc_2D_SI}
\end{equation}
while the effective magnetic contribution is
\begin{equation}
    \epsilon_{\rm mag}^{2D}(r)
    =
    \frac{1}{2\mu_0}
    \int_{-\infty}^{\infty}dz\,
    \left[
    B_r(r,z)^2+
    \left(B_z(r,z)-B_{\rm ext}\right)^2
    \right].
    \label{eq:eps_mag_2D_SI}
\end{equation}
The induced rather than total vertical field appears because the energy of the
imposed field has already been subtracted in Eq.~\eqref{eq:Gibbs_Pearl_SI}.
Choose the natural energy-per-area scale
\begin{equation}
    \epsilon_0
    =
    \frac{B_{\rm ext}^2R}{2\mu_0}.
    \label{eq:epsilon0_SI}
\end{equation}
Then
\begin{equation}
    \frac{\epsilon_{\rm sc}^{2D}(s)}{\epsilon_0}
    =
    2\gamma\widetilde A_\phi(s)^2\Theta(1-s),
    \label{eq:eps_sc_dimensionless_SI}
\end{equation}
and, by reflection symmetry,
\begin{equation}
    \frac{\epsilon_{\rm mag}^{2D}(s)}{\epsilon_0}
    =
    2\int_0^\infty d\zeta\,
    \left[
    \left(\frac{B_r}{B_{\rm ext}}\right)^2
    +
    \left(\frac{B_z-B_{\rm ext}}{B_{\rm ext}}\right)^2
    \right].
    \label{eq:eps_mag_dimensionless_SI}
\end{equation}

For $\gamma\ll1$, Eq.~\eqref{eq:Atilde_weak_leading} gives
\begin{equation}
    \frac{\epsilon_{\rm sc}^{2D}}{\epsilon_0}
    \simeq
    \frac{\gamma}{2}s^2,
\end{equation}
whereas Eq.~\eqref{eq:Atilde_weak_first} implies
\begin{equation}
    \frac{B_{\rm ind}}{B_{\rm ext}}=O(\gamma),
\end{equation}
and therefore
\begin{equation}
    \frac{\epsilon_{\rm mag}^{2D}}{\epsilon_0}=O(\gamma^2).
\end{equation}
The phase-stiffness contribution therefore dominates parametrically. The
physical reason is that the superconducting sheet still carries nearly the bare
vector potential $A_\phi^{\rm ext}\sim B_{\rm ext}R$, while only a weak stray
field is generated in vacuum.

For $\gamma\gg1$, Eq.~\eqref{eq:Atilde_strong} instead gives, away from the edge,
\begin{equation}
    \frac{\epsilon_{\rm sc}^{2D}}{\epsilon_0}
    \simeq
    \frac{8}{\pi^2\gamma}
    \frac{s^2}{1-s^2}.
    \label{eq:eps_sc_strong}
\end{equation}
Thus the stiffness energy is suppressed by $1/\gamma$ at fixed interior
position. The ideal edge singularity requires a finite-$\Lambda$ boundary layer
for a quantitative treatment arbitrarily close to $s=1$, but it does not alter
the fact that the integrated stiffness contribution is subleading in the
strong-screening limit. At the same time,
Eqs.~\eqref{eq:a_3D_from_Aplane_SI} and \eqref{eq:Atilde_strong} show that the induced vector potential is
$O(B_{\rm ext}R)$ in the surrounding vacuum, so
\begin{equation}
    B_{\rm ind}=O(B_{\rm ext}),
\end{equation}
and the magnetic contribution in
Eq.~\eqref{eq:eps_mag_dimensionless_SI} is $O(1)$. The dominant field-induced
energy is therefore the three-dimensional magnetostatic energy required to bend
and expel the applied flux.

This difference also explains the change in the size scaling. In weak
screening, $A_\phi\sim B_{\rm ext}R$ and the stiffness energy scales as
\begin{equation}
    G_{\rm sc}
    \sim
    \frac{B_{\rm ext}^2R^4}{\mu_0\Lambda},
\end{equation}
so that $G_{\rm sc}/(\pi R^2)\sim B_{\rm ext}^2R^2/(\mu_0\Lambda)$. In strong
screening, the induced field is $O(B_{\rm ext})$ throughout a three-dimensional
region whose linear extent is set by $R$. The corresponding magnetostatic
energy scales as
\begin{equation}
    G_{\rm mag}
    \sim
    \frac{B_{\rm ext}^2}{\mu_0}R^3,
\end{equation}
and therefore
\begin{equation}
    \frac{G_{\rm mag}}{\pi R^2}
    \sim
    \frac{B_{\rm ext}^2}{\mu_0}R.
\end{equation}
The crossover from an $R^2/\Lambda$ scale to an $R$ scale is therefore a
crossover from a two-dimensional stiffness-dominated energy to a
three-dimensional stray-field-dominated energy.

\begin{figure}[t]
    \centering
    \includegraphics[width=0.8\linewidth]
    {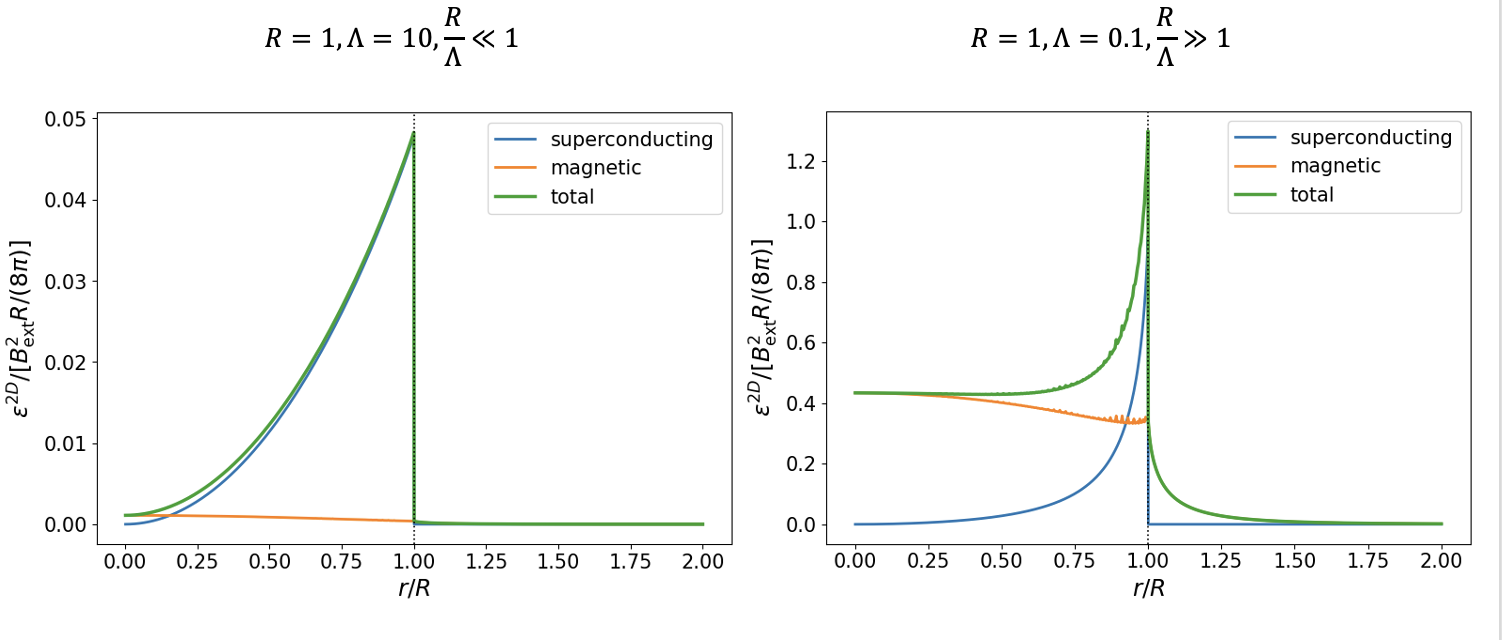}
    \caption{
    Effective two-dimensional phase-stiffness, magnetic, and total energy
    densities for weak and strong Pearl screening. The weak-screening energy is
    dominated by the superconducting stiffness term, whereas in the
    strong-screening regime the magnetic stray-field contribution becomes the
    leading term and extends outside the disk.
    }
    \label{fig:energy_density_SI}
\end{figure}

We next turn from the spatial distribution of the energy to its dependence
on the applied magnetic field. The normalized Pearl equation is linear and
contains no explicit $B_{\rm ext}$, so self-consistent screening changes the
curvature of the field-dependent energy but not its quadratic dependence on
$B_{\rm ext}$. The weak- and strong-screening limits derived above determine the
corresponding asymptotic curvatures.

Because the phase-only theory keeps the magnitude of the order parameter fixed,
it does not include the condensation energy associated with forming the
superconducting state. We therefore isolate the field-dependent diamagnetic
contribution,
\begin{equation}
    \Delta G_{\rm dia}(B_{\rm ext})
    \equiv
    G(B_{\rm ext})-G(0),
\end{equation}
and define
\begin{equation}
    \Delta g_{\rm dia}
    \equiv
    \frac{\Delta G_{\rm dia}}{\pi R^2}.
\end{equation}
In the no-vortex phase-only problem considered here, the retained terms vanish
at $B_{\rm ext}=0$, so $\Delta G_{\rm dia}=G_{\rm min}$.

At the stationary solution, a uniform rescaling of the induced vector potential
gives
\begin{equation}
    \frac{1}{2\mu_0}
    \int d^3r\,|\mathbf B_{\rm ind}|^2
    =
    -\frac{1}{\mu_0\Lambda}\int_{r<R}d^2r\,
    A_\phi(r,0)a_\phi(r,0).
\end{equation}
Substituting this identity back into the Gibbs functional yields
\begin{equation}
    G_{\rm min}
    =
    \frac{1}{\mu_0\Lambda}\int_{r<R}d^2r\,
    A_\phi(r,0)A_\phi^{\rm ext}(r).
    \label{eq:Gmin_stationary_SI}
\end{equation}
Using $\widetilde A_\phi(s)=A_\phi(Rs,0)/(B_{\rm ext}R)$,
\begin{equation}
    \Delta g_{\rm dia}
    =
    \frac{B_{\rm ext}^2R^2}{\mu_0\Lambda}
    \int_0^1ds\,s^2\widetilde A_\phi(s)
    =
    \frac{B_{\rm ext}^2R}{\mu_0}
    \gamma
    \int_0^1ds\,s^2\widetilde A_\phi(s).
    \label{eq:dia_energy_B_SI}
\end{equation}
Thus the field-induced energy remains exactly quadratic in $B_{\rm ext}$.

For weak screening, Eq.~\eqref{eq:Atilde_weak_leading} gives
\begin{equation}
    \Delta g_{\rm dia}
    \simeq
    \frac{B_{\rm ext}^2R\gamma}{8\mu_0}
    =
    \frac{B_{\rm ext}^2R^2}{8\mu_0\Lambda},
    \qquad \gamma\ll1.
    \label{eq:weak_energy_SI}
\end{equation}
This is the same result obtained by neglecting electromagnetic back-reaction,
because the induced vector potential is only a perturbative correction in this
limit.

For strong screening, Eq.~\eqref{eq:Atilde_strong} gives
\begin{align}
    \gamma
    \int_0^1ds\,s^2\widetilde A_\phi(s)
    &\longrightarrow
    \frac{2}{\pi}
    \int_0^1ds\,
    \frac{s^3}{\sqrt{1-s^2}}
    \nonumber\\
    &=
    \frac{4}{3\pi}.
\end{align}
Hence
\begin{equation}
    \Delta g_{\rm dia}
    \xrightarrow[\gamma\rightarrow\infty]{}
    \frac{4R}{3\pi\mu_0}B_{\rm ext}^2.
    \label{eq:strong_energy_SI}
\end{equation}
The strong-screening energy is independent of the microscopic phase stiffness to leading
order. Once the disk is already close to perfect screening, increasing the
stiffness further cannot substantially change the magnetic response. The
remaining scale is set by the three-dimensional magnetostatic geometry of the
disk.

If the field dependence is written as
\begin{equation}
    \Delta g_{\rm dia}
    =
    \frac{1}{2}\kappa_{\rm dia} B_{\rm ext}^2,
\end{equation}
then
\begin{equation}
    \kappa_{\rm dia}
    =
    \frac{2R}{\mu_0}
    \gamma
    \int_0^1ds\,s^2\widetilde A_\phi(s),
\end{equation}
with limiting forms
\begin{equation}
    \kappa_{\rm dia}
    \simeq
    \frac{R^2}{4\mu_0\Lambda},
    \qquad \gamma\ll1,
\end{equation}
and
\begin{equation}
    \kappa_{\rm dia}
    \longrightarrow
    \frac{8R}{3\pi\mu_0},
    \qquad \gamma\gg1.
\end{equation}
The change from $R^2/\Lambda$ to $R$ is the global energetic counterpart of the
local crossover derived above: weak screening stores most of the energy in the
two-dimensional superconducting stiffness, whereas strong screening stores the
dominant energy in the three-dimensional magnetic stray field.

\begin{figure}[t]
    \centering
    \includegraphics[width=0.8\linewidth]{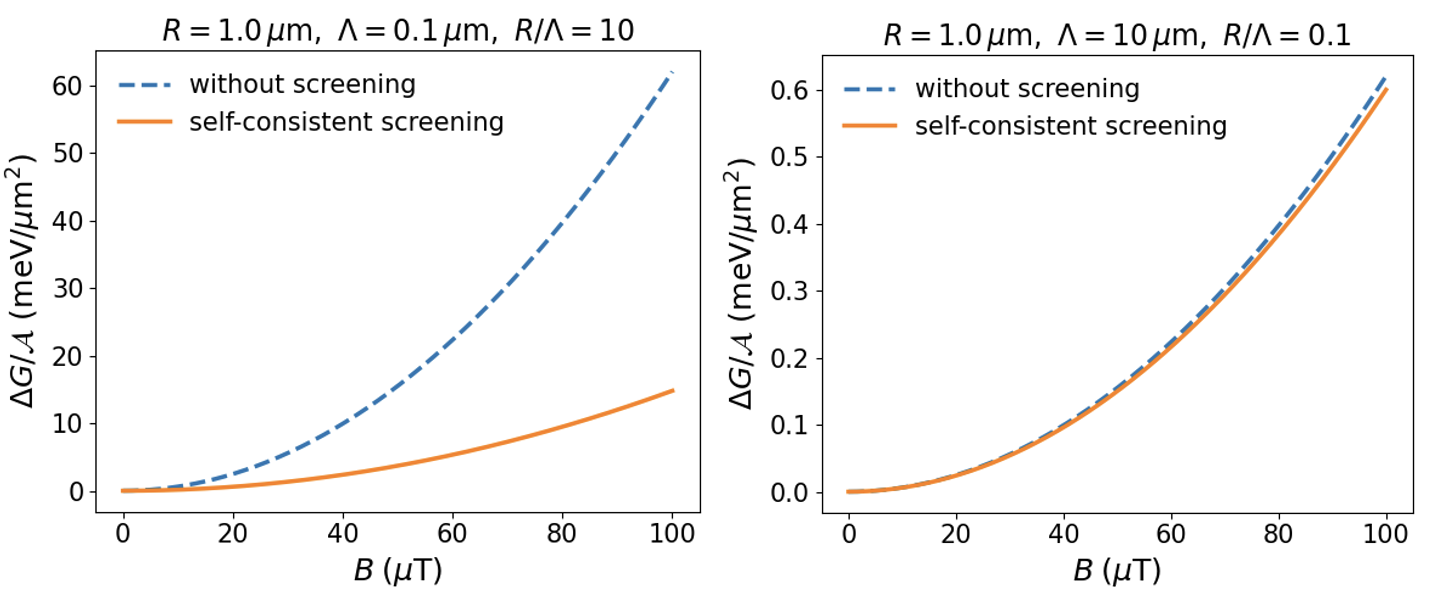}
    \caption{
    Field-induced diamagnetic Gibbs energy per unit area as a function of
    $B_{\rm ext}$. The weak-screening curve is close to the unscreened London
    result, while strong screening reduces the curvature to the ideal-disk
    magnetostatic value. In both limits the field dependence remains quadratic
    within the linear London theory.
    }
    \label{fig:energy_density_B_screening_SI}
\end{figure}

\subsection{Pearl screening of a single vortex}

We now extend the same electromagnetic kernel to a centered vortex with winding $\eta=1$. In this subsection we separate the field-driven Meissner response from the vortex-driven component, solve the latter in the regime where the Pearl crossover fits inside the sample, and use the screened profile to determine the zero-field vortex self-energy. Outside the microscopic core, $r>\xi$, the condensate phase obeys $\nabla\theta=\hat{\boldsymbol\phi}/r$. It is convenient to represent this winding by
\begin{align}
\mathbf A_v(\mathbf r)
\equiv\frac{\Phi_0}{2\pi}\nabla\theta
=\frac{\Phi_0}{2\pi r}\hat{\boldsymbol\phi}.
\label{eq:Av_single_vortex}
\end{align}
The phase-only Gibbs functional is then
\begin{align}
G[a_\phi]
={}&
\frac{1}{\mu_0\Lambda}
\int_{\xi<r<R}d^2r\,
\left[A_\phi(r,0)-A_v(r)\right]^2
+\frac{1}{2\mu_0}\int d^3r\,|\nabla\times\mathbf a|^2
+E_{\rm core},
\label{eq:G_single_vortex_SI}
\end{align}
where $A_\phi(r,z)=B_{\rm ext}r/2+a_\phi(r,z)$. The cutoff at $r=\xi$ only separates the long-distance London region from the microscopic vortex core; its cutoff dependence is absorbed into $E_{\rm core}$.

We begin by separating the self-consistent response into Meissner and vortex components. Define the gauge-invariant azimuthal potential
\begin{align}
U_\phi(r)\equiv A_\phi(r,0)-A_v(r),
\qquad
K_\phi(r)=-\frac{2}{\mu_0\Lambda}U_\phi(r).
\label{eq:U_total_definition}
\end{align}
Using the same magnetostatic kernel $Q$ introduced in Eq.~\eqref{eq:Q_3D_SI},
\begin{align}
a_\phi(r,z)
=-\frac{1}{2\pi\Lambda}
\int_\xi^Rdr'\,r'Q(r,r';z)U_\phi(r').
\label{eq:a_from_U_single_vortex}
\end{align}
It is useful to define the linear Pearl operator
\begin{align}
\widehat{\mathcal L}X(r)
\equiv
\frac{1}{2\pi\Lambda}
\int_\xi^Rdr'\,r'Q(r,r';0)X(r'),
\label{eq:single_vortex_Pearl_operator}
\end{align}
so that the sheet equation reads $(1+\widehat{\mathcal L})U_\phi=A_\phi^{\rm ext}-A_v$. By linearity,
\begin{align}
U_\phi&=U_M+U_v,
\label{eq:U_decomposition}\\
(1+\widehat{\mathcal L})U_M&=A_\phi^{\rm ext},
\label{eq:UM_operator}\\
(1+\widehat{\mathcal L})U_v&=-A_v.
\label{eq:Uv_operator}
\end{align}
The Meissner component $U_M$ is driven by exactly the same external-field source as the vortex-free sector. The only difference is that the hard-core model removes the microscopic region $r<\xi$ from the vortex problem. For $\xi/R\ll1$, this changes the global Meissner response only by subleading core corrections, so to leading order $U_M$ is the vortex-free solution obtained from Eq.~\eqref{eq:Atilde_Pearl_gamma}. In particular,
\begin{align}
U_M(r)\simeq\frac{B_{\rm ext}r}{2},\quad R\ll\Lambda;
\qquad
U_M(Rs)\simeq\frac{2B_{\rm ext}\Lambda}{\pi}\frac{s}{\sqrt{1-s^2}},\quad R\gg\Lambda,
\label{eq:UM_screening_limits}
\end{align}
away from the outer-edge boundary layer.

Thus only the vortex-driven component requires a new dimensionless equation. Using $s=r/R$, $s_\xi=\xi/R$, $\gamma=R/\Lambda$, and $\widetilde U_v(s)=U_v(Rs)/[\Phi_0/(2\pi R)]$, Eq.~\eqref{eq:Uv_operator} becomes
\begin{align}
\widetilde U_v(s)
+\frac{\gamma}{2\pi}
\int_{s_\xi}^{1}ds'\,s'\mathcal Q(s,s';0)\widetilde U_v(s')
=-\frac{1}{s}.
\label{eq:Uv_dimensionless}
\end{align}
The control parameter is therefore the same $\gamma=R/\Lambda$ as in the vortex-free problem; the new physics is the singular $1/s$ source associated with the phase winding.

We now solve Eq.~\eqref{eq:Uv_operator}. The vortex problem contains spatial
structure on all scales between the microscopic core size and the disk radius,
so the relevant question is which spatial scales are larger or smaller than
$\Lambda$. The most transparent way to expose this scale dependence is to work
in two-dimensional momentum space. After the finite boundary is neglected, the
magnetostatic Green function depends only on $\mathbf r-\mathbf r'$, so the
electromagnetic integral operator becomes a convolution and Fourier
transformation diagonalizes it mode by mode.

The finite disk itself is not translationally invariant because the current is
restricted to $r<R$. Therefore the momentum-space solution derived below is
not an exact solution of the finite-disk problem. The approximation consists
of replacing the finite superconducting domain by an infinite uniform film
when evaluating the local vortex screening.

This approximation is controlled mode by mode. A Fourier component with
wavevector $k$ probes a spatial scale of order $1/k$. It is insensitive to the
outer boundary when
\begin{equation}
    kR\gg1.
    \label{eq:kR_condition}
\end{equation}
The Pearl crossover occurs at $k\sim1/\Lambda$. Therefore the entire crossover
can be described by the infinite-film solution when
\begin{equation}
    \frac{R}{\Lambda}\gg1.
    \label{eq:Fourier_strong_condition}
\end{equation}
Indeed, the crossover modes then satisfy $kR\sim R/\Lambda\gg1$ and do not
resolve the disk boundary. Near the outer edge the finite-disk solution must
again be used.

The same Fourier solution is still useful conceptually when $R\ll\Lambda$,
but it is no longer a controlled global solution of the finite disk. In that
case the smallest wavevector supported by the sample is of order $1/R$, and
\begin{equation}
    k_{\min}\Lambda
    \sim
    \frac{\Lambda}{R}
    \gg1.
\end{equation}
The disk reaches its boundary before modes with $k\Lambda\lesssim1$ can
develop. The vortex is therefore approximately unscreened throughout most of
the sample, but the infrared cutoff and the detailed behavior near $r\sim R$
must be determined by the finite disk rather than by the infinite-film
solution.

At short distances there is a second limitation. The London theory ceases to
be valid for $r\lesssim\xi$, corresponding to wavevectors $k\gtrsim1/\xi$.
Thus the momentum-space treatment is a London description of modes between the
infrared scale set by the finite sample and the ultraviolet scale set by the
core. In the strong-screening hierarchy $\xi\ll\Lambda\ll R$, the physically
important crossover modes $k\sim1/\Lambda$ lie parametrically inside this
window.

Within the infinite-film approximation, the vector version of
Eq.~\eqref{eq:a_from_U_single_vortex} is
\begin{equation}
    \mathbf a_v(\mathbf r,0)
    =
    -\frac{1}{2\pi\Lambda}
    \int d^2r'\,
    \frac{\mathbf U_v(\mathbf r')}
    {|\mathbf r-\mathbf r'|}.
    \label{eq:av_infinite_realspace}
\end{equation}
The vortex source itself is not translationally invariant; it singles out the
vortex position. Translational invariance is required only of the linear
electromagnetic operator. In the infinite film the kernel depends only on the
coordinate difference, so Eq.~\eqref{eq:av_infinite_realspace} is a
convolution even though the source and solution are spatially nonuniform.

Using
\begin{equation}
    \mathbf U_v(\mathbf k)
    =
    \int d^2r\,
    e^{-i\mathbf k\cdot\mathbf r}\mathbf U_v(\mathbf r),
\end{equation}
together with
\begin{equation}
    \int d^2r\,
    \frac{e^{-i\mathbf k\cdot\mathbf r}}{r}
    =
    \frac{2\pi}{k},
\end{equation}
Eq.~\eqref{eq:av_infinite_realspace} becomes
\begin{equation}
    \mathbf a_v(\mathbf k,0)
    =
    -\frac{1}{k\Lambda}\mathbf U_v(\mathbf k).
    \label{eq:av_Uv_fourier}
\end{equation}
Since
\begin{equation}
    \mathbf U_v
    =
    \mathbf a_v-\mathbf A_v,
\end{equation}
we obtain
\begin{equation}
    \left(1+\frac{1}{k\Lambda}\right)
    \mathbf U_v(\mathbf k)
    =
    -\mathbf A_v(\mathbf k),
\end{equation}
and therefore
\begin{equation}
    \mathbf U_v(\mathbf k)
    =
    -\frac{k\Lambda}{1+k\Lambda}
    \mathbf A_v(\mathbf k),
    \label{eq:Uv_fourier_solution}
\end{equation}
while
\begin{equation}
    \mathbf a_v(\mathbf k,0)
    =
    \frac{1}{1+k\Lambda}
    \mathbf A_v(\mathbf k).
    \label{eq:av_fourier_solution}
\end{equation}

These equations exhibit the screening mode by mode. For $k\Lambda\gg1$,
\begin{equation}
    \mathbf U_v(\mathbf k)
    \simeq
    -\mathbf A_v(\mathbf k),
    \qquad
    \mathbf a_v(\mathbf k,0)
    \ll
    \mathbf A_v(\mathbf k),
\end{equation}
so short-wavelength vortex structure is essentially unscreened. For
$k\Lambda\ll1$,
\begin{equation}
    \mathbf U_v(\mathbf k)
    \simeq
    -k\Lambda\mathbf A_v(\mathbf k),
    \qquad
    \mathbf a_v(\mathbf k,0)
    \simeq
    \mathbf A_v(\mathbf k),
\end{equation}
so the long-wavelength gauge-invariant vortex potential is strongly
suppressed.

For the azimuthal vortex source it is convenient to use the order-one Hankel
transform,
\begin{equation}
    F(r)
    =
    \int_0^\infty dk\,kJ_1(kr)F(k),
\end{equation}
for which
\begin{equation}
    A_v(k)
    =
    \frac{\Phi_0}{2\pi k}.
\end{equation}
Equation~\eqref{eq:Uv_fourier_solution} gives
\begin{equation}
    U_v(r)
    =
    -\frac{\Phi_0\Lambda}{2\pi}
    \int_0^\infty dk\,
    \frac{kJ_1(kr)}{1+k\Lambda}.
    \label{eq:Uv_Hankel}
\end{equation}
Using
\begin{equation}
    \frac{1}{1+k\Lambda}
    =
    \int_0^\infty dt\,e^{-t}e^{-k\Lambda t},
\end{equation}
and performing the $k$ integral,
\begin{equation}
    U_v(r)
    =
    -\frac{\Phi_0}{2\pi r}
    \int_0^\infty dt\,e^{-t}
    \frac{t}{\sqrt{t^2+(r/\Lambda)^2}}.
    \label{eq:Uv_realspace_crossover}
\end{equation}
This expression gives the real-space Pearl crossover directly.

For $\xi\ll r\ll\Lambda$, the integral approaches unity, so
\begin{equation}
    U_v(r)
    \simeq
    -\frac{\Phi_0}{2\pi r}
    =
    -A_v(r).
    \label{eq:Uv_inner}
\end{equation}
Thus the induced electromagnetic vector potential is small compared with the
phase-winding source in the inner region,
\begin{equation}
    a_\phi^v
    =
    A_v+U_v
    \ll
    A_v,
\end{equation}
and the vortex current retains the unscreened London form
\begin{equation}
    K_\phi^v(r)
    \simeq
    \frac{\Phi_0}{\pi\mu_0\Lambda r}.
\end{equation}

For $\Lambda\ll r\ll R$, the integral in
Eq.~\eqref{eq:Uv_realspace_crossover} behaves as $\Lambda/r$, giving
\begin{equation}
    U_v(r)
    \simeq
    -\frac{\Phi_0\Lambda}{2\pi r^2}.
    \label{eq:Uv_outer}
\end{equation}
Equivalently,
\begin{equation}
    a_\phi^v(r,0)
    =
    A_v(r)+U_v(r)
    \simeq
    \frac{\Phi_0}{2\pi r}
    \left(1-\frac{\Lambda}{r}\right).
\end{equation}
At long distance the induced electromagnetic vector potential approaches the
phase-winding vector potential, leaving only the smaller gauge-invariant
residual $U_v\propto\Lambda/r^2$. The sheet current correspondingly crosses
over to
\begin{equation}
    K_\phi^v(r)
    \simeq
    \frac{\Phi_0}{\pi\mu_0r^2}.
\end{equation}

The screening profile is therefore
\begin{equation}
    U_v(r)
    \sim
    \begin{cases}
        \displaystyle
        -\frac{\Phi_0}{2\pi r},
        &
        \xi\ll r\ll\Lambda,
        \\[10pt]
        \displaystyle
        -\frac{\Phi_0\Lambda}{2\pi r^2},
        &
        \Lambda\ll r\ll R.
    \end{cases}
    \label{eq:Uv_crossover_summary}
\end{equation}
The strong-screening limit is nonuniform in space. At a fixed scaled position
$r/R=O(1)$, $U_v$ becomes small when $R/\Lambda\rightarrow\infty$, but the
inner region $r\lesssim\Lambda$ retains the unscreened $1/r$ structure.

It is important that this momentum-space simplification applies to the local vortex screening problem, but not to the global uniform-field Meissner problem. The preceding calculation works because, in the strong-screening hierarchy
$\xi\ll\Lambda\ll R$, the physically important vortex crossover occurs at
$k\sim1/\Lambda$, for which
\begin{equation}
    kR
    \sim
    \frac{R}{\Lambda}
    \gg1.
\end{equation}
The crossover modes therefore do not resolve the outer boundary, and the
finite disk can be replaced locally by an infinite film.

The uniform perpendicular-field problem is qualitatively different. The
applied field is a global, zero-wavevector source. In symmetric gauge,
$A_\phi^{\rm ext}=B_{\rm ext}r/2$ grows linearly with $r$ and its Fourier
transform is distributional, concentrated at $k=0$. More importantly, the
leading Meissner-current pattern is set by the finite disk geometry itself.
Even for $R\gg\Lambda$, the ideal-disk current is a function of $r/R$ and is
organized by the outer edge. The relevant source therefore never satisfies
the local condition $kR\gg1$ that allowed the boundary to be discarded in the
vortex problem.

A Fourier--Bessel representation of the induced Meissner field is still
possible, but it does not diagonalize the finite-disk problem. The conditions
inside and outside the disk become a pair of dual integral equations rather
than a simple algebraic factor such as $k\Lambda/(1+k\Lambda)$. Thus the
momentum-space method is a local and controlled simplification for the
centered vortex when the Pearl crossover fits inside the sample, whereas the
finite boundary remains part of the leading physics of the uniform
perpendicular-field Meissner state.

With the screened vortex profile in hand, we can determine the constant term in the single-vortex Gibbs energy. At $B_{\rm ext}=0$,
\begin{equation}
    G_v(0)
    =
    E_{\rm core}
    +
    G_v^{\rm sc}
    +
    G_v^{\rm mag},
    \label{eq:Gv0_decomposition}
\end{equation}
where
\begin{equation}
    G_v^{\rm sc}
    =
    \frac{1}{\mu_0\Lambda}
    \int d^2r\,U_v^2,
    \label{eq:Gv_sc_definition}
\end{equation}
and
\begin{equation}
    G_v^{\rm mag}
    =
    \frac{1}{2\mu_0}
    \int d^3r\,|\mathbf B_v|^2.
    \label{eq:Gv_mag_definition}
\end{equation}
Both terms are required once the electromagnetic field is treated
self-consistently. In particular, the zero-field vortex energy is no longer
only a phase-stiffness cost.

The real-space asymptotics already determine the logarithmic structure. In
the inner region $\xi<r<\min(R,\Lambda)$,
Eq.~\eqref{eq:Uv_inner} gives
\begin{equation}
    U_v(r)
    \simeq
    -\frac{\Phi_0}{2\pi r},
\end{equation}
and therefore
\begin{equation}
    G_v^{\rm sc}
    \supset
    \frac{\Phi_0^2}{2\pi\mu_0\Lambda}
    \int_\xi^{\min(R,\Lambda)}\frac{dr}{r}.
\end{equation}
The logarithmic stiffness contribution is consequently
\begin{equation}
    G_v^{\rm sc}
    \sim
    \frac{\Phi_0^2}{2\pi\mu_0\Lambda}
    \ln\frac{\min(R,\Lambda)}{\xi}.
    \label{eq:Gv_log_scaling}
\end{equation}
For $r>\Lambda$, the screened tail $U_v\propto\Lambda/r^2$ gives
$rU_v^2\propto\Lambda^2/r^3$, so the outer stiffness contribution is
convergent and produces no additional logarithm.

The magnetic contribution in Eq.~\eqref{eq:Gv_mag_definition} can be
evaluated with the same infinite-film solution. For a self-generated
magnetostatic field,
\begin{equation}
    \frac{1}{2\mu_0}
    \int d^3r\,|\mathbf B_v|^2
    =
    \frac{1}{2}
    \int d^2r\,K_\phi^v(r)a_\phi^v(r,0).
    \label{eq:Bv2_Ka_identity}
\end{equation}
Using the order-one Hankel transform and Parseval's identity,
\begin{equation}
    G_v^{\rm mag}
    =
    \pi
    \int_0^\infty dk\,k\,
    K_\phi^v(k)a_\phi^v(k).
\end{equation}
From Eqs.~\eqref{eq:Uv_fourier_solution} and
\eqref{eq:av_fourier_solution},
\begin{equation}
    K_\phi^v(k)
    =
    -\frac{2}{\mu_0\Lambda}U_v(k)
    =
    \frac{2k}{\mu_0(1+k\Lambda)}A_v(k),
\end{equation}
and
\begin{equation}
    a_\phi^v(k)
    =
    \frac{1}{1+k\Lambda}A_v(k).
\end{equation}
Since $A_v(k)=\Phi_0/(2\pi k)$,
\begin{equation}
    G_v^{\rm mag}
    =
    \frac{\Phi_0^2}{2\pi\mu_0}
    \int dk\,
    \frac{1}{(1+k\Lambda)^2}.
    \label{eq:Gv_mag_k}
\end{equation}
For comparison, Parseval's identity gives
\begin{equation}
    G_v^{\rm sc}
    =
    \frac{\Phi_0^2}{2\pi\mu_0}
    \int dk\,
    \frac{k\Lambda}{(1+k\Lambda)^2}.
    \label{eq:Gv_sc_k}
\end{equation}

Introduce
\begin{equation}
    x=k\Lambda,
    \qquad
    E_\Lambda
    \equiv
    \frac{\Phi_0^2}{2\pi\mu_0\Lambda}.
    \label{eq:E_Lambda_definition}
\end{equation}
To use the infinite-film formulas as estimates for the finite disk, we
introduce the physical cutoffs
\begin{equation}
    x_{\min}\sim\frac{\Lambda}{R},
    \qquad
    x_{\max}\sim\frac{\Lambda}{\xi}.
    \label{eq:x_cutoffs}
\end{equation}
This cutoff procedure is not an exact finite-disk calculation. It captures
the leading logarithms and parametric scaling when there is good separation
between $R$, $\Lambda$, and $\xi$, but the $O(1)$ constants depend on the
microscopic core regularization and on the actual finite-disk edge solution.

With this understanding,
\begin{equation}
    G_v^{\rm sc}
    \simeq
    E_\Lambda
    \int_{x_{\min}}^{x_{\max}}dx\,
    \frac{x}{(1+x)^2},
    \label{eq:Gv_sc_fourier}
\end{equation}
and
\begin{equation}
    G_v^{\rm mag}
    \simeq
    E_\Lambda
    \int_{x_{\min}}^{x_{\max}}dx\,
    \frac{1}{(1+x)^2}.
    \label{eq:Gv_mag_fourier}
\end{equation}
The two integrals are
\begin{align}
    G_v^{\rm sc}
    &\simeq
    E_\Lambda
    \left[
        \ln(1+x)+\frac{1}{1+x}
    \right]_{x_{\min}}^{x_{\max}},
    \label{eq:Gv_sc_integrated}
    \\
    G_v^{\rm mag}
    &\simeq
    E_\Lambda
    \left[
        \frac{1}{1+x_{\min}}
        -
        \frac{1}{1+x_{\max}}
    \right].
    \label{eq:Gv_mag_integrated}
\end{align}
Their sum is particularly simple,
\begin{equation}
    G_v(0)-E_{\rm core}
    \simeq
    E_\Lambda
    \ln
    \left[
        \frac{1+\Lambda/\xi}
             {1+\Lambda/R}
    \right].
    \label{eq:Gv0_interpolation}
\end{equation}
The interpolation in Eq.~\eqref{eq:Gv0_interpolation} should be understood
with the same cutoff qualification. The limiting logarithms and the scaling
of the magnetic contribution are robust; the detailed constant terms are not
universal.

For weak screening, $\xi\ll R\ll\Lambda$, one has
$x_{\min}\gg1$ and $x_{\max}\gg x_{\min}$. The stiffness energy is
\begin{equation}
    G_v^{\rm sc}
    \simeq
    E_\Lambda
    \ln\frac{R}{\xi},
\end{equation}
while
\begin{equation}
    G_v^{\rm mag}
    \sim
    E_\Lambda\frac{R}{\Lambda}
\end{equation}
is parametrically smaller than the logarithmic stiffness term. Thus
\begin{equation}
    G_v(0)
    \simeq
    E_{\rm core}
    +
    \frac{\Phi_0^2}{2\pi\mu_0\Lambda}
    \ln\frac{R}{\xi}.
    \label{eq:Gv0_weak}
\end{equation}
The sample ends before Pearl screening develops, so the finite radius supplies
the infrared cutoff.

For strong screening, $\xi\ll\Lambda\ll R$, one has
$x_{\min}\ll1$ and $x_{\max}\gg1$. Within the sharp-cutoff approximation,
\begin{equation}
    G_v^{\rm sc}
    \simeq
    E_\Lambda
    \left(
        \ln\frac{\Lambda}{\xi}-1
    \right),
    \label{eq:Gv_sc_strong}
\end{equation}
whereas
\begin{equation}
    G_v^{\rm mag}
    \simeq
    E_\Lambda.
    \label{eq:Gv_mag_strong}
\end{equation}
Therefore
\begin{equation}
    G_v(0)
    \simeq
    E_{\rm core}
    +
    \frac{\Phi_0^2}{2\pi\mu_0\Lambda}
    \ln\frac{\Lambda}{\xi}.
    \label{eq:Gv0_strong}
\end{equation}
The logarithm survives because the vortex remains essentially unscreened
between $\xi$ and $\Lambda$, but it no longer grows with $R$. The magnetic
self-energy is finite, of order $E_\Lambda$, and is an essential part of the
full zero-field vortex energy. The separate constant terms in
Eqs.~\eqref{eq:Gv_sc_strong} and \eqref{eq:Gv_mag_strong} depend on the cutoff
implementation, whereas the replacement
\begin{equation}
    \ln\frac{R}{\xi}
    \longrightarrow
    \ln\frac{\Lambda}{\xi}
\end{equation}
is the robust strong-screening result.

\subsection{Vortex energy and orbital magnetization}

Having determined the screened vortex profile and zero-field self-energy, we now restore the applied field. We first derive the general field expansion, then evaluate the weak- and strong-screening limits of the vortex energy and orbital moment, and finally compare these asymptotic results with the numerical finite-disk solution. Because $U_M$ and $\mathbf B_M$ are linear in $B_{\rm ext}$ while
$U_v$ and $\mathbf B_v$ are independent of $B_{\rm ext}$ for a fixed vortex,
the Gibbs energy has the exact quadratic structure
\begin{equation}
    G_{1v}(B_{\rm ext})
    =
    G_v(0)
    +
    G_{Mv}(B_{\rm ext})
    +
    G_M(B_{\rm ext}),
    \label{eq:G1v_field_expansion}
\end{equation}
where
\begin{equation}
    G_v(0)\propto B_{\rm ext}^0,
    \qquad
    G_{Mv}\propto B_{\rm ext},
    \qquad
    G_M\propto B_{\rm ext}^2.
\end{equation}
Equivalently,
\begin{equation}
    G_{1v}(B_{\rm ext})
    =
    G_0+G_1B_{\rm ext}+G_2B_{\rm ext}^2.
\end{equation}
The constant term contains the vortex stiffness energy, vortex magnetic
self-energy, and core energy. The quadratic term is the ordinary Meissner
energy and is the same as in the no-vortex state when the same screening
operator is used. In the hard-core implementation it differs only through the
small change of the superconducting domain. The new quantity that controls
the vortex orbital moment is the term linear in $B_{\rm ext}$.

The full functional can be written as
\begin{equation}
    G
    =
    \frac{1}{\mu_0\Lambda}
    \int d^2r\,
    \left(U_M+U_v\right)^2
    +
    \frac{1}{2\mu_0}
    \int d^3r\,
    \left|\mathbf B_M+\mathbf B_v\right|^2
    +
    E_{\rm core}.
    \label{eq:G_decomposed_fields}
\end{equation}
The field-linear contribution from the stiffness energy is
\begin{equation}
    G_{Mv}^{\rm sc}
    =
    \frac{2}{\mu_0\Lambda}
    \int d^2r\,U_MU_v
    =
    \frac{2}{\mu_0\Lambda}
    \int d^2r\,
    U_M\left(a_\phi^v-A_v\right),
    \label{eq:Gcross_sc_revised}
\end{equation}
and the magnetic cross term is
\begin{equation}
    G_{Mv}^{\rm mag}
    =
    \frac{1}{\mu_0}
    \int d^3r\,
    \mathbf B_M\cdot\mathbf B_v.
    \label{eq:Gcross_mag_revised}
\end{equation}

To combine them, write
\begin{equation}
    \mathbf B_v
    =
    \nabla\times\mathbf a^v.
\end{equation}
Using
\begin{equation}
    \nabla\cdot
    \left(\mathbf a^v\times\mathbf B_M\right)
    =
    \mathbf B_M\cdot
    \left(\nabla\times\mathbf a^v\right)
    -
    \mathbf a^v\cdot
    \left(\nabla\times\mathbf B_M\right),
\end{equation}
and discarding the surface term at infinity,
\begin{equation}
    G_{Mv}^{\rm mag}
    =
    \frac{1}{\mu_0}
    \int d^3r\,
    \mathbf a^v\cdot
    \left(\nabla\times\mathbf B_M\right).
\end{equation}
Amp\`ere's law gives
\begin{equation}
    \nabla\times\mathbf B_M
    =
    \mu_0\mathbf J_M,
\end{equation}
so
\begin{equation}
    G_{Mv}^{\rm mag}
    =
    \int d^3r\,
    \mathbf a^v\cdot\mathbf J_M
    =
    \int d^2r\,
    a_\phi^vK_\phi^M.
\end{equation}
Using
\begin{equation}
    K_\phi^M
    =
    -\frac{2}{\mu_0\Lambda}U_M,
\end{equation}
we obtain
\begin{equation}
    G_{Mv}^{\rm mag}
    =
    -\frac{2}{\mu_0\Lambda}
    \int d^2r\,
    U_Ma_\phi^v.
    \label{eq:Gcross_mag_cancellation}
\end{equation}
Adding Eqs.~\eqref{eq:Gcross_sc_revised} and
\eqref{eq:Gcross_mag_cancellation}, the terms containing the vortex-induced
vector potential cancel exactly:
\begin{equation}
    G_{Mv}
    =
    -\frac{2}{\mu_0\Lambda}
    \int d^2r\,
    A_v(r)U_M(r).
    \label{eq:Gcross_final_U_M}
\end{equation}

The cancellation has a direct variational interpretation. The Meissner field
$a_\phi^M$ is already a stationary solution of the field-driven Gibbs
functional. An additional electromagnetic variation in the direction
$a_\phi^v$ therefore cannot change that relaxed energy to first order. The
linear change of the stiffness energy is exactly compensated by the linear
change of the magnetic field energy. The only surviving term linear in both
the applied field and the vortex sector is the explicit coupling of the
vortex source $A_v$ to the self-consistent Meissner potential $U_M$.

This stationarity statement is exact when the Meissner and vortex components
are solved with the same superconducting domain and stiffness profile. In the
hard-core model it is exact for the annular operator. Replacing the annular
$U_M$ by the full-disk uniform-state solution is the additional
$\xi/R\ll1$ approximation discussed above.

For the centered vortex,
\begin{equation}
    A_v(r)
    =
    \frac{\Phi_0}{2\pi r},
\end{equation}
so
\begin{align}
    G_{Mv}
    &=-\frac{2}{\mu_0\Lambda}
    \int_\xi^R2\pi r\,dr\,
    \frac{\Phi_0}{2\pi r}U_M(r)
    \nonumber\\
    &=-\frac{2}{\mu_0\Lambda}\Phi_0
    \int_\xi^Rdr\,U_M(r).
    \label{eq:Gcross_radial_U_M}
\end{align}
Thus the orbital magnetic moment is controlled entirely by the same self-consistent Meissner solution as in the vortex-free state; no separate solution for $a_\phi^v$ is needed to determine the coefficient linear in $B_{\rm ext}$. The weak- and strong-screening scalings follow directly from the two limits in Eq.~\eqref{eq:UM_screening_limits}.

The preceding analysis separates the roles of $U_v$ and $U_M$: the vortex
component determines the zero-field self-energy, while the Meissner component
determines the coupling linear in the applied field. We now assemble the two
limiting regimes.

\noindent\textit{Weak screening: $\xi\ll R\ll\Lambda$.}
When the sample is smaller than the Pearl length, the boundary is reached
before the vortex screening crossover can develop. The vortex sector is
approximately unscreened through most of the disk,
\begin{equation}
    U_v(r)
    \simeq
    -A_v(r)
    =
    -\frac{\Phi_0}{2\pi r}.
\end{equation}
Consequently,
\begin{equation}
    G_v(0)
    \simeq
    E_{\rm core}
    +
    \frac{\Phi_0^2}{2\pi\mu_0\Lambda}
    \ln\frac{R}{\xi},
\end{equation}
with a magnetic self-energy that is parametrically smaller than the
logarithmic stiffness term.

The Meissner component is also weakly screened,
\begin{equation}
    U_M(r)
    \simeq
    A_\phi^{\rm ext}(r)
    =
    \frac{B_{\rm ext}r}{2}.
\end{equation}
Equation~\eqref{eq:Gcross_radial_U_M} therefore gives
\begin{align}
    G_{Mv}^{\rm weak}
    &=-\frac{2}{\mu_0\Lambda}\Phi_0
    \int_\xi^Rdr\,
    \frac{B_{\rm ext}r}{2}
    \nonumber\\
    &=-\frac{\Phi_0B_{\rm ext}}{2\mu_0\Lambda}
    \left(R^2-\xi^2\right).
\end{align}
For $R\gg\xi$,
\begin{equation}
    G_{Mv}^{\rm weak}
    \simeq
    -\frac{\Phi_0B_{\rm ext}}{2\mu_0\Lambda}R^2.
    \label{eq:Gcross_weak_final}
\end{equation}
The vortex orbital moment,
\begin{equation}
    m_v
    \equiv
    -\frac{\partial G_{Mv}}{\partial B_{\rm ext}},
\end{equation}
is therefore
\begin{equation}
    m_v^{\rm weak}
    \simeq
    \frac{\Phi_0}{2\mu_0\Lambda}R^2.
    \label{eq:mv_weak_final}
\end{equation}
The area law follows because $A_v\propto1/r$ while
$U_M\simeq A_\phi^{\rm ext}\propto r$, so their product is approximately
constant over the disk.

The quadratic term $G_M(B_{\rm ext})$ is the weak-screening Meissner energy
already obtained in the uniform sector. Thus in this regime the full energy
has the structure
\begin{equation}
    G_{1v}(B_{\rm ext})
    =
    E_{\rm core}
    +
    \frac{\Phi_0^2}{2\pi\mu_0\Lambda}
    \ln\frac{R}{\xi}
    -
    \frac{\Phi_0R^2}{2\mu_0\Lambda}B_{\rm ext}
    +
    G_M^{\rm weak}(B_{\rm ext})
\end{equation}
to leading order in $\xi/R$ and $R/\Lambda$.

\noindent\textit{Strong screening: $\xi\ll\Lambda\ll R$.}
When the Pearl length lies well inside the disk, the vortex sector develops
the full screening crossover
\begin{equation}
    U_v(r)
    \simeq
    \begin{cases}
        \displaystyle
        -\frac{\Phi_0}{2\pi r},
        &
        \xi\ll r\ll\Lambda,
        \\[10pt]
        \displaystyle
        -\frac{\Phi_0\Lambda}{2\pi r^2},
        &
        \Lambda\ll r\ll R.
    \end{cases}
\end{equation}
The zero-field energy therefore saturates with system size,
\begin{equation}
    G_v(0)
    \simeq
    E_{\rm core}
    +
    \frac{\Phi_0^2}{2\pi\mu_0\Lambda}
    \ln\frac{\Lambda}{\xi},
\end{equation}
up to nonuniversal $O(E_\Lambda)$ constants. The logarithm is generated by the
inner interval $\xi<r<\Lambda$. The magnetic field energy
$\frac{1}{2\mu_0}\int|\mathbf B_v|^2$ contributes a finite amount of order
$E_\Lambda$ but no additional logarithm.

The Meissner component is instead controlled by the global disk geometry. For
$R\gg\Lambda$, the uniform ideal-disk solution gives
\begin{equation}
    U_M(Rs)
    \simeq
    \frac{2B_{\rm ext}\Lambda}{\pi}
    \frac{s}{\sqrt{1-s^2}},
    \label{eq:UM_strong_dimensional}
\end{equation}
away from the edge boundary layer. Substituting into
Eq.~\eqref{eq:Gcross_radial_U_M},
\begin{align}
    G_{Mv}^{\rm strong}
    &=-\frac{2}{\mu_0\Lambda}\Phi_0
    \int_\xi^Rdr\,
    \frac{2B_{\rm ext}\Lambda}{\pi}
    \frac{r/R}{\sqrt{1-r^2/R^2}}
    \nonumber\\
    &=-\frac{4\Phi_0B_{\rm ext}R}{\pi\mu_0}
    \sqrt{1-\frac{\xi^2}{R^2}}.
    \label{eq:Gcross_strong_final}
\end{align}
For $R\gg\xi$,
\begin{equation}
    G_{Mv}^{\rm strong}
    \simeq
    -\frac{4\Phi_0B_{\rm ext}}{\pi\mu_0}R,
\end{equation}
and
\begin{equation}
    m_v^{\rm strong}
    \simeq
    \frac{4\Phi_0}{\pi\mu_0}R.
    \label{eq:mv_strong_final}
\end{equation}

The change from the weak-screening area law to the strong-screening linear law follows directly from the Meissner scale: $U_M\sim B_{\rm ext}R$ for $R\ll\Lambda$, whereas $U_M\sim B_{\rm ext}\Lambda$ for $R\gg\Lambda$. Substitution into Eq.~\eqref{eq:Gcross_radial_U_M} therefore gives $m_v\propto R^2/\Lambda$ in weak screening and $m_v\propto R$ in strong screening. Thus
\begin{equation}
    m_v
    \sim
    \begin{cases}
        \displaystyle
        \frac{\Phi_0}{\mu_0\Lambda}R^2,
        &
        R\ll\Lambda,
        \\[8pt]
        \displaystyle
        \frac{\Phi_0}{\mu_0}R,
        &
        R\gg\Lambda,
    \end{cases}
\end{equation}
up to the numerical coefficients derived above.

The complete strong-screening energy therefore has the schematic structure
\begin{equation}
    G_{1v}(B_{\rm ext})
    =
    E_{\rm core}
    +
    \frac{\Phi_0^2}{2\pi\mu_0\Lambda}
    \ln\frac{\Lambda}{\xi}
    -
    \frac{4\Phi_0R}{\pi\mu_0}B_{\rm ext}
    +
    G_M^{\rm strong}(B_{\rm ext}),
\end{equation}
where $G_M^{\rm strong}\propto B_{\rm ext}^2$ is the same leading Meissner
energy as in the uniform disk when the same screening operator is used.

Finally, dividing the orbital moment by the disk area gives
\begin{equation}
    M_v^{2D}
    =
    \frac{m_v}{\pi R^2}
    \simeq
    \begin{cases}
        \displaystyle
        \frac{\Phi_0}{2\pi\mu_0\Lambda},
        &
        R\ll\Lambda,
        \\[10pt]
        \displaystyle
        \frac{4\Phi_0}{\pi^2\mu_0R},
        &
        R\gg\Lambda.
    \end{cases}
\end{equation}
The two crossovers in the single-vortex problem therefore have different
origins. The zero-field vortex self-energy changes
$\ln(R/\xi)\rightarrow\ln(\Lambda/\xi)$ because $U_v$ crosses from a $1/r$
tail to a $\Lambda/r^2$ tail at the Pearl length. The orbital moment changes
from $R^2/\Lambda$ to $R$ because the field-driven Meissner potential $U_M$
is reduced from the scale $B_{\rm ext}R$ to $B_{\rm ext}\Lambda$ in the
strong-screening disk.

\begin{figure}[t]
    \centering
    \includegraphics[width=1\linewidth]{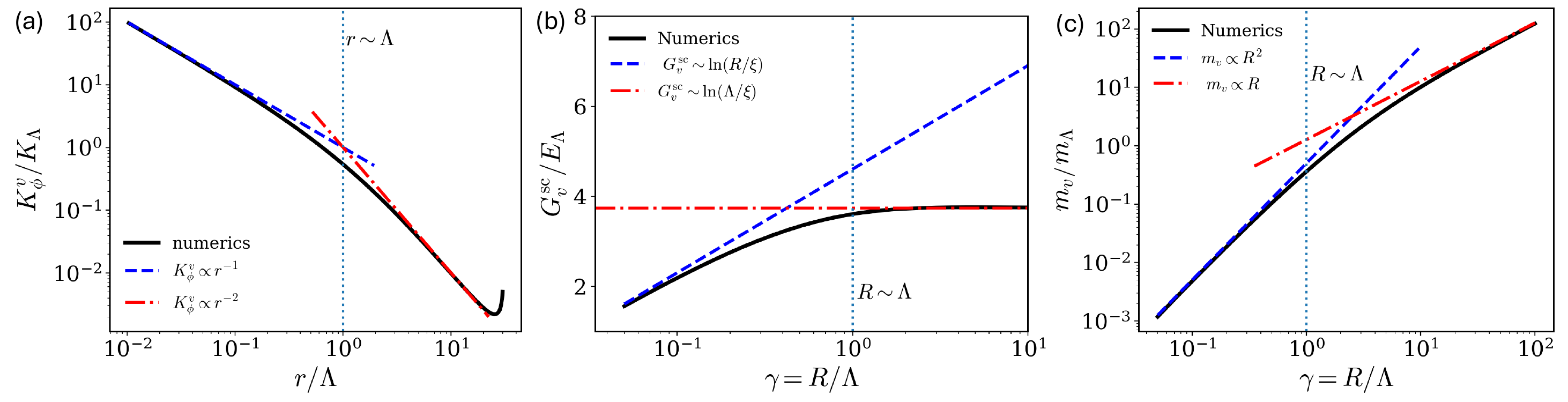}
    \caption{
    Numerical benchmark of the self-consistent Pearl screening of a centered
    vortex.
    (a) Vortex sheet current for $R=30\Lambda$. The numerical result crosses
    over from the short-distance behavior $K_\phi^v\propto r^{-1}$ to the
    screened long-distance behavior $K_\phi^v\propto r^{-2}$ around
    $r\sim\Lambda$, in agreement with the analytical results derived above.
    The deviation close to $r\sim R$ originates from the finite disk edge.
    (b) Vortex stiffness energy as a function of
    $\gamma=R/\Lambda$ at fixed $\Lambda$ and $\xi$. For $R\ll\Lambda$, the
    energy follows the unscreened logarithmic growth
    $G_v^{\rm sc}\sim\ln(R/\xi)$, while for $R\gg\Lambda$ it saturates because
    the long-distance vortex current is screened beyond the Pearl length.
    (c) Vortex orbital magnetic moment as a function of $R/\Lambda$. The
    numerical solution reproduces the crossover from
    $m_v\propto R^2$ in the weak-screening regime to
    $m_v\propto R$ in the strong-screening regime. Vertical dotted lines mark
    the crossover scale $R\sim\Lambda$.
    }
    \label{fig:vortex_screening}
\end{figure}

We finally benchmark the analytical results obtained above against the numerical
solution of the finite-disk Pearl equation. The main screening crossovers are
summarized in Fig.~\ref{fig:vortex_screening}.

Figure~\ref{fig:vortex_screening}(a) directly confirms the real-space
screening behavior derived above. At distances shorter than the Pearl length,
the electromagnetic back-reaction is weak and the vortex retains its
unscreened $1/r$ current profile. For $r\gtrsim\Lambda$, the induced vector
potential increasingly compensates the phase-winding contribution, leading to
the $1/r^2$ screened tail. The small deviation near $r\simeq R$ reflects the
finite sample boundary and is therefore absent in the infinite-film
approximation.

The same crossover is reflected in the vortex stiffness energy,
Fig.~\ref{fig:vortex_screening}(b). When $R\ll\Lambda$, the entire sample lies
inside the unscreened region and the stiffness energy grows logarithmically
with the system size, as obtained above. Once $R$ exceeds $\Lambda$, the
$1/r^2$ long-distance current renders the additional contribution from
$r>\Lambda$ convergent, so the stiffness energy approaches an
$R$-independent value controlled by $\Lambda$ rather than by the disk radius.

Finally, Fig.~\ref{fig:vortex_screening}(c) verifies the distinct crossover
of the vortex orbital response. In the weak-screening regime the
field-driven vector potential is essentially unscreened, giving the area-law
behavior $m_v\propto R^2$. In the strong-screening regime the self-consistent
Meissner response suppresses the field-driven vector potential, and the
orbital moment crosses over to the linear behavior $m_v\propto R$ derived
above.

The numerical results therefore confirm the two complementary consequences
of Pearl screening discussed in the previous sections: screening of the
vortex component $U_v$ changes the long-distance current and cuts off the
logarithmic vortex stiffness energy at the scale $\Lambda$, while screening
of the field-driven component $U_M$ changes the vortex orbital moment from an
$R^2$ to an $R$ dependence.

\end{document}